\documentclass[sn-nature]{sn-jnl}

\usepackage{graphicx}%
\usepackage{multirow}%
\usepackage{amsmath,amssymb,amsfonts}%
\usepackage{amsthm}%
\usepackage{mathrsfs}%
\usepackage[title]{appendix}%
\usepackage{xcolor}%
\usepackage{textcomp}%
\usepackage{manyfoot}%
\usepackage{booktabs}%
\usepackage{algorithm}%
\usepackage{algorithmicx}%
\usepackage{algpseudocode}%
\usepackage{listings}%
\usepackage{subfig}
\usepackage{todonotes}
\usepackage{array}
\usepackage{soul}    

\newcolumntype{C}[1]{>{\centering\let\newline\\\arraybackslash\hspace{0pt}}m{#1}}

\begin{document}

\title[AtmoRep: a stochastic model of atmosphere dynamics using large scale representation learning]{AtmoRep: A stochastic model of atmosphere dynamics using large scale representation learning}

\author[1*]{\fnm{Christian} \sur{Lessig}}\email{christian.lessig@ovgu.de}

\author[2]{\fnm{Ilaria} \sur{Luise}}\email{ilaria.luise@cern.ch}
\author[3]{\fnm{Bing} \sur{Gong}}\email{b.gong@fz-juelich.de}
\author[3]{\fnm{Michael} \sur{Langguth}}\email{m.langguth@fz-juelich.de}
\author[3]{\fnm{Scarlet} \sur{Stadtler}}\email{s.stadtler@fz-juelich.de}
\author[3]{\fnm{Martin} \sur{Schultz}}\email{m.schultz@fz-juelich.de}

\affil[1]{\orgname{Department of Computer Science, Otto-von-Guericke-Universit{\"a}t Magdeburg, Universitätsplatz 2, Magdeburg, Germany}}
\affil[*]{\orgname{now at the European Centre for Medium Range Weather Forecasting, Robert-Schumann-Platz, Bonn, Germany}}

\affil[2]{\orgname{CERN, European Center for Nuclear Research, Esplanade des Particules 1, Meyrin, Switzerland}}

\affil[3]{\orgdiv{Jülich Supercomputing Centre, Forschungszentrum Jülich, Wilhelm-Johnen-Str., Jülich, Germany}}



\abstract{
  The atmosphere affects humans in a multitude of ways, from loss of life due to adverse weather effects to long-term social and economic impacts on societies.
  Computer simulations of atmospheric dynamics are, therefore,  
  of great importance for the well-being of our and future generations~\cite{Bauer2015,Rockstrom2023aa}.
  Classical numerical models of the atmosphere, however, exhibit biases due to incomplete process descriptions and they are computationally highly demanding~\cite{Bauer2015}. 
  Very recent AI-based weather forecasting models~\cite{Pathak2022,bi2023accurate,Lam2022,chen_fengwu_2023,chen2023fuxi} reduce the computational costs but they lack the versatility of conventional models and do not provide probabilistic predictions.
  Here, we propose AtmoRep, a novel, task-independent stochastic computer model of atmospheric dynamics that can provide skillful results for a wide range of applications.
  AtmoRep uses large-scale representation learning from artificial intelligence~\cite{Devlin2019,Radford2018} to determine a general description of the highly complex, stochastic dynamics of the atmosphere from the best available estimate of the system's historical trajectory as constrained by observations~\cite{Hersbach2020}.
  This is enabled by a novel self-supervised learning objective and a unique ensemble that samples from the stochastic model with a variability informed by the one in the historical record.
  The task-independent nature of AtmoRep enables skillful results for a diverse set of applications without specifically training for them and
  we demonstrate this for nowcasting, temporal interpolation, model correction, and counterfactuals. 
  We also show that AtmoRep can be improved with additional data, for example radar observations, and that it can be extended to tasks such as downscaling. 
  Our work establishes that large-scale neural networks can provide skillful, task-independent models of atmospheric dynamics.
  With this, they provide a novel means to make the large record of atmospheric observations accessible for applications and for scientific inquiry, complementing existing simulations based on first principles.
}





\maketitle

\section*{Main}
\label{sec:main}

The atmosphere and its dynamics have a significant impact on human well-being.
Adverse weather effects led to the loss of over 2 million lives in the last 50 years and caused economic damages of more than $4.3$ trillion dollars~\cite{WMO_2023}.
The weather also influences many daily aspects of our societies, such as agricultural decision making, the efficiency of industrial processes,
or the availability of renewable energies such as solar and wind power.
The atmosphere, furthermore, plays a critical role for Earth's climate and hence for our understanding of and adaptation to climate change. 
An accurate and equitable modeling of atmospheric dynamics is consequently of critical importance to allow for evidence-based decision making that improves human well-being and minimizes adverse impacts for current and future generations~\cite{UN-SGD-2015}.
\begin{figure}[t]
    \centering
    \includegraphics[width=\textwidth]{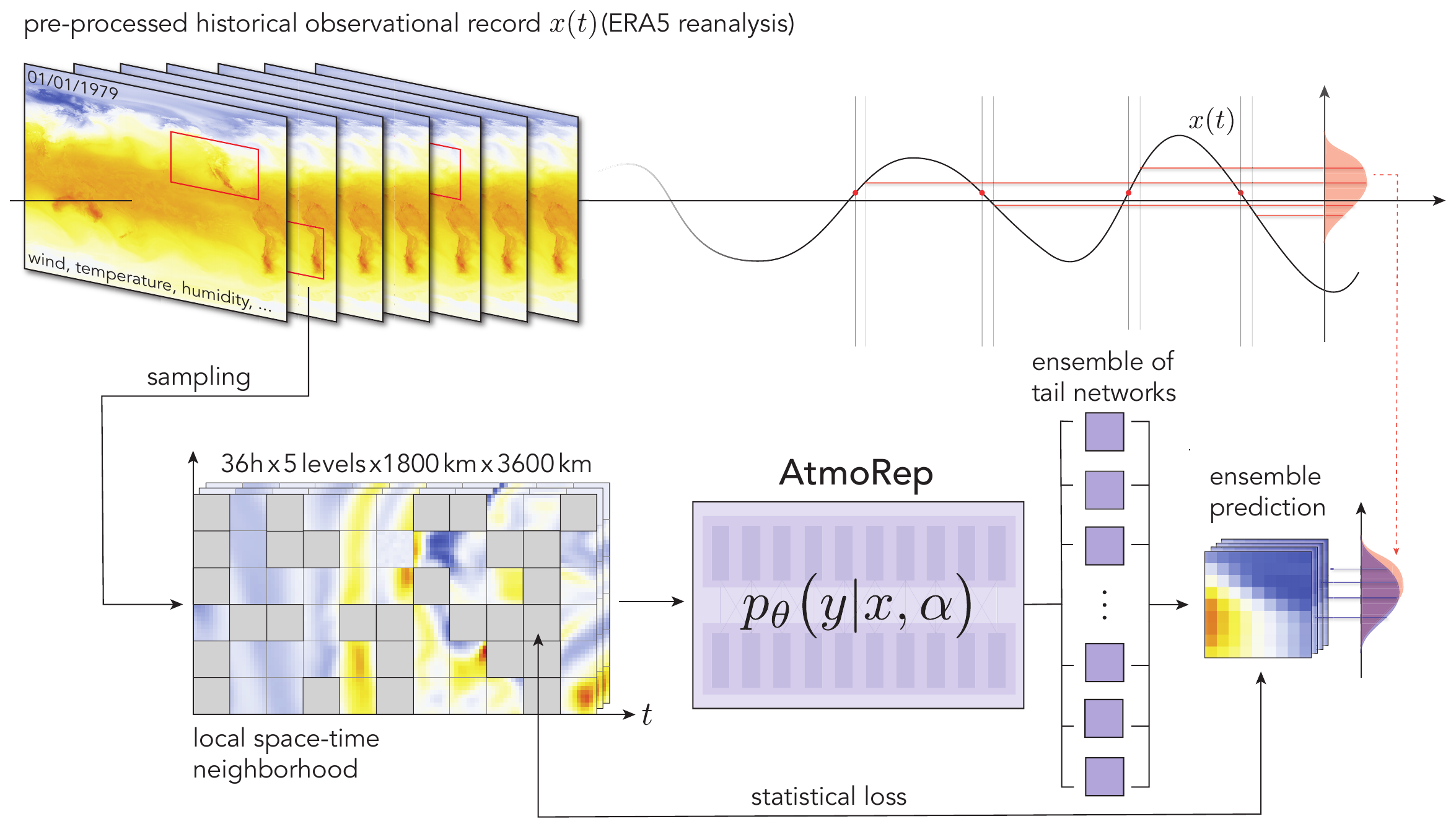}
    \caption{\small The AtmoRep model provides a numerical representation $p_{\theta}( y \vert x , \alpha)$ of the conditional probability $p(y \vert x , \alpha)$ for atmospheric states $x$,$y$ subject to external conditions $\alpha$, e.g. the time of $x$,$y$ or their location on the globe. It is implemented as a transformer neural network with $3.5$ billion parameters and trained from the ERA5 reanalysis (top left). For training, local space-time neighborhoods are randomly sampled. The neighborhoods are subdivided into smaller patches, called tokens, and the self-supervised learning task is to reconstruct randomly masked or distorted patches (bottom left, gray patches). An ensemble of prediction heads is used to sample from the AtmoRep core model 
    and provide probabilistic predictions for possible states consistent with the un-masked tokens (bottom right). 
The ensemble spread that is learned during training arises from the intrinsic variability of the data, i.e. that similar atmospheric states $x(t)$ have different associated states $y$, for example with a fixed offset in time (top right).}
    \label{fig:atmorep-concept}
\end{figure}

Classical models for atmospheric dynamics are based on the fundamental laws of physics, e.g. conservation of mass and energy~\cite{Holm2002,Durran2010}.
Because the resulting equations cannot be solved analytically, computer simulations play a central role in describing the dynamics~\cite{Charney1950b,Silberman1954,Platzman1960,Arakawa1981}.
Despite tremendous progress in the last decades~\cite{Bauer2015}, current simulations still exhibit deficiencies in describing relevant physical processes~\cite{Palmer2019,Palmer2019b,Bauer2021,Balaji2022}, leading, for example, to inaccurate representations of extreme events with strong adverse impacts. 
Current simulations also suffer from high computational and energy costs.
%

In our work, we develop a novel yet powerful approach to modeling atmospheric dynamics that combines the large observational record with the rapid advances in artificial intelligence~\cite{LeCun2015,Bengio2021}.
We use large scale representation learning~\cite{Devlin2019,Radford2018}, a state-of-the-art methodology in machine learning, to train a statistical model of atmospheric dynamics that is task-agnostic and can be used for a wide range of applications with either no or only little task-specific additional training.
The model, called AtmoRep, is given by a large generative neural network with $3.5$ billion parameters and trained using the ERA5 reanalysis~\cite{Hersbach2020}, 
which provides the most complete assimilation of observations into an estimate of the historical trajectory of the atmosphere that is available.
Through the training with observation-based data, AtmoRep can learn effects and dynamics that are present in these but very complex or computationally expensive to model using traditional approaches.
We demonstrate the versatility and utility of AtmoRep through skillful nowcasting, temporal interpolation, and model correction as well as the generation of counterfactuals, for example how an atmospheric state would have evolved in a different year or region.
These intrinsic capabilities of AtmoRep can be achieved without task-specific training and they hence provide an analogue of the zero-shot abilities that have first been observed for foundational models in natural language processing~\cite{Brown2020}. AtmoRep generalizes these for the first time to Earth system science.
We also demonstrate how our model can be extended for other tasks, e.g. with a task-specific tail network, by using it for downscaling where we achieve highly competitive results.
Finally, we show that AtmoRep can be bias corrected using observational data to further improve the representation of the dynamics in the network.

AtmoRep uses a flexible and versatile neural network architecture that can be employed regionally or globally and with different physical fields. 
This improves over existing large-scale AI-based weather forecasting models that are inherently global and use a fixed number of variables.
An accurate and robust representation of the dynamics is learned in AtmoRep using a novel self-supervised training protocol that extends existing ones~\cite{Devlin2019,He2022} to four-dimensional space-time and that is one of the keys to AtmoRep's intrinsic capabilities.
A further innovation is AtmoRep's ensemble that has a variability that derives from those in the training data.
Through this, it differs in an essential way from existing, perturbation-based ensemble methods in both conventional and AI-based models, e.g.~\cite{Bauer2015,Palmer2019b,Weyn2021,Bi2023aa}.

AtmoRep demonstrates, for the first time, the principle and potential of AI-based, task-agnostic atmospheric models as a complement to traditional ones, such as general circulation models. 
As a statistical approach, AtmoRep generates samples from the learned distribution, which is derived from the available observational record and hence can include effects and phenomena that are not modeled by existing equation-based simulations.
We believe that a further development of AtmoRep will allow for the methodology to become an important tool in a wide range of applications where atmospheric dynamics play a role.

\subsection*{Stochastic modeling of atmospheric dynamics}

For our work, we build on the description of the atmosphere as a stochastic dynamical system, cf.~\cite{Lorenz1969,Hasselmann1976,palmer_nonlinear_2001,Palmer2008,Palmer2008_1,Palmer2019}. 
The dynamics are, in principal, determined by the deterministic laws of classical mechanics and thermodynamics.
A stochastic modeling is, however, appropriate because of the strong sensitivity of the time evolution on the initial conditions, practical limits on the availability of observations to constrain these~\cite{Carrassi2018}, and a wide range of small scale process whose feedback is best represented stochastically~\cite{Hasselmann1976,palmer_nonlinear_2001}.

Given an approximate atmospheric input state $\tilde{x}$, we thus model physically consistent states $\tilde{y}$ with the probability distribution 
\begin{align}
  \label{eq:stochastic}
  \tilde{p}\big( \tilde{y} \vert \tilde{x} , \tilde{\alpha} \big) .
\end{align}
For example, $\tilde{y}$ can be a future state for a given initial condition $\tilde{x}$; alternatively, $\tilde{y}$ and $\tilde{x}$ can be local states defined at the same time but at different locations.
The external conditions $\tilde{\alpha}$ complement $\tilde{x}$ and can describe, for instance, its year or boundary conditions such as global forcings.
Eq.~\ref{eq:stochastic} is more abstract than, for example, models based on partial differential equations.
However, this allows the model to also include processes that are difficult to capture with other approaches.

Since Eq.~\ref{eq:stochastic} is a highly complex, instationary probability distribution with no known analytic description, we introduce the approximation
\begin{align}
  \label{eq:model:numerical}
  p_{\theta}\big( y \vert x , \alpha \big) \approx  \tilde{p}\big( \tilde{y} \vert \tilde{x} , \tilde{\alpha} \big)
\end{align}
where $p_{\theta}( y \, \vert x , \alpha )$ is a large, generative transformer neural network~\cite{Vaswani2017} with $3.5$ billion parameters $\theta$.
The network provides a general, task-agnostic, stochastic model of atmospheric dynamics that we refer to as AtmoRep, see Fig.~\ref{fig:atmorep-concept} for an overview.

The AtmoRep model $p_{\theta}( y \, \vert x , \alpha )$ is determined by pre-training on observation-based data, specifically the ERA5 reanalysis~\cite{Hersbach2020}.
This enables $p_{\theta}( y \, \vert x , \alpha)$ to include effects and dynamical behavior that are contained in the data but are difficult or computationally expensive to model using first principles.
To learn a physical and stochastically consistent model, AtmoRep provides ensemble predictions for the state $y$.
The ensemble is trained from only the single, high-resolution trajectory in ERA5 so that its spread reflects the intrinsic variability in the training data (Fig.~\ref{fig:atmorep-concept} top-right).
%
For AtmoRep to be an unbiased estimate of the true distribution, we employ a self-supervised pre-training objective that minimizes the distance $\mathcal{D}( \tilde{p}( y , x , \alpha ) , p_{\theta}( y , x , \alpha ) )$ between the data distribution $\tilde{p}( y , x )$ and $p_{\theta}( y , x , \alpha )$ with a Monte Carlo estimate over the training data set.

The input to AtmoRep is an atmospheric state $x$ given by wind velocity (or vorticity and divergence), vertical velocity, temperature, specific humidity and total precipitation
in a local space-time neighborhood of, for example, $36 \, \mathrm{h} \times 5 \, \textrm{vertical levels} \times 1800 \, \mathrm{km} \times 3600 \, \mathrm{km}$, respectively (Fig.~\ref{fig:atmorep-concept}, left).
In applications, the network can operate on different neighborhood sizes than during pre-training and the modular design of AtmoRep allows for task-specific configurations with different physical fields, see the Methods section.
For processing by the transformer-based neural network, the space-time neighbourhood is tiled into smaller patches, which are known as tokens.
The label-free, self-supervised pre-training objective is to provide ensemble predictions for a randomly selected subset of the tokens that are masked or distorted (see Fig.~\ref{fig:atmorep-concept}, bottom-left).
The $4$-dimensional masking with large masking ratios of up to $0.9$ enables the network to learn the general relationship of local atmospheric information in space and time, and hence of atmospheric dynamics.
%
Further details on the network architecture of AtmoRep and its training are presented in the Methods section and in the supplementary material.

%
%
%
%
%

\subsection*{Intrinsic Capabilities}

%

\begin{figure}[h!]
    \centering
    \includegraphics[width=\textwidth]{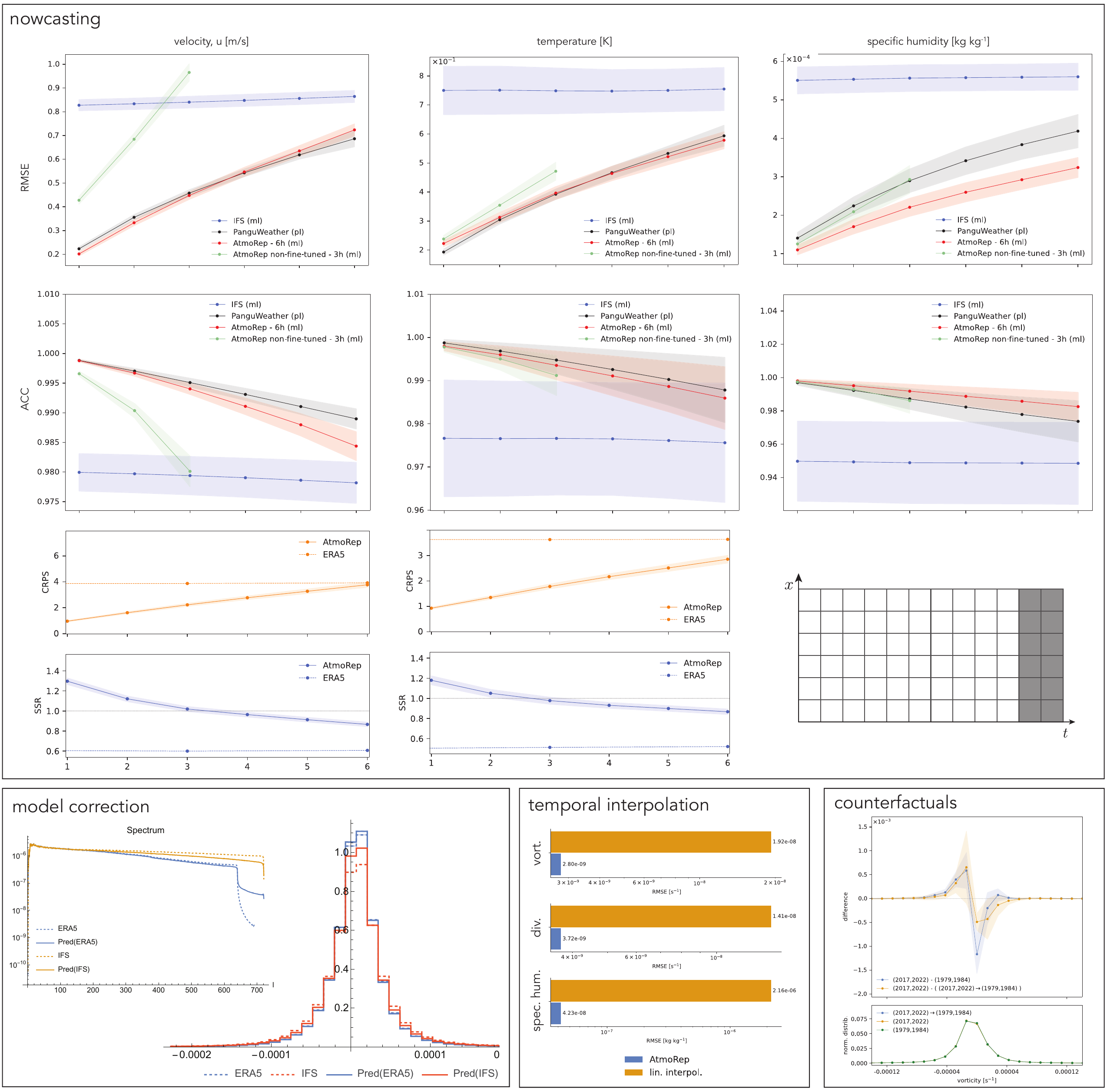}
   \caption{\small 
   AtmoRep can be used for a diverse set of applications without task-specific training 
   (shaded areas depict one standard deviation).
     \emph{Nowcasting:} Short-term forecasting can be realized by masking tokens at the future-most time step(s) (bottom right inset). Skill is compared to Pangu-Weather and ECWMF's IFS for zonal velocity, temperature and specific humidity. 
     AtmoRep results are shown for a pre-trained model and one with modest fine-tuning for the task.
    \emph{Model correction:} AtmoRep is robust for out-of-distribution input. We exploit it for model correction by using output from IFS as input to AtmoRep. Our model faithfully handles the data, preserving the higher frequency content (top left), and shifts the distribution towards the ERA5 one (right).
    %
    \emph{Temporal interpolation:} Temporal interpolation is accomplished by masking tokens in the middle of the temporal domain. Performance is compared to linear interpolation.
    \emph{Counterfactuals:} Using initial conditions from, e.g., the period $(2017,2022)$ but prescribed as being from $(1979,1984)$ by using the external conditions $\alpha$ allows for the generation of counterfactuals. The plot shows the difference between the original and the counterfactual distributions, as well as the shape of the full distributions.
      }
    \label{fig:intrinsic}
\end{figure}

AtmoRep's model formulation as $p_{\theta}\big( y \, \vert x , \alpha \big)$, i.e. as a numerical approximation for the probability distribution $\tilde{p}\big( \tilde{y} \vert \tilde{x} , \tilde{\alpha} \big)$ over atmospheric states, intrinsically includes a variety of relevant applications that can be implemented directly using a pre-trained model.
For example, when $y$ is in the future with respect to $x$ then $p_{\theta}\big( y \, \vert x , \alpha \big)$ becomes a forecasting model; when $y$ corresponds to missing information within $x$ in space or time, then the model performs spatio-temporal interpolation; and when $x$ is output from an equation-based simulation, AtmoRep can be used to correct it towards the observationally better-constrained ERA5.
The task is in each case implemented through the masked token model by specifying the tokens corresponding to the sought after information $y$ as masked in the input to AtmoRep, see cf. Fig.~\ref{fig:atmorep-concept}, bottom left. The model's prediction then provides the estimate for $y$.
When an input state $x$ is used with incorrect but statistically consistent external information $\alpha$, then AtmoRep also allows for the generation of counterfactuals, that is, for example, a prediction of how $x$ would have evolved in a different historical regime or at a different location. 
Atmorep serves in this case as a statistical sample generator whose distribution is controlled by the external conditions $\alpha$.
The foregoing applications can be realized with AtmoRep with only a pre-trained model and without task-specific training since they are implicitly contained in the pre-training objective, which is designed to learn $p_{\theta}( y \, \vert x , \alpha )$ using the extended masked token model.
We therefore refer to these tasks as intrinsic capabilities.
They are the analogue of the zero-shot abilities of large language models~\cite{Brown2020} that are also tasks implicitly contained in the training objective (e.g. translation or text completion).
%
A summary of AtmoRep's skill for different intrinsic capabilities is presented in Fig.~\ref{fig:intrinsic} and they are discussed below. 
Experimental protocols and more detailed evaluations are provided in the Extended Data section and the supplementary material.

\paragraph{Nowcasting}

AtmoRep can be used for probabilistic nowcasting, i.e. short-term forecasting, when $y$ is a future state with respect to $x$. 
This is implemented by masking all tokens at the last time step(s) in the space-time cube that forms the input to the network, see Fig.~\ref{fig:intrinsic}.
AtmoRep has skill for the task directly after pre-training through the masked token model but the skill can be improved by fine-tuning. 
To quantify AtmoRep's nowcasting abilities, we compared to ECMWF's Integrated Forecasting System (IFS) and Pangu-Weather~\cite{bi2023accurate}.
For deterministic forecasting skill, root mean square error (RMSE) and the anomaly correlation coefficient (ACC) were computed.
Fig.~\ref{fig:intrinsic} shows that AtmoRep attains performance comparable to Pangu-Weather with better performance in particular for very short forecast horizons and in selected variables such as specific humidity.
Zero-shot performance is thereby worse than after fine-tuning but still improves over the IFS at very short times.
 Fig.~\ref{fig:intrinsic} also shows the continuous ranked probability score (CRPS) for AtmoRep and ERA5, the latter computed using the ERA5 ensemble that is available at 3-hour time resolution.
The results demonstrate that AtmoRep has comparable or slightly better probabilistic nowcasting skill.
Results for other variables as well as for spread-skill ratio (SSR) are available in the Extended Data section in Figs.~\ref{fig:ext-fceval-RMSE}-~\ref{fig:ext-fcmap} where we also show visualizations of a forecast.
Overall, our results demonstrate that AtmoRep has state-of-the-art nowcasting performance with no or very little task specific training.
Compared to the IFS, AtmoRep has computational and energy costs that are significantly lower.

\paragraph{Temporal interpolation}
Temporal interpolation refers to the task of (re-)creating atmospheric state data with a higher temporal resolution than the input. It is of importance, for example, for the compression of weather and climate datasets. 
With AtmoRep, it can be realized by masking tokens within the space-time cube that is the input to the network.
As presented in Fig.~\ref{fig:intrinsic}, the model shows substantially better skill, quantified as one order of magnitude lower RMSE, in reconstructing the $3$ time steps within a $3\, \mathrm{h}$ wide token compared to linear interpolation. In the supplementary material we show a comparison for additional variables and also for different Multiformer configurations.

\paragraph{Model correction}

AtmoRep's internal representation of atmospheric dynamics is sufficiently robust and general that the model can take as input data from a related but different distribution than those seen during pre-training. 
We demonstrate this with data from ECMWF's operational Integrated Forecast System (IFS), which has a substantially higher resolution than the training data and whose distribution differs also in other aspects.
Since AtmoRep is trained to predict ERA5, it will as output provide data that is consistent with ERA5 given the input. 
This amounts to model correction of IFS data towards the observationally better constrained ERA5 reanalysis.
In Fig.~\ref{fig:intrinsic}, bottom left, we show that the AtmoRep prediction with IFS input is corrected towards ERA5. 
The correction has deficiencies, e.g. due to the imprint of the initial conditions and since the training is imperfect, but a clear trend can be observed.
The figure also shows the substantially higher frequency content of IFS data and that AtmoRep partially reproduces these higher frequencies, despite not having encountered such high-frequency content during pre-training.
Further results are provided in the Extended Data section in Fig.~\ref{fig:ext-ifs-correct}.


\paragraph{Counterfactuals}

In weather and climate research, counterfactuals are a methodology to answer ``what if'' questions. They play a central role for example for the attribution of human impacts on extreme weather events~\cite{vautard_attribution_2016} or to obtain more robust statistics on the possible evolution and outcomes of such events~\cite{woo_counterfactual_2021}.
In AtmoRep, next to an initial condition $x$ also the external conditions $\alpha$ are provided to the model. 
This can be used for the generation of counterfactual scenarios by using together with a given physical $x$ (e.g. from ERA5) an alternative external information $\hat{\alpha}$.
For AtmoRep $p_{\theta}(y \vert x , \alpha)$ to be applicable, the initial condition has to be statistically consistent with $\hat{\alpha}$, i.e. it should be possible that $x$ occurred in, for example, the year specified by $\hat{\alpha}$.
Furthermore, the AtmoRep network must have learned a robust representation of the dependence of atmospheric dynamics on $\alpha$, which is not an explicit training objective.

To demonstrate the generation of counterfactuals with AtmoRep, we consider vorticity close to the surface, i.e. at model level $137$. This variable shows a clear distributional shift in the ERA5 dataset between the early ERA5 years, i.e. 1979-1984, and the later ones, i.e. 2017-2022, but without a fundamental change in the support of the distribution.
We perform the counterfactual experiment with nowcasting with initial conditions from the late years and $\hat{\alpha}$ that prescribes the earlier time range. 
We denote this as $(2017,2022) \rightarrow (1979,1984)$.
As control experiment, we also perform nowcasting for both time ranges with the correct $\alpha$, see Extended Data Fig.~\ref{fig:ext-counterfac} for a visualization of the methodology.
If AtmoRep had not learned a dependence on $\alpha$, the result of the counterfactual experiments would be statistically identical to the control experiments for $(2017,2022)$, i.e. no distributional shift could be observed; in a histogram difference plot, as shown, the difference would vanish.
In contrast, Fig.~\ref{fig:intrinsic}, bottom right, shows that the counterfactual experiment leads to a distributional shift which is similar to the actual one in the ERA5 data, albeit with a smaller magnitude.
The deficiencies are likely due to the imprint from the initial conditions that cannot be fully removed in a short-term forecast and from the learned model that not perfectly captures the target distribution.
Nonetheless, our results demonstrate, for the first time, that AI-based models can be used for counterfactuals within the learned distribution. 
\subsection*{Extension of AtmoRep to other applications}

Next to the tasks that AtmoRep can perform intrinsically, the model can also be extended to accomplish other applications, e.g. by adding a task-specific tail network. 
This still allows to exploit the pre-trained model and its skill and, for example, reduces the task-specific training time.
To demonstrate the principle, we consider downscaling, i.e. mapping a low resolution spatio(-temporal) distribution to a higher resolution one.
With AtmoRep we can realize downscaling by factoring the target distribution as
\begin{align}
  p_{\theta}\big( y_{h} \vert x \big) = p_{\theta'}\big( y_{h} \vert y \big) \, p_{\theta}\big( y \vert x , \alpha \big) .
\end{align}
where $p_{\theta}\big( y \vert x , \alpha \big)$ is the pre-trained AtmoRep model and $p_{\theta'}\big( y_{h} \vert y \big)$ is the downscaling-specific network that maps to samples $y_h$ from the high-resolution target distribution.
We also use a transformer for $p_{\theta'}\big( y_{h} \vert y \big)$, see the supplement for details.

\begin{figure}
    \centering
    \includegraphics[width=\textwidth]{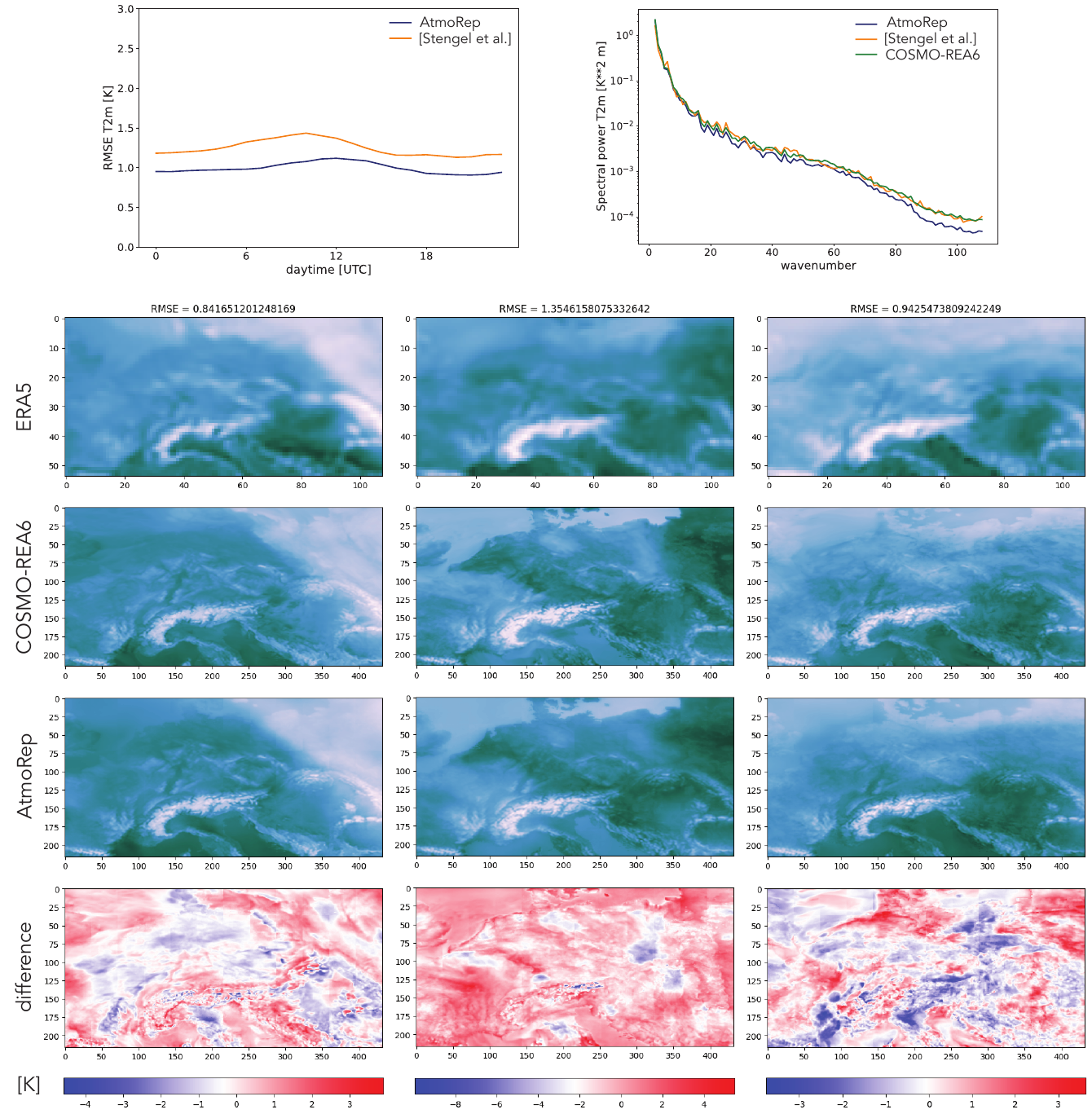}
    \caption{Results for downscaling from ERA5 to temperature at $2\,\mathrm{m}$ in the COSMO REA6 dataset for a region in central-eastern Europe. 
    For AtmoRep, zonal and meridional velocities as well as temperature were used as input (at model level $137$, approximately $1000 \, \mathrm{hPa}$). 
    The top row shows the RMSE as well as the spectrum compared to the results obtained with the GAN proposed by Stengel et. al~\cite{stengel_adversarial_2020} and retrained for our setup (see the supplementary material for details). At the bottom we show three examples for downscaled fields (third row) as well as the ERA5 input (top row) and the COSMO REA6 reference (second row). Also shown is the difference between COSMO REA6 and the downscaled field provided by AtmoRep (bottom row).}
    \label{fig:ext-downscaling}
\end{figure}

To demonstrate the approach, we consider $2\,\mathrm{m}$ temperature from the COSMO REA6 reanalysis \cite{Bollmeyer2015} that has $4$-times the resolution of ERA5 and that shows improved physics in particular over steep terrain.
As baseline we use the GAN-based statistical downscaling model by Stengel et al.~\cite{stengel_adversarial_2020}. 
The results in Fig.~\ref{fig:ext-downscaling} show that AtmoRep outperforms the GAN by Stengel et al. substantially in RMSE although its spectrum is slightly too low for very high frequencies.
The three examples in the figure, furthermore, establish that AtmoRep not only increases the resolution but also adjusts the distribution, see e.g. the left most example in Fig.~\ref{fig:ext-downscaling} where the front over Eastern Europe is substantially further East in ERA5 than in COSMO REA6 and AtmoRep corrects this to high accuracy.

\subsection*{Bias correction of AtmoRep with observational data}
 
\begin{figure}
    \centering
    \includegraphics[width=\textwidth]{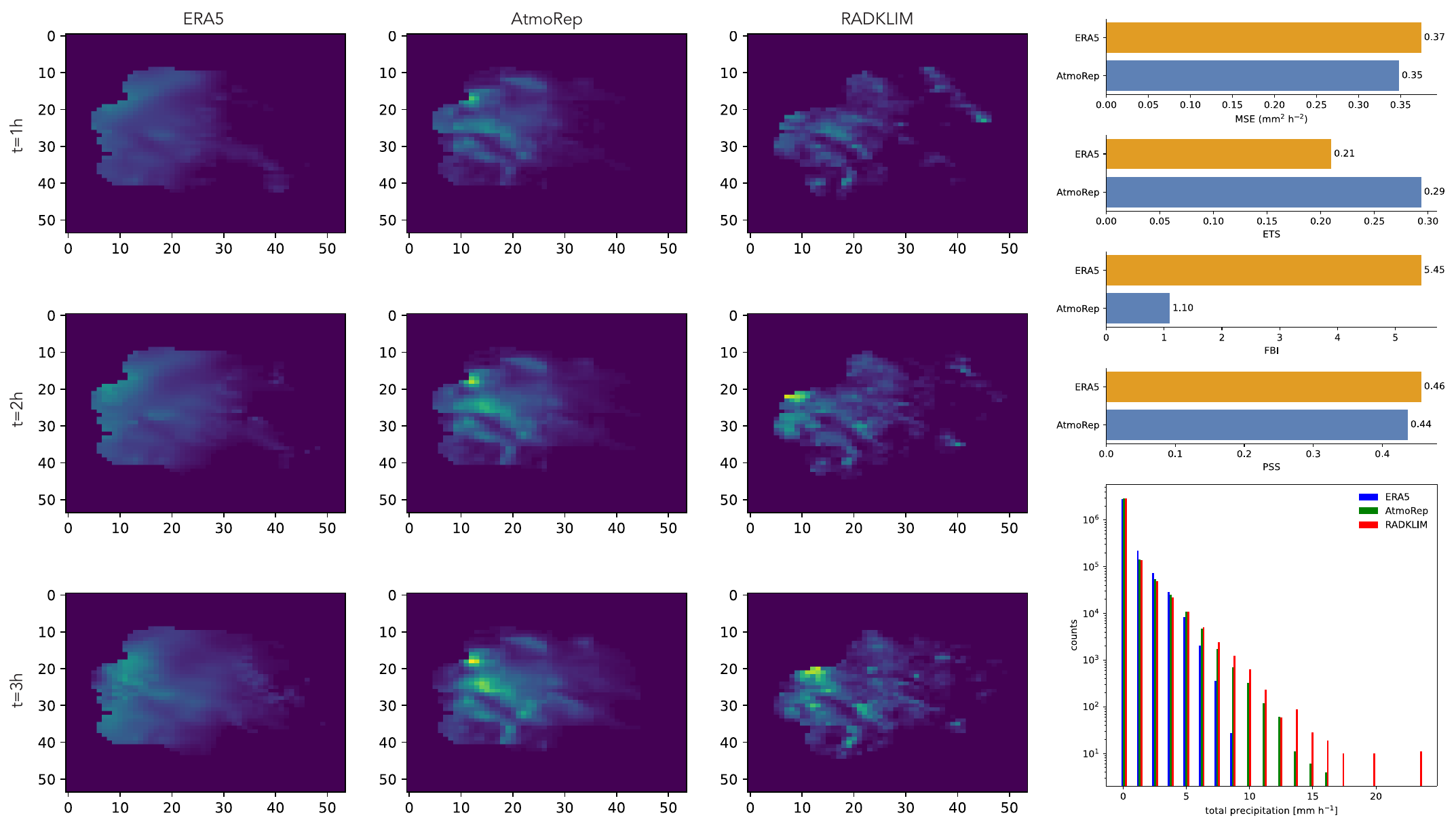}
    \caption{\textit{Left:} precipitation forecast for ERA5 (left), AtmoRep fine-tuned (center) and RADKLIM (right) for a 3h forecast in 2019. \textit{Right:} Comparison between the mean square error (MSE), Equitable Threat Score (ETS), Peirce Skill Score (PSS) and Frequency Bias Indicator (FBI) in ERA5 and the fine-tuned AtmoRep, using RADKLIM data as ground truth obtained averaging yearly predictions from 2019. The bottom right part shows the distribution of hourly accumulated total precipitations for AtmoRep, ERA5 and RADKLIM. 
    }
    \label{fig:atmorep-radklim}
\end{figure}

ERA5 has known deficiencies and biases, e.g.~\cite{Hersbach2020,foli_evaluation_2022,hoffmann_assessment_2022,choudhury_evaluation_2023}, and this limits the potential of AI-based models that are trained on reanalyses or simulation data~\cite{ben-bouallegue_rise_2023}. 
With AtmoRep, we can use observational data to improve the model and remove biases introduced through the ERA5 training distribution. 
To demonstrate this, we use precipitation radar data from the RADKLIM dataset \cite{Winterrath2018}, pre-processed to match the ERA5 resolution.
Since the RADKLIM domain is smaller than the one used during pre-training, we use $12 \times 6 \times 6$ tokens instead of $12 \times 6 \times 12$ as input per level, exploiting the flexibility of AtmoRep's token-ized input.
Missing values in the observational data were ignored in the loss computation for the bias correction and the training task was the prediction of the RADKLIM precipitation field given ERA5 data as input.
This is equivalent to training for precipitation nowcasting.
We evaluate the model by comparing the output distribution to the RADKLIM one using different metrics. 
As shown in Fig.~\ref{fig:atmorep-radklim}, left, the precipitation fields of AtmoRep show enhanced skill compared to the original ERA5 predictions and the areal extent and shape of the precipitation fields that are forecast by AtmoRep are much closer to RADKLIM than the original ERA5 data.
The maximum rainfall intensity remains lower than in RADKLIM, but it is improved compared to ERA5.

\subsection*{Conclusion}
\label{sec:conclusion}

AtmoRep is a novel, task-independent stochastic model of atmospheric dynamics.
It provides an alternative methodology to make the observational record available and has skill for a range of applications with no or only little task-specific training.
It is realized by a large generative neural network pre-trained on the ERA5 reanalysis using a new self-supervised training objective.
AtmoRep, furthermore, employs a novel ensemble that samples from the stochastic model and whose spread reflects the variability of the training data distribution.
We demonstrated the intrinsic zero-shot capabilities of AtmoRep with nowcasting, temporal interpolation, and model correction.
With moderate fine-tuning, our nowcasting performance is comparable to existing forecasting models, including ECMWF's IFS and Pangu-Weather.
We also demonstrated, for the first time, the ability to perform counterfactuals with an AI-based model, exemplifying AtmoRep's ability to serve as sample generator from the highly complex and instationary learned distribution.

AtmoRep opens up many avenues for future work. 
We believe that, through their generality and computational efficiency, large scale representation models can play a significant role in the next generations of Earth system models, complementing existing ones for example when large ensembles are needed or for tasks such as counterfactuals.
AtmoRep can also be extended as a parameterization for general circulation models where, through its training on observation-based data, it has the potential to help address the closure problem that is a major source of uncertainties in existing weather and climate simulations~\cite{Bauer2015,Balaji2022}.
Its training on data makes AtmoRep also amenable for the assimilation of different datasets into a coherent representation.
This is a particular promising direction when AtmoRep is developed further so that learning from nearly unprocessed observations becomes possible, extending what we already demonstrated for precipitation bias correction.
The masked token model used for pre-training requires AtmoRep to fill in missing values in the physical fields that are input to the model. When the masking is modified so that it is no longer per-token, this amounts to data assimilation as required, for example, for the initialisation of numerical weather prediction models.
We also believe that AtmoRep can become an important tool for scientific inquiry~\cite{Wang:2023aa}, similar to how general circulation models are currently a central tool in atmospheric science.
For example, the counterfactuals introduced in the present work provide a means to study how the temporal distributional shift in the training dataset affects specific weather patterns.

AtmoRep demonstrates the potential of large scale representation learning in atmospheric science and the results in the present work are, in our opinion, only a first step towards the possibilities that are enabled by the methodology.

\newpage

\section{Methods}
\label{sec:methods}

\subsection{Datasets}
To train the AtmoRep model, the ERA5 reanalysis dataset~\cite{Hersbach2020} was used with an hourly temporal resolution and ERA5's default equi-angular grid with $721 \times 1440$ grid points in space. 
In the vertical dimension, we employed model levels $96$, $105$, $114$, $123$, $137$, corresponding approximately to pressure levels 546, 693, 850, 947, and 1012~hPa. We used model levels so that the physical fields are valid everywhere and they do not cut through orographic features.
As variables, zonal and meridional wind components, vorticity, divergence, vertical velocity, temperature, specific humidity, and total precipitation were employed.

For model correction, we employed output of the operational Integrated Forecasting System (IFS) by ECMWF as of 2020.
The COSMO REA6~\cite{Bollmeyer2015} provided higher resolutions data for downscaling. 
This dataset was remapped onto an equiangular grid with $4$-times the resolution of the ERA5 reanalysis dataset.
For precipitation bias correction, we employed the RADKLIM dataset~\cite{winterrath_overview_2019}, which represents gauge-adjusted precipitation estimates from the German radar network.
It was remapped to the ERA5 equiangular grid on its domain with an hourly temporal resolution.
For the remappings we employed a first order, conservative re-mapping method.

All datasets were normalized to zero mean and unit variance either globally per month or on a per grid point basis. 
See the supplementary material for further details on data handling and pre-processing.

\subsection{Model formulation}

%

AtmoRep provides a task-independent, numerical stochastic model $p_{\theta}( y \vert x , \alpha)$ of the dynamics of the atmosphere, i.e. a foundational model~\cite{Bommasani2022} for atmospheric dynamics.
It is realized by an encoder-decoder transformer neural network with dense attention~\cite{Vaswani2017}, see Extended Data Fig.~\ref{fig:multiformer_overview} for an overview of the architecture.
The input to the model is an atmospheric state $x$ in a local neighborhood in $4D$ space-time, subdivided into smaller sub-regions that form the tokens the transformer operates on (Fig.~\ref{fig:atmorep-concept}, bottom-left).
Working with a local input from anywhere on the globe allows the model to learn position-independent, general principles of atmospheric dynamics.
Through the external information $\alpha$ that include the global position, the network is, nonetheless, able to also learn location-dependent effects, see Extended Data Fig.~\ref{fig:ext-error-season} and Fig.~\ref{fig:stdev_map} for specific examples of such local effects.
The temporal information in $\alpha$ enables the model to, furthermore, learn instationary behavior in time, for example shifts in the training data distribution or seasonal effects.
Because the network input consists of a set of tokens, the trained AtmoRep model can be flexibly used for space-time regions that are smaller or larger than the training ones. 
For example, the spatial extent of the RADKLIM radar dataset 
is smaller than those spanned by the $6 \times 12$ tokens used during pre-training. 
Therefore, when fine-tuning for bias correction we use $6 \times 6$ tokens in space.
Similarly, since different vertical levels correspond to different tokens, also the number of levels at evaluation time can differ from that during training. 
We exploit this for downscaling where we only employ the model level closest to the surface.
The supplementary material contains a quantitative evaluation of how changes in the neighborhood size effect the model performance; improvements are possible by fine-tuning for a change in size. 
Global forecasts, such as the ones shown in Extended Data Fig.~\ref{fig:ext-fcmap}, can be realized with the local AtmoRep model by tiling the globe. With overlap between adjacent regions, we can avoid artifacts in forecasts due to the tiling.
The flexibility to employ AtmoRep as a local or a global model is a unique feature of our approach compared to other large scale AI-based forecasting models in the literature. 

In the AtmoRep network architecture, we use one encoder-decoder transformer per physical field to respect the different properties of the fields that comprise an atmospheric state. 
For instance, temperature in ERA5 changes much more slowly and has less spatial variability than vorticity so that we use a larger token size and smaller embedding dimension for it.
The individual per-field transformers are coupled through cross-attention to allow for interactions between fields in the model (Extended Data Fig.~\ref{fig:multiformer_overview}).
We call this architecture the Multiformer. 
Our approach to couple individual per-field transformers provides the advantage that fields can be pre-trained independently and then combined into a multi-field model.
This is more efficient than training a multi-field one from the outset because the computational costs of dense attention scale quadratically with the number of tokens. 
The Multiformer design also creates flexibility to combine pre-trained per-field transformers into application-specific models. 
For example, for downscaling we use a $3$-field configuration with only the wind components and temperature, which are the most relevant variables for this problem.
Usually a few training epochs are sufficient for individually pre-trained per-field transformers to cohere into a skillful multi-former model.
%
Further details on the model can be found in the supplementary material.

\paragraph{Ensemble}
AtmoRep uses an ensemble to provide an nonparametric representation of the conditional probability distribution over the output state $y$.
For each physical field, the ensemble is generated by a set of prediction heads, each consisting of only a linear layer.
These heads maps from the latent, internal representation in the AtmoRep core model $p_{\theta}( y \vert x , \alpha )$ to the grid representation of the physical field in space-time, see Fig.~\ref{fig:atmorep-concept}, bottom right. 
Conceptually, the prediction heads hence sample states $y$ from $p_{\theta}( y \vert x , \alpha )$ to provide the nonparametric representation of their distribution.
In all of our experiments, we used an ensemble size of $16$ except for downscaling where it was $4$.
The ensemble is trained with a novel statistical loss function (see below) on only the single, deterministic ERA5 high-resolution trajectory.
The ensemble spread consequently derives solely from the spatio-temporal variability in the ERA5 training data and hence, at least partially, from the intrinsic one in the observational record, see Fig.~\ref{fig:atmorep-concept}, top-right.
The training methodology with an ensemble is an integral aspect of our approach to obtain a stochastic model that provides an unbiased estimator for the probability distribution of the physical system.
AtmoRep's ensemble differs in an essential way from existing ones in numerical weather prediction where perturbations of the initial conditions~\cite{Bauer2015,Pathak2022,Bi2023aa} or the model parameters are employed~\cite{Bauer2015,Weyn2021}.
The AtmoRep ensemble is thereby computationally inexpensive in both training and inference since it is only generated in the prediction heads, which are defined as simple linear layers. 
This is in contrast to many ensemble methods in machine learning where a set of different models is combined.

\paragraph{Training and Loss}
AtmoRep's training task is the prediction of randomly masked and distorted tokens, see Fig.~\ref{fig:atmorep-concept}, bottom left.
This is an extension of masked token models used in natural language processing~\cite{Devlin2019, Radford2018} and computer vision~\cite{He2022}.
In addition to complete masking, we distort some tokens by either adding noise or reducing their resolution.
The distortions were inspired by~\cite{Devlin2019} and they encourage the model to not rely on the physical correctness of the non-masked tokens and instead learn a robust and probabilistic representation of the relationship between atmospheric states $x$ and $y$.
The masking ratio was increased during training from $0.25$ to values between $0.5$ and $0.9$ depending on the field. 
The increase led to a more difficulty training task over time and to better representations, improving, e.g., the zero-shot forecasting performance of models.

The loss used to train the AtmoRep model is a distance function $\mathcal{D}( \tilde{p}( y , x ; \alpha ) , p_{\theta}( y , x ; \alpha) )$ between the instationary data distribution $\tilde{p}( y , x ; \alpha)$ and the distribution modeled by AtmoRep.
With a Monte Carlo estimate over the training data $\bar{\mathcal{X}} = \{ (\bar{x},\bar{y}) \}$, it can be approximated by
\begin{align}
  \label{eq:loss:divergence}
  \mathcal{D}\big( \tilde{p}( y , x ; \alpha) , p_{\theta}( y , x ; \alpha) \big) 
  \approx \frac{1}{N} \sum_{\bar{x} \in \bar{\mathcal{X}}} d\big( \bar{y} , p_{\theta}( y \vert \bar{x}, \alpha) \big) 
\end{align}
where AtmoRep's ensemble prediction provides an nonparametric estimate of $p_{\theta}( y | x)$, see the supplementary material for the derivation and details.
The distance function $d( \cdot \, , \cdot )$ above measures the quality of the model's predictions for each individual training example. It includes our novel statistical loss function $d_s( \cdot \, , \cdot )$ given by
\begin{align}
  d_s\big( \bar{y} , \hat{y} \big) 
  = \Big\vert 1 - \int \delta_{\bar{y}}(y) \, G_{\mu,\sigma}(y) \, dy \Big\vert^2
  = \Big\vert 1 - G_{\mu,\sigma}(\bar{y}) \Big\vert^2
\end{align} 
where $\bar{y}$ is the single observed value for each field and grid point, formally described by the Dirac-delta $\delta_{\bar{y}}(y)$, and $G_{\mu,\sigma}$ is the unnormalized Gaussian whose mean $\mu$ and variance $\sigma$ is given by those of AtmoRep's ensemble prediction $\hat{y}$.
We hence currently only consider the first two statistical moments of the ensemble in the loss computation.
We complement the statistical loss with a regularization term that controls the variance $\sigma$ as well as an MSE loss term per ensemble member $\hat{y}_k$.
Therefore
\begin{align}
  d \big( \bar{y} , \hat{y} \big) 
  = \sum_k \vert \bar{y} - \hat{y}_k \vert^2 + d_s\big( y , \hat{y} ) + \sqrt{ \sigma } .
\end{align}

Training was first performed on individual fields with field-specific transformers.
Subsequently, when individual fields were largely converged, the fields were combined into the Multiformer and training was continued. 
We thereby trained three different Multiformer configurations, one with velocity and all other fields, one with vorticity, divergence and all other fields, and a $3$-field configuration with wind and temperature.
The improvement through the coupling of the single-field transformers is quantified in the supplementary material.

\subsection{Evaluation}
\label{sec:evaluation}
Each application has been analysed with common and suitable metrics for quantifying AtmoRep's skill.
Where applicable, comparisons with existing approaches have been included to relate our results to the state-of-the-art in literature.

\paragraph{Nowcasting} 
Results have been obtained using forecasts at $0$ and $12$ UTC for the entire year 2018. The predictions are compared to the IFS as standard reference for classical numerical weather prediction models as well as Pangu-Weather~\cite{Bi2023aa} as an example for a state-of-the-art AI-based model.     
ERA5 on model levels was used as ground truth for AtmoRep and IFS and ERA5 on pressure levels for Pangu-Weather.
For the ensemble analysis, we compared against ERA5 (since IFS ensemble data was not available to us) using the CRPS and assuming a Gaussian distribution.
For AtmoRep we employed the $5$-field velocity Multiformer fine-tuned for $6\,\mathrm{h}$ forecasting as described in the supplementary material as well as the $5$-field velocity Multiformer without fine-tuning, the latter to determine the zero-shot performance.
To avoid tiling artifacts, an overlap of $18$ and $54$ grid points between adjacent neighborhoods has been used for both Multiformer configurations.


\paragraph{Counterfactuals}
For the counterfactual experiment, we randomly sampled from the early and late time range, generating in total approximately 8 billion samples for each of the early and late year range and the counterfactual run. 
The experiments were run with the vorticity-divergence 5-field Multiformer with short term forecasts with $3 \, \mathrm{h}$ lead time (no fine-tuning for forecasting was applied).
The difference histogram in Fig.~\ref{fig:intrinsic}, bottom right, is with respect to normalized distributions while the distributions shown below are unnormalized.

\paragraph{Downscaling} The downscaling analysis has been performed using hourly predictions for the year 2018 and with the entire downscaled domain, i.e. $[-1.25^{\circ}, 25.75^{\circ}] \times [42.125^{\circ}, 55.625^{\circ}]$ in latitude and longitude.
The AtmoRep network was the $3$-field Multiformer with both velocity components as well as temperature and using only model level $137$ as input. 
The downscaling network had 6 transformer blocks and used an embedding dimension twice the size as for ERA5 for temperature due to the much larger token size in terms of grid points in the predictions.


\paragraph{Bias corrections} 
For bias correction, we evaluate the metrics as shown in Fig.~\ref{fig:atmorep-radklim}. 
These have been computed averaging hourly spaced predictions from 2019. The data from 2018 was used as validation set to determine the best bias corrected model. As reported above, the generator architecture is the $5$-field vorticity-divergence Multiformer used with $6 \times 6$ tokens, which aligns well with the spatial extent of the RADKLIM dataset $([44.5^{\circ}, 57.75^{\circ}] \times [3.5^{\circ}, 16.75^{\circ}]$ latitude-longitude).

\section{Code availability}
\label{sec:code-availability}
The AtmoRep model code is Open Source under an MIT license (\url{https://opensource.org/license/mit/}). The code used to generate the results presented in the paper as well as pre-trained model weights and analysis code for generating plots will be made publicly available (with persistent identifier) upon acceptance of this manuscript.

\section{Data availability}
\label{sec:data-availability}
ERA5 data are openly and freely available from ECMWF (\url{https://cds.climate.copernicus.eu/cdsapp#!/dataset/reanalysis-era5-complete}). For use with the pre-trained AtmoRep model, hourly data for the selected variables and model levels is required; a pre-computed subset of ERA5 data in the format that is used in AtmoRep can also be obtained from the Jülich meteocloud (upon acceptance of the paper).
The code for data normalization is provided with the AtmoRep source code. 
IFS data that has been used for model correction experiments were retrieved from ECMWF's MARS archive (\url{https://confluence.ecmwf.int/display/UDOC/MARS+user+documentation}).
Pangu-Weather data used for the comparisons have been generated using the ai-models tool from ECMWF (\url{https://github.com/ecmwf-lab/ai-models})
COSMO REA6 data were obtained from the open data archive of the German Weather Service (DWD) (\url{https://opendata.dwd.de/climate_environment/REA/COSMO_REA6/}). This open data archive also provides the RADKLIM data (\url{http://dx.doi.org/10.5676/DWD/RADKLIM_YW_V2017.002}). 
Pre-processing scripts for COSMO REA6 and RADKLIM are also available with the AtmoRep source code.



\section*{Acknowledgments} 
IL acknowledges funding by the CERN Knowledge Transfer Fund and the CERN Initiative for Environmental Applications (CIPEA). MGS and SS acknowledge funding from the EU under grant ERC-Adv-787576 "IntelliAQ". BG and ML received funding from the EuroHPC project MAELSTROM (EU grant id 955513 and BMBF grant id 6HPCO29). This research was supported in part by the National Science Foundation under Grant No. NSF PHY-1748958 through the CLIMATE21 workshop in November/December 2021 at the Kavli Institute for Theoretical Physics where initial ideas for the project were developed. Compute time was provided by the Jülich Supercomputing Centre under the project \texttt{atmo-rep}. David Hidary contributed the visualisations of the attention maps. The authors are grateful to Matthew Chantry, Mariana Clare, Yi Deng, Robert Brunstein, Maike Sonnewald, Olaf Stein, Aneesh Subramanian, and Max for providing valuable data and discussions. We thank ECMWF for producing ERA5 and the German Weather Service DWD for generating the COSMO REA6 reanalysis and the RADKLIM dataset. Kaifeng Bi is acknowledged for providing the Pangu-Weather source code and data and ECMWF for the ai-models tool.


\newpage

\begin{appendices}

\clearpage
\section*{Extended Data}\label{sec:extdata}


\begin{figure}[h]
    \centering
    \includegraphics[width=0.8\textwidth]{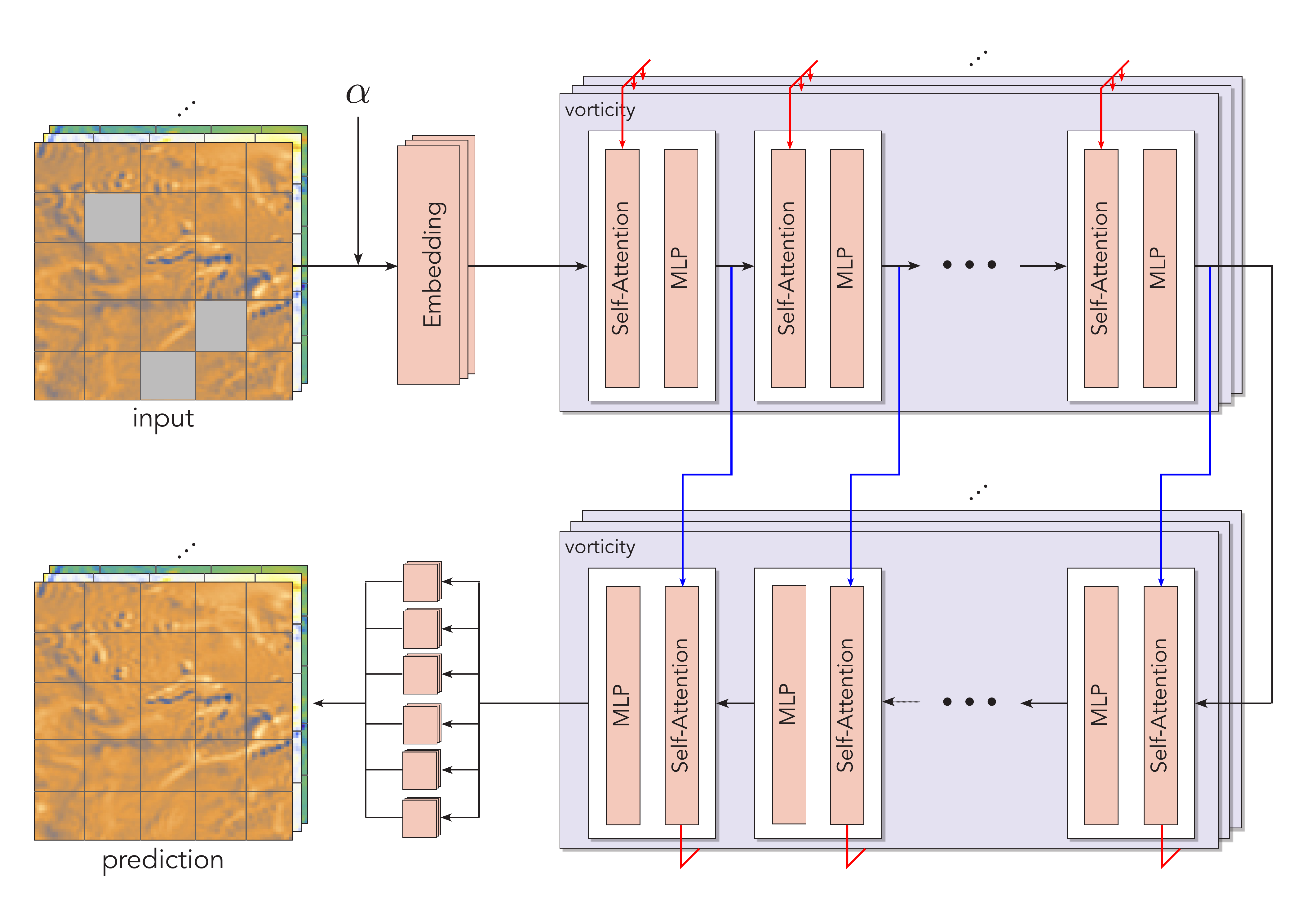}
    \caption{Overview of the AtmoRep neural network architecture. Its core (right, bluish) consists of a stack of encoder-decoder transformers with one per physical field and coupled through cross-attention.
    UNet-like connections (dark blue) between encoder and decoder are used to facilitate that a multi-resolution representation is learned.
    The external conditions $\alpha$ are encoded using a linear layer and appended to the network input.
    The embedding network is a linear layer and it is preceded by a local positional encoding (not shown) based on the sine/cosine one from the original transformer work but extended to the four-dimensional domain considered in AtmoRep.
   The members of the ensemble of prediction heads consist each of a linear layer. A different random initialization of each linear layer and training with our novel ensemble loss is sufficient to prevent mode collapse.
   Further details on the network architecture can be found in the supplementary material.
   }
    \label{fig:multiformer_overview}
\end{figure}

\begin{figure}
    \centering
    \includegraphics[trim={1.5cm 1.5cm 1.0cm 1.5cm},clip,width=\textwidth]{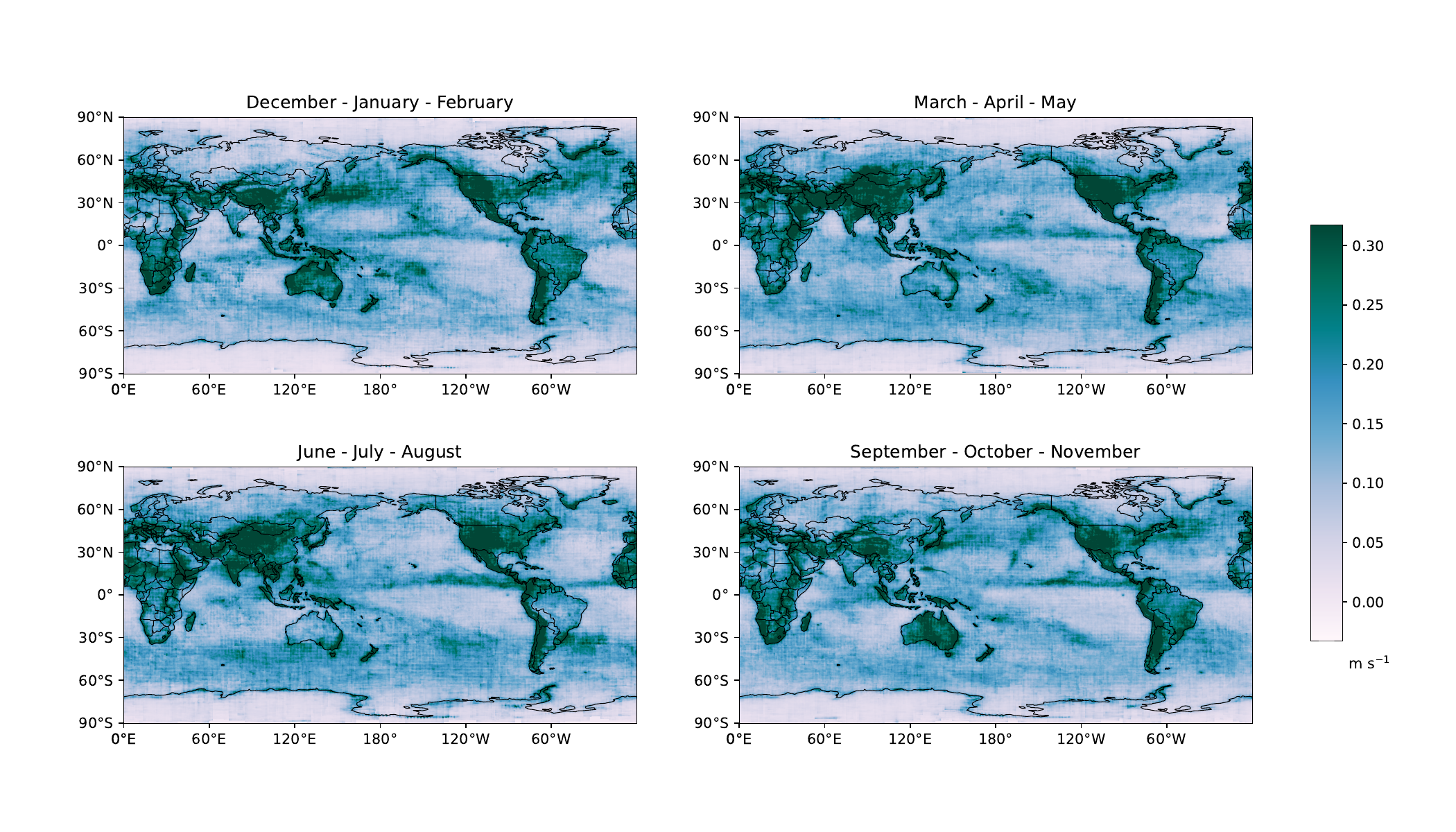}
    
    \vspace{0.3cm}
    
    \includegraphics[trim={1.5cm 1.5cm 1.0cm 1.5cm},clip,width=\textwidth]{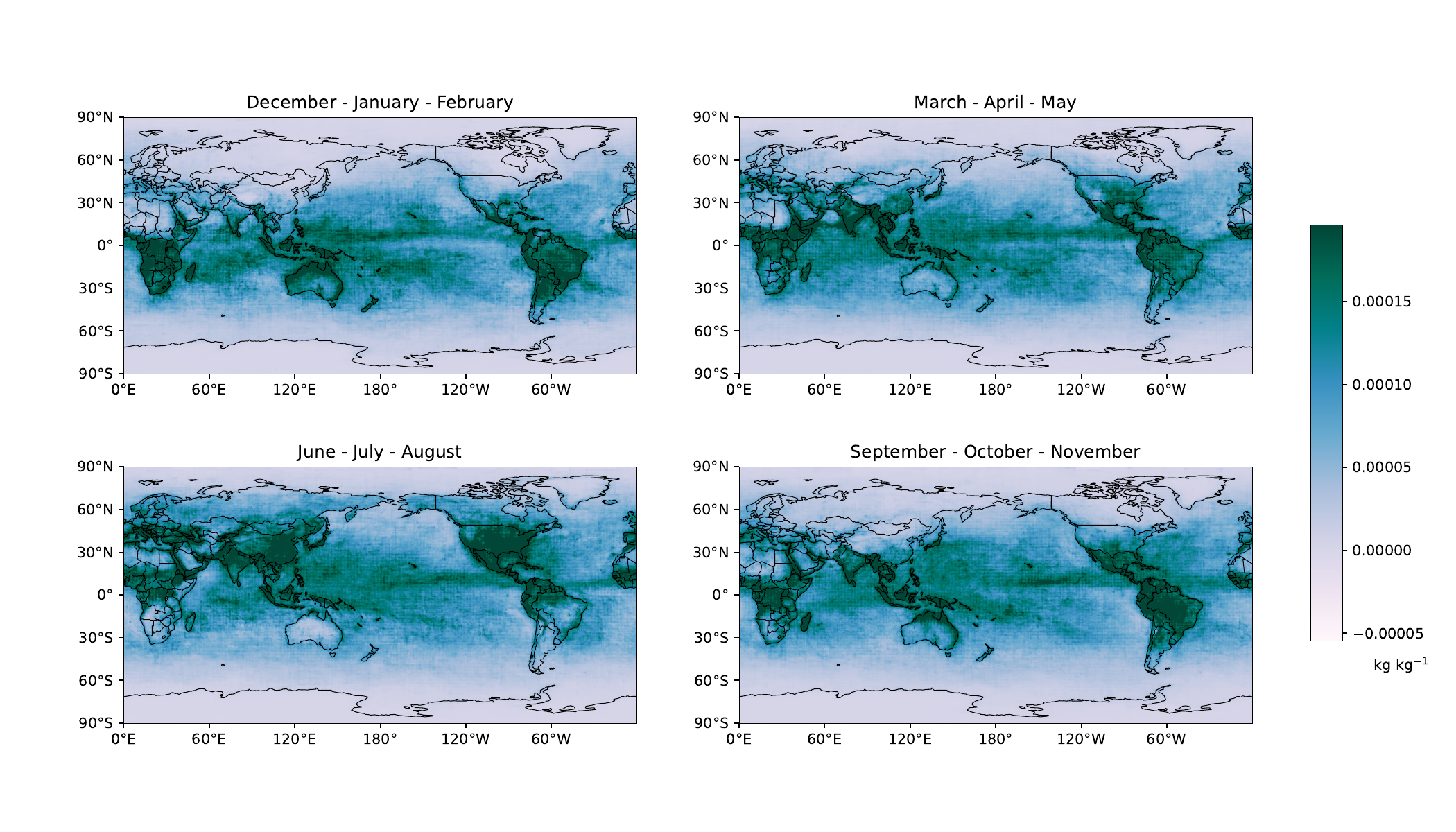}
    \caption{Spatial distribution of the mean error for 2018 as a function of the season for zonal velocity (top) and specific humidity (bottom). 
    Seasonal changes of the error can be observed and these are well aligned with changes in the atmospheric dynamics. For example, in the North Atlantic the errors are larger in the winter months where one has larger wind velocities compared to the summer months.
    Also topographic features are apparent in the error maps.
    The detailed structures were learned solely from the dynamics in the training data and no land-sea mask or orographic information was provided to the network. 
    Comparing to Fig.~\ref{fig:stdev_map}, a strong correlation between the error and the standard deviation can be observed.
    }
    \label{fig:ext-error-season}
\end{figure}

\begin{figure}[t]
    \centering
     \includegraphics[trim={1.5cm 1.5cm 1.0cm 1.5cm},clip,width=\textwidth]{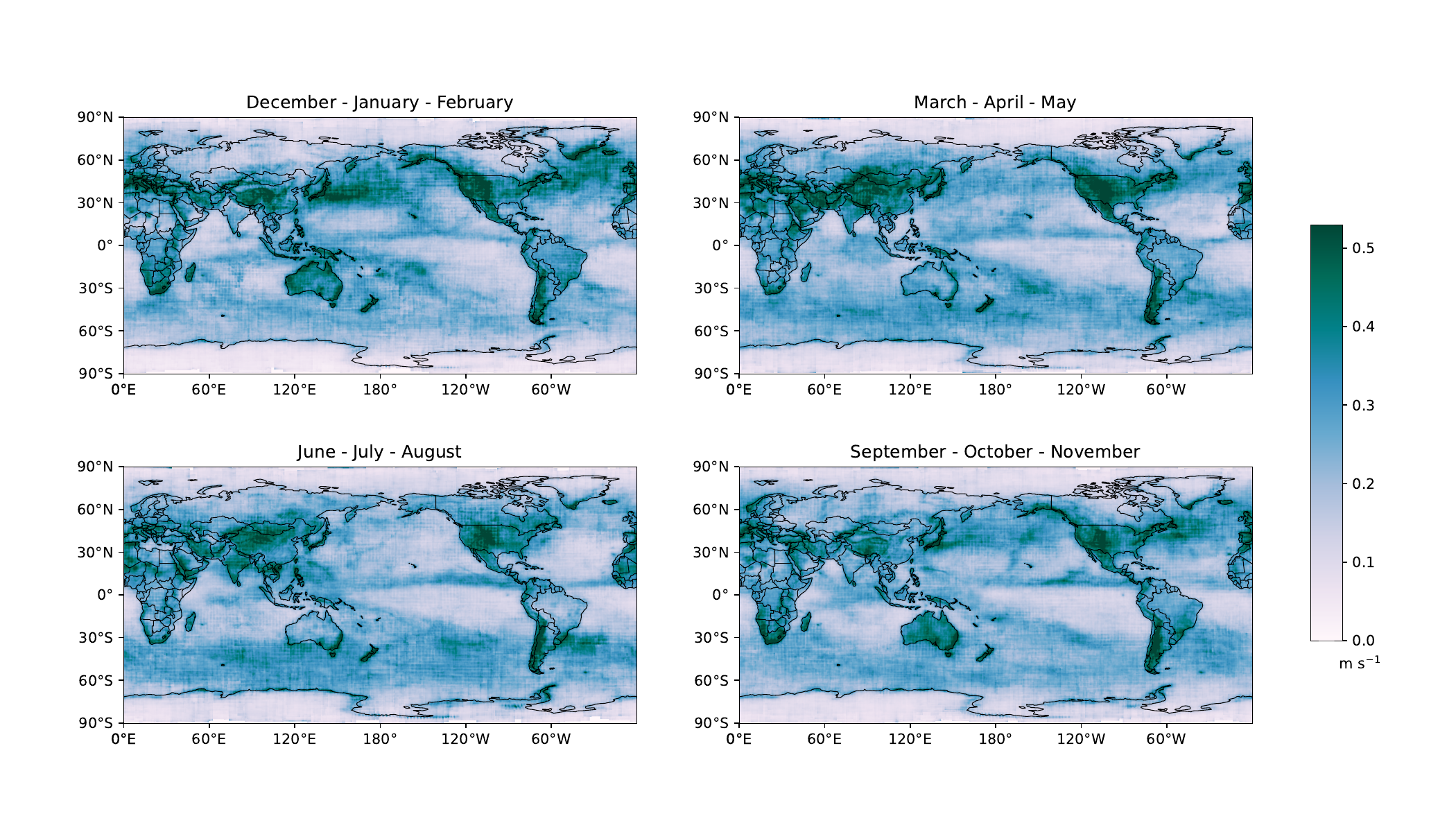}
    
    \vspace{0.3cm}
    
    \includegraphics[trim={1.5cm 1.5cm 1.0cm 1.5cm},clip,width=\textwidth]{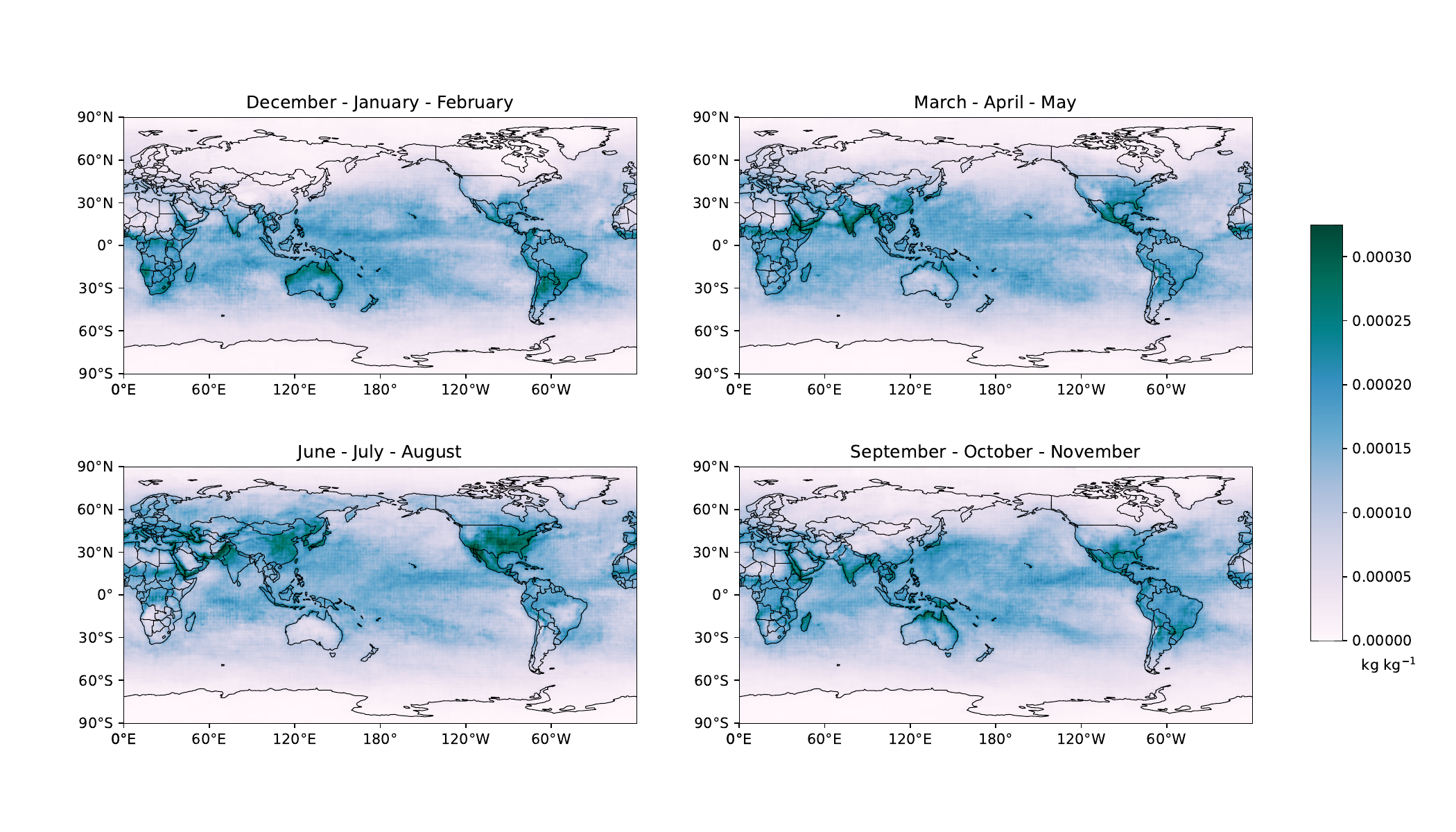}
    \caption{Spatial distribution of the AtmoRep ensemble standard deviation for 2018 as a function of the season for zonal velocity (top) and specific humidity (bottom). The ensemble spread shows a strong correlation with the physical variability of the system, for example with seasonal storms in the North Atlantic. The ensemble spread is also strongly correlated with the prediction error, cf. Fig.~\ref{fig:ext-error-season}.}
    \label{fig:stdev_map}
\end{figure}

\begin{figure}
    \centering
    \includegraphics[width=\textwidth]{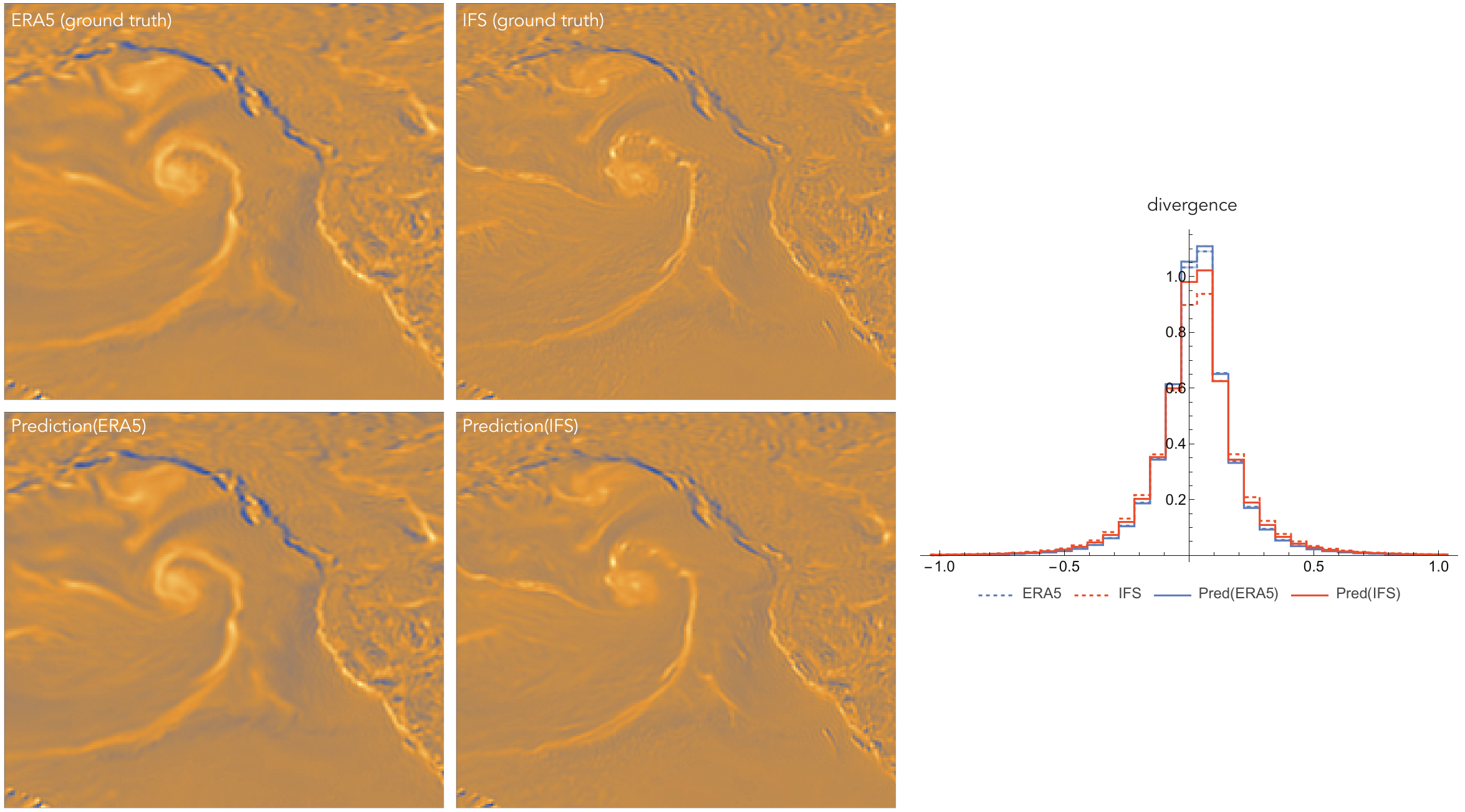}
    \caption{Additional results for model correction. \emph{Left: } Zoom-ins for AtmoRep predictions with ERA5 and IFS model data as input. The AtmoRep predictions for the IFS data have much finer details than with ERA5, see for example the filaments originating from the storm in the center. This reflects the higher frequency content in the input (top-right), see also the inset spectrum in Fig.~\ref{fig:intrinsic}, bottom left. \emph{Right: } Histogram for divergence. Analogous to the results for vorticity, a clear shift in the model output towards ERA5 can be observed when the input is IFS data.}
    \label{fig:ext-ifs-correct}
\end{figure}

\begin{figure}
    \centering
    \includegraphics[width=0.9\textwidth]{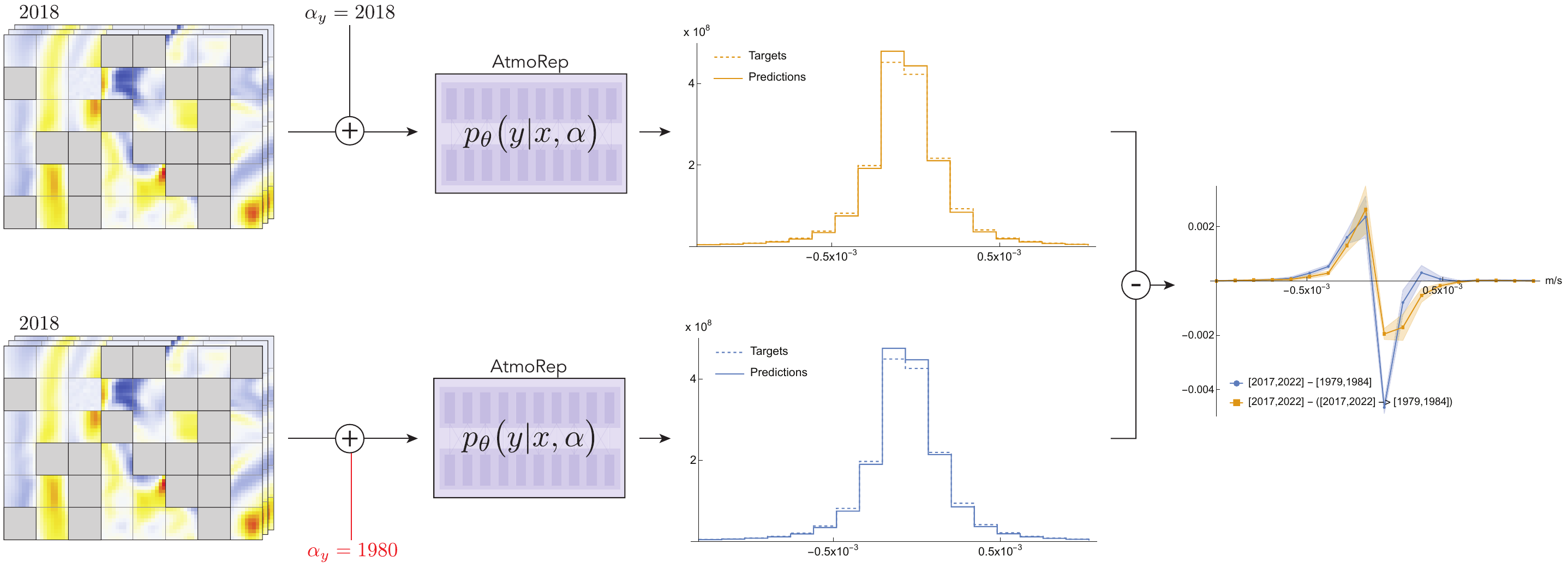}
    \caption{Depiction of the experimental setup for counterfactuals. A large set of fixed initial conditions, e.g. from the year $2018$, are the input to AtmoRep. These are used once with the physical external conditions, i.e. $\alpha_y = 2018$, and once with modified but statistically plausible once, e.g. $\alpha_y = 1980$. The output by AtmoRep are two different distributions, since the model learned the instationary behavior of the training data. For analysis, we show a difference of the histograms since this makes distributional shifts more apparent (in particular, the difference of the normalized histograms is used so that initial conditions can be sampled).}
    \label{fig:ext-counterfac}
\end{figure}

%
%
%
%

\begin{figure}
    \centering
    \includegraphics[width=0.4\textwidth]{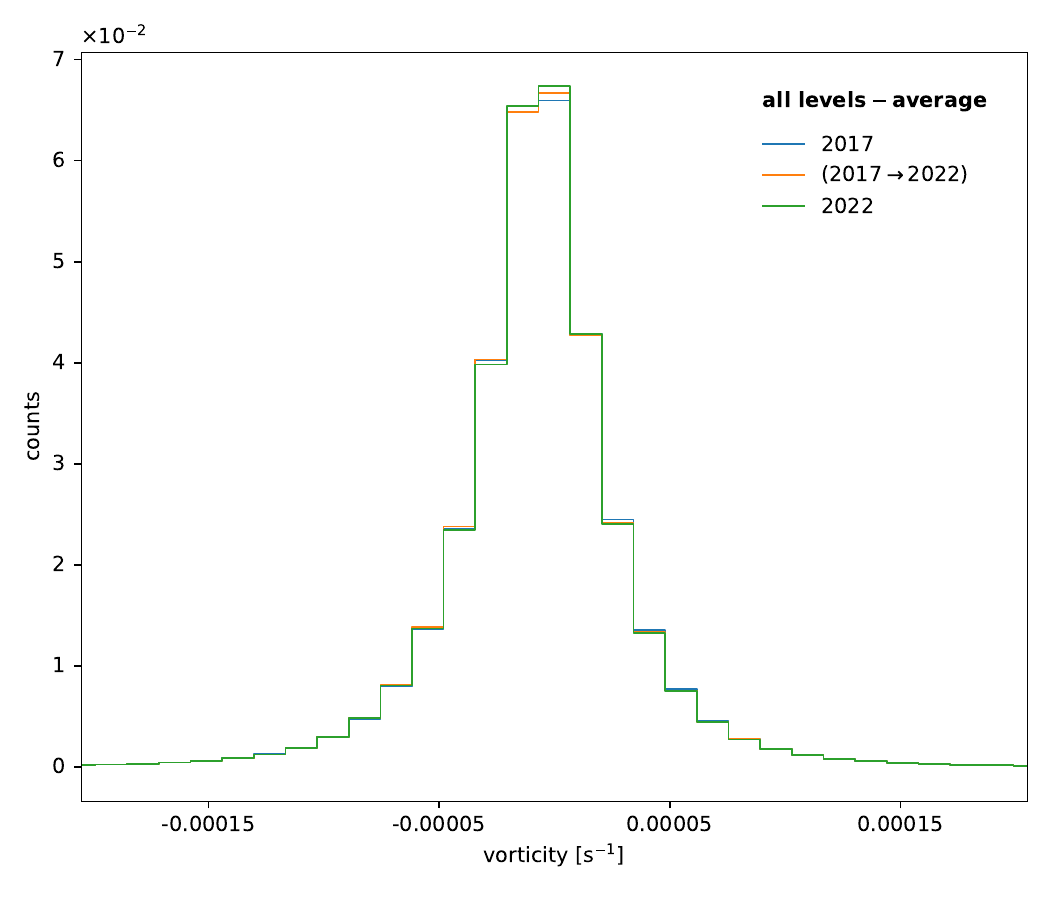}
    \includegraphics[width=0.4\textwidth]{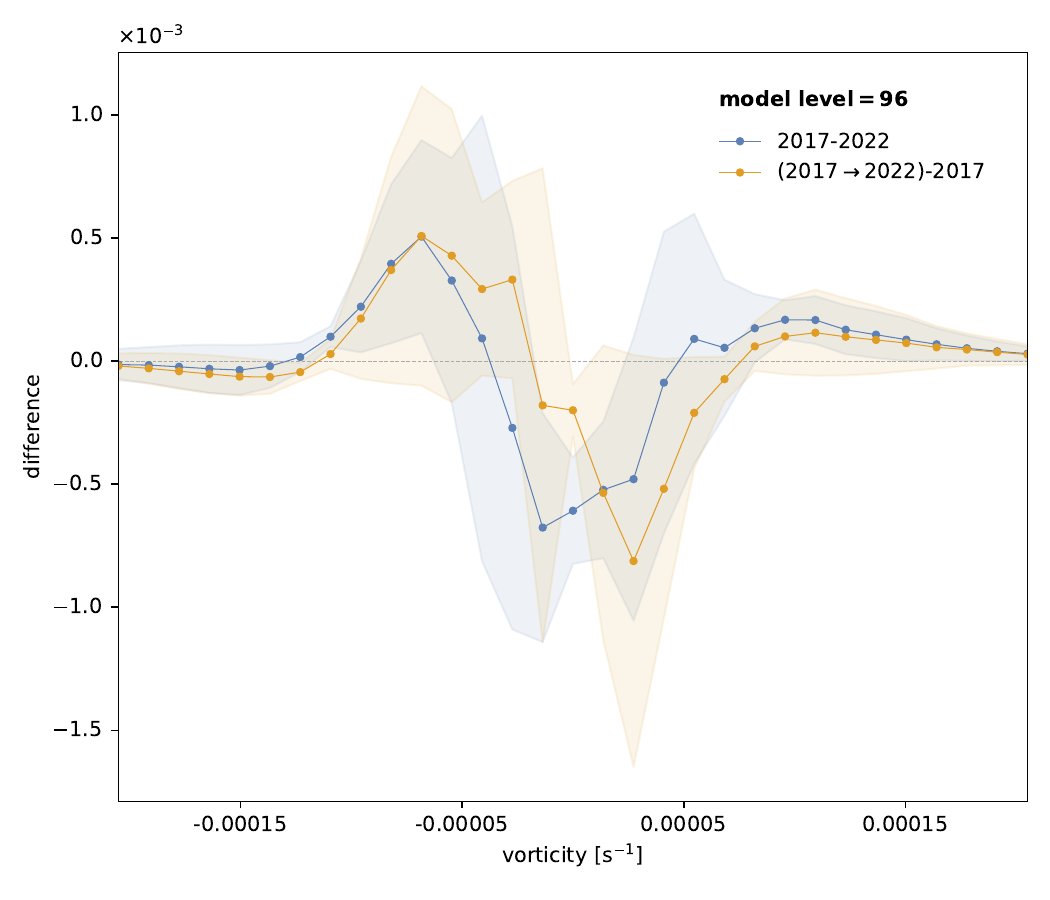}
    \includegraphics[width=0.4\textwidth]{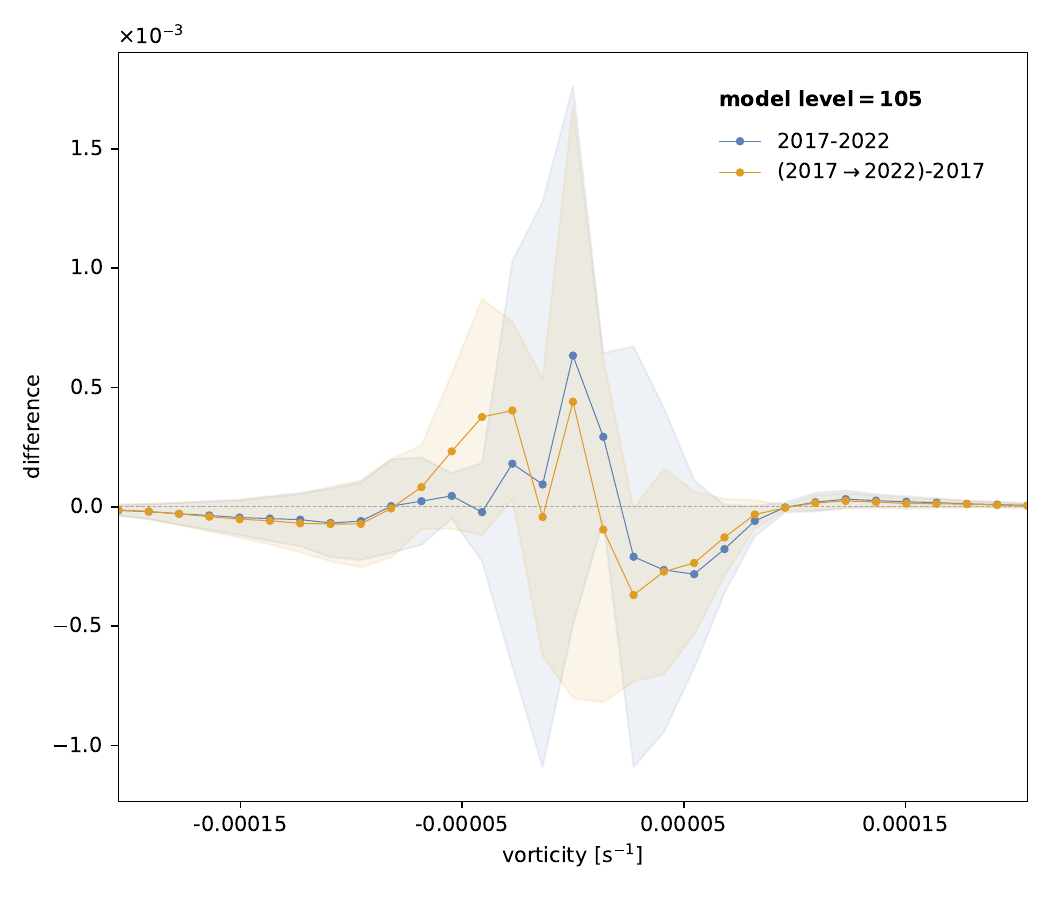}
    \includegraphics[width=0.4\textwidth]{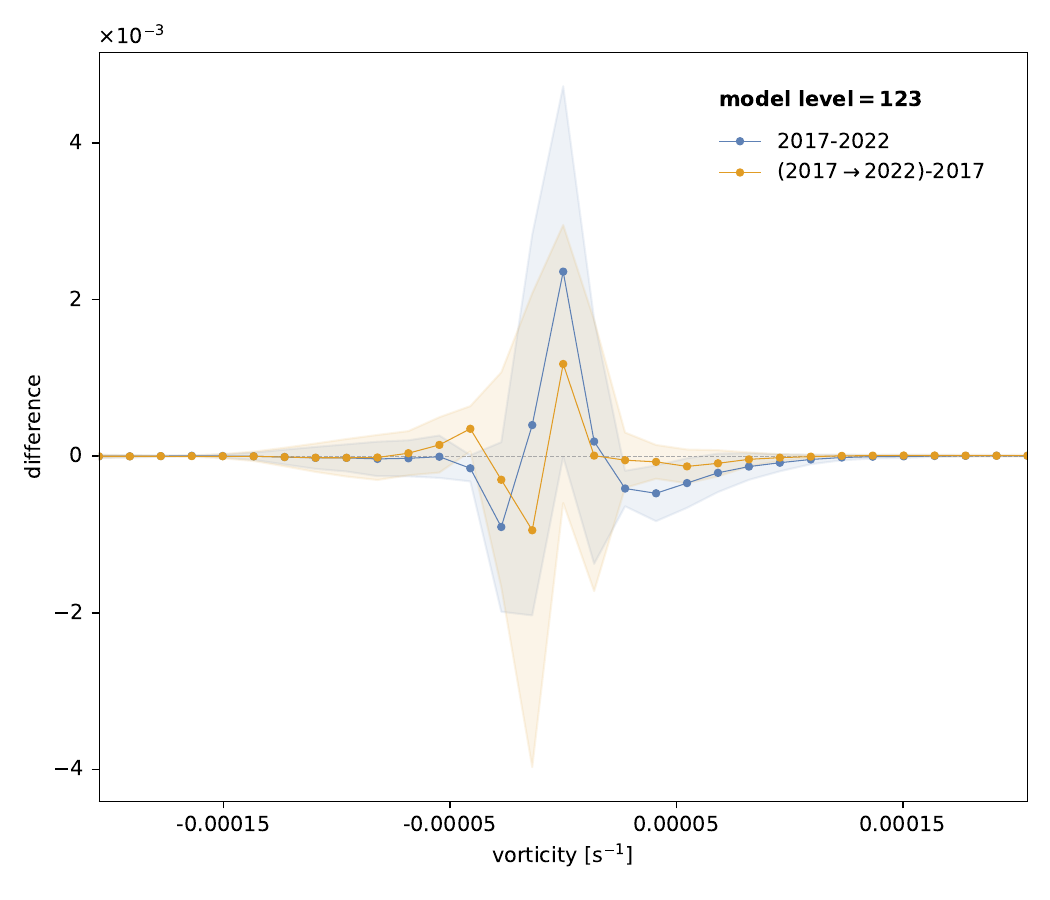}
    \includegraphics[width=0.4\textwidth]{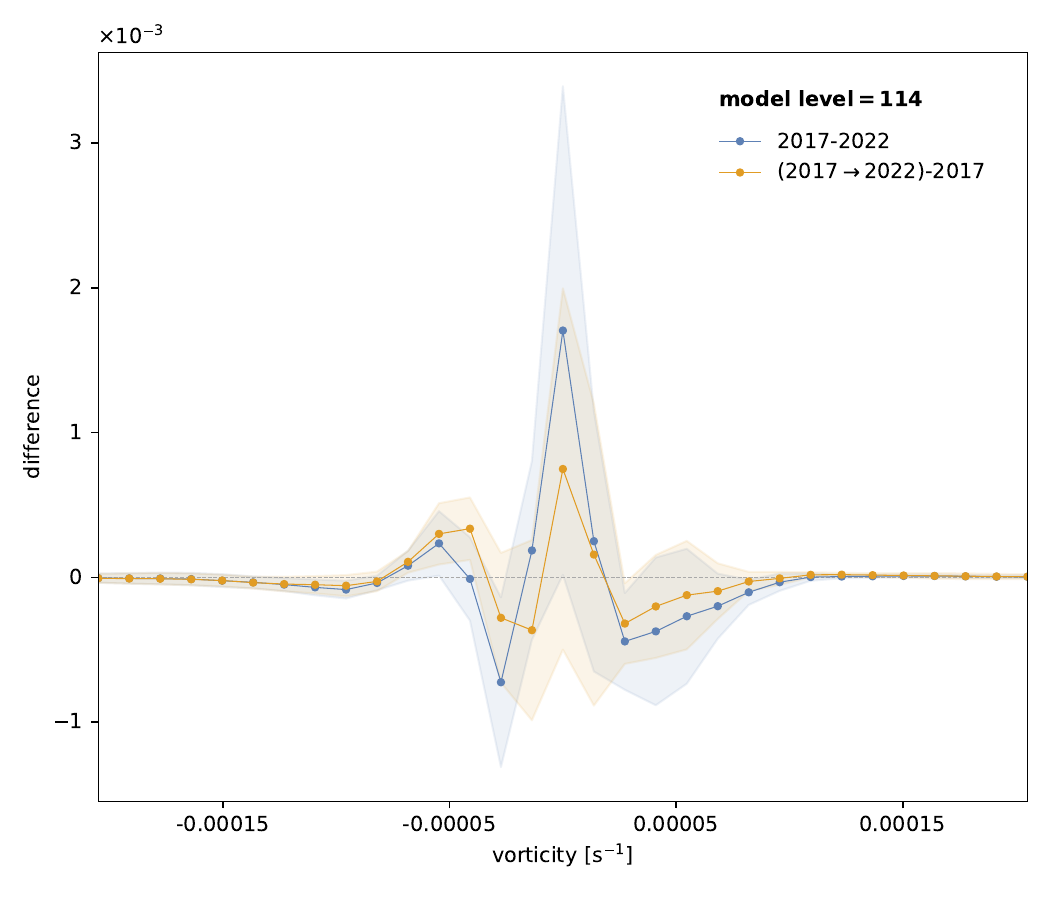}
    \includegraphics[width=0.4\textwidth]{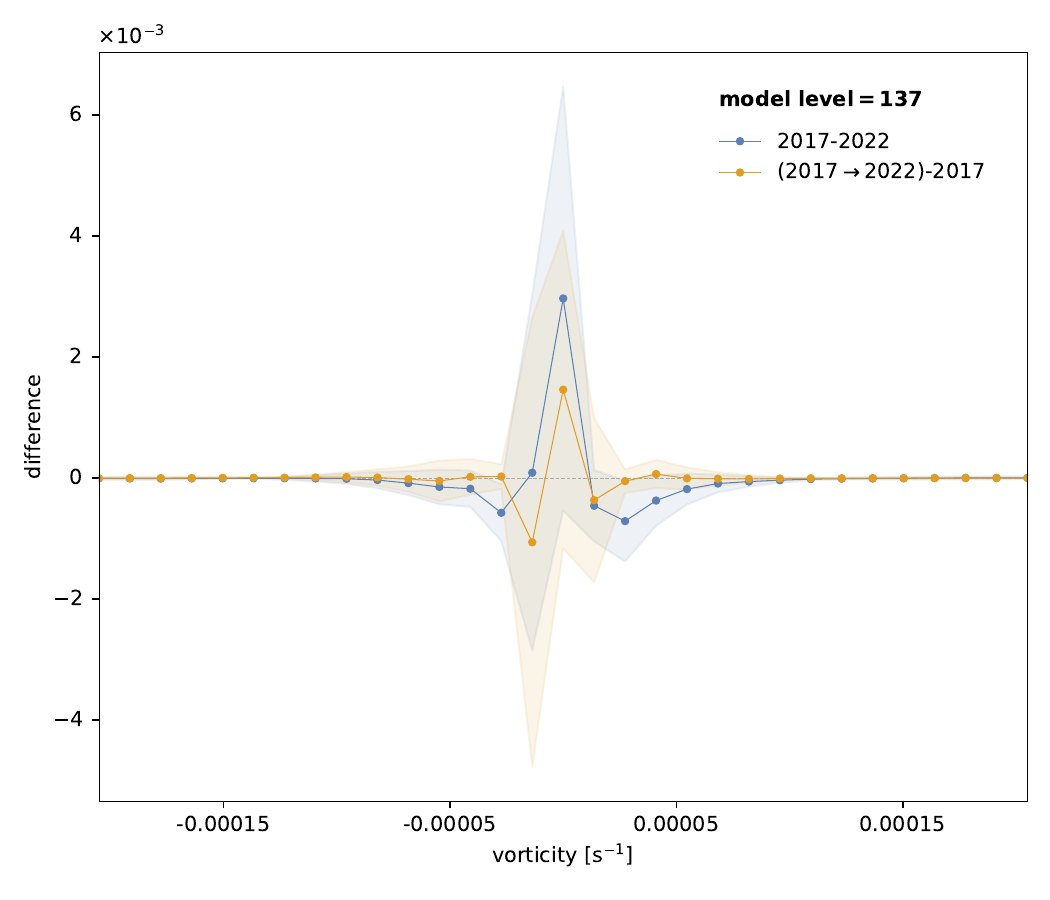}
    \caption{The same methodology as for counterfactuals can be used to study the temporal extrapolation abilities of AtmoRep $p_{\theta}(y \vert x , \alpha)$, i.e. to what extent it can provide predictions beyond the training period which are consistent with the ERA5 distribution for the extended time range.
    For this, we have sampled the AtmoRep vorticity/divergence Multiformer configuration with initial conditions $x$ from $2017$ and the external conditions $\alpha$ prescribed as $2022$ (denoted as '2017 $\rightarrow$ 2022'). The distribution of the predictions is compared to the true ERA5 distribution in $2022$.
    Top left: the averaged distributions of vorticity for the three evaluations: $2017$,  $2022$,  and $2017 \rightarrow 2022 $.  The other plots show the difference between the $2022$ (or simulated $2022$) and $2017$ distributions for each vertical level.
    Shaded areas depict one standard deviation.
    No perfect match between distributions can be observed but a clear trend is visible for all vertical levels. 
    The results show the robustness and generality of AtmoRep's learned distribution $p_{\theta}(y \vert x , \alpha)$. 
    Note that AtmoRep would not able to extrapolate under more significant and nonlinear shifts in the data distribution, e.g. those that can be expected from climate change on longer time horizons.
    }
    \label{fig:ext-extrapolation}
\end{figure}

\newpage
\begin{figure}[htpb!]
  \centering
 \includegraphics[width=0.19\textwidth]{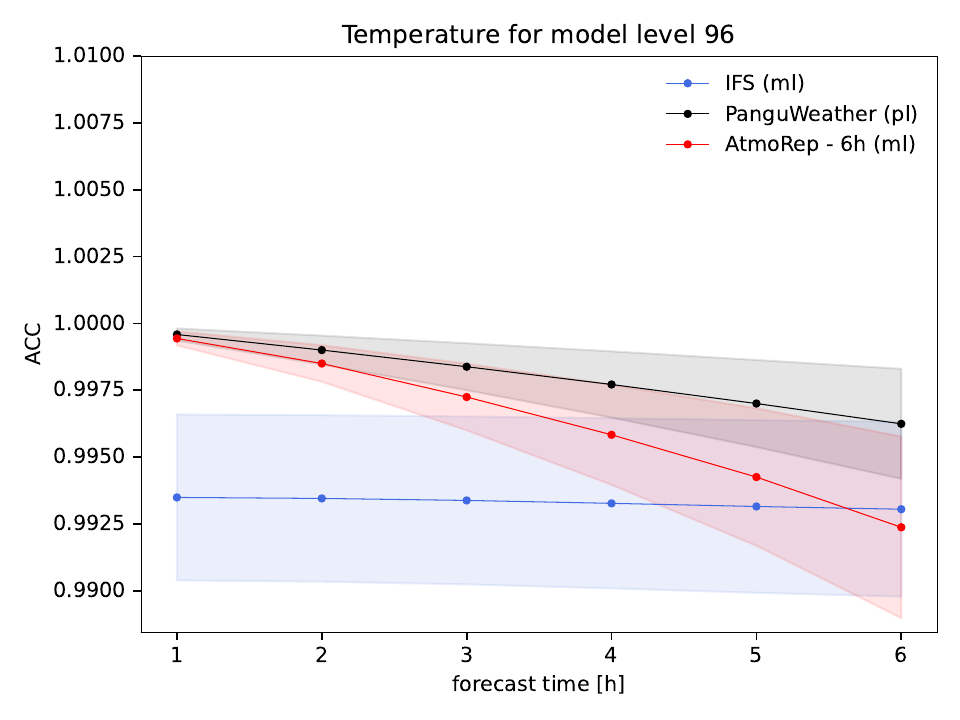}
 \includegraphics[width=0.19\textwidth]{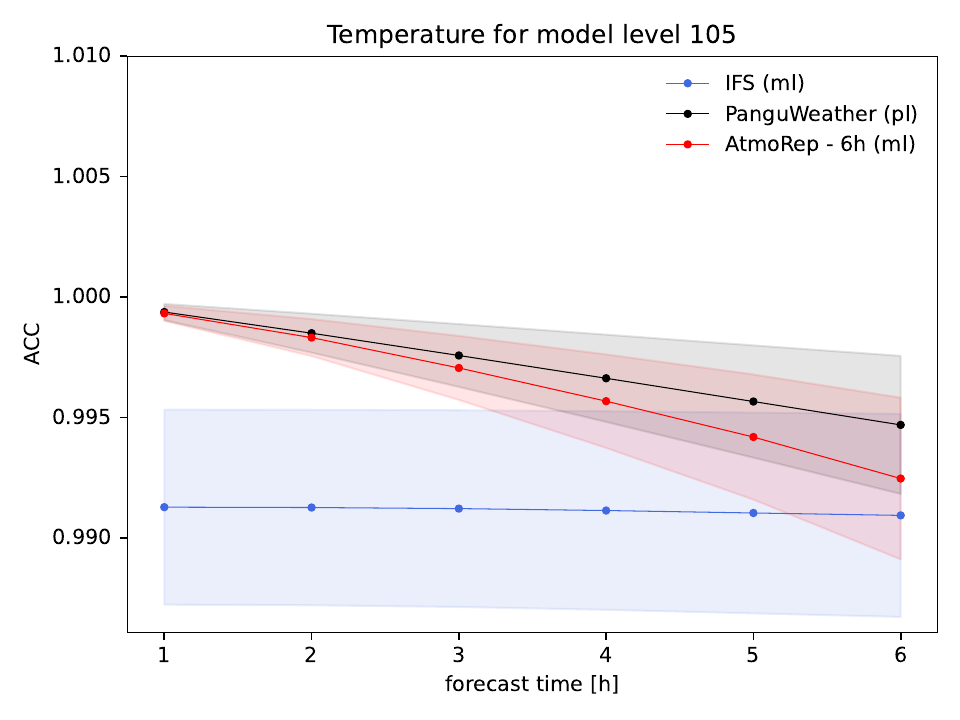}
 \includegraphics[width=0.19\textwidth]{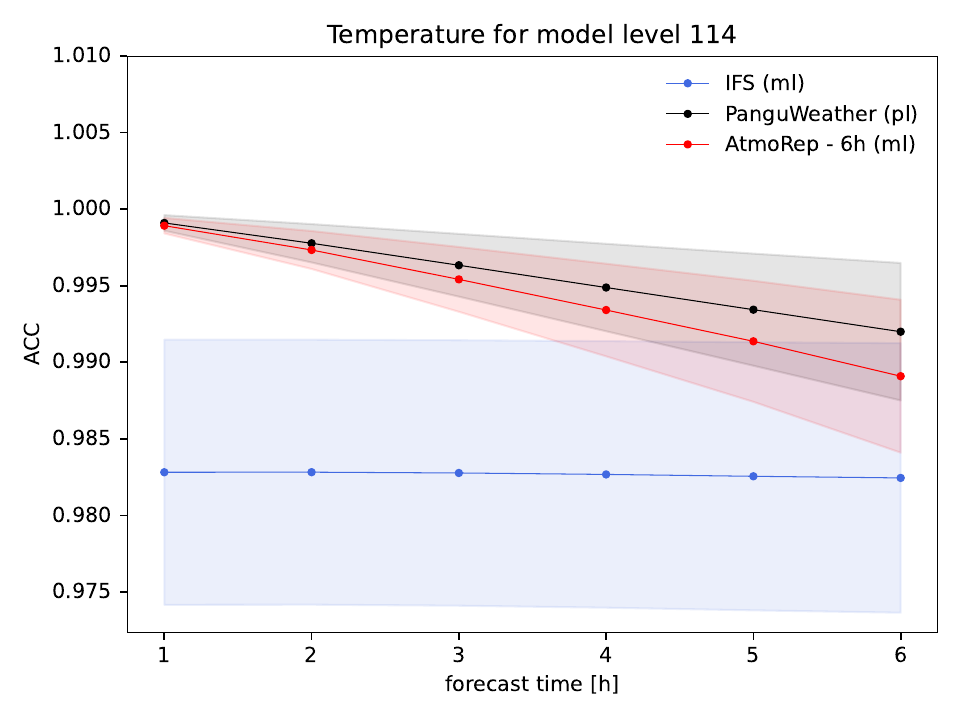}
 \includegraphics[width=0.19\textwidth]{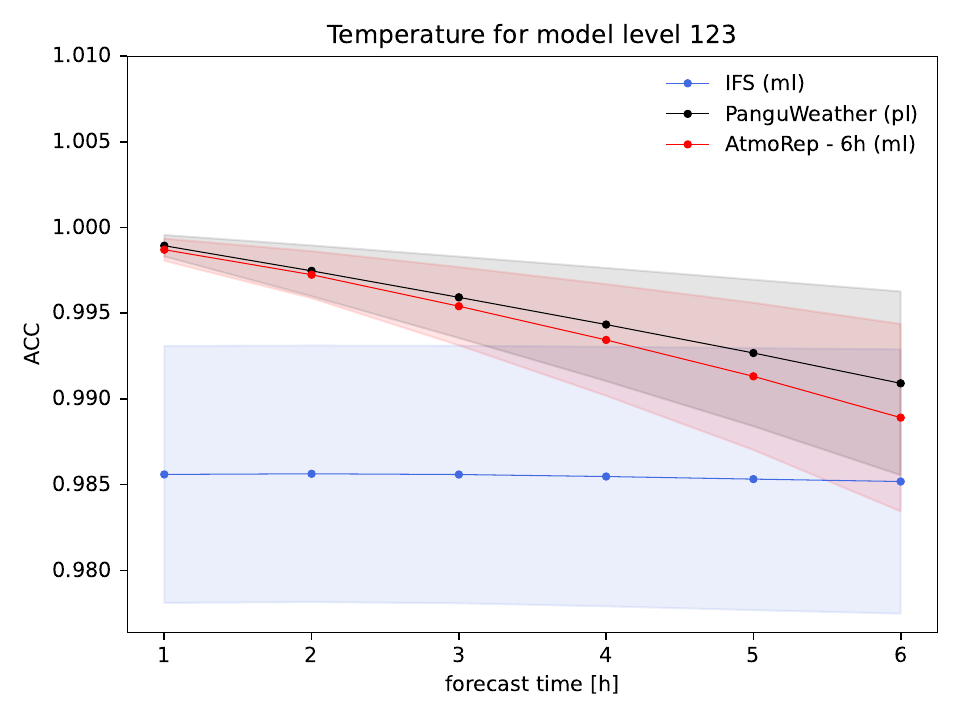}
 \includegraphics[width=0.19\textwidth]{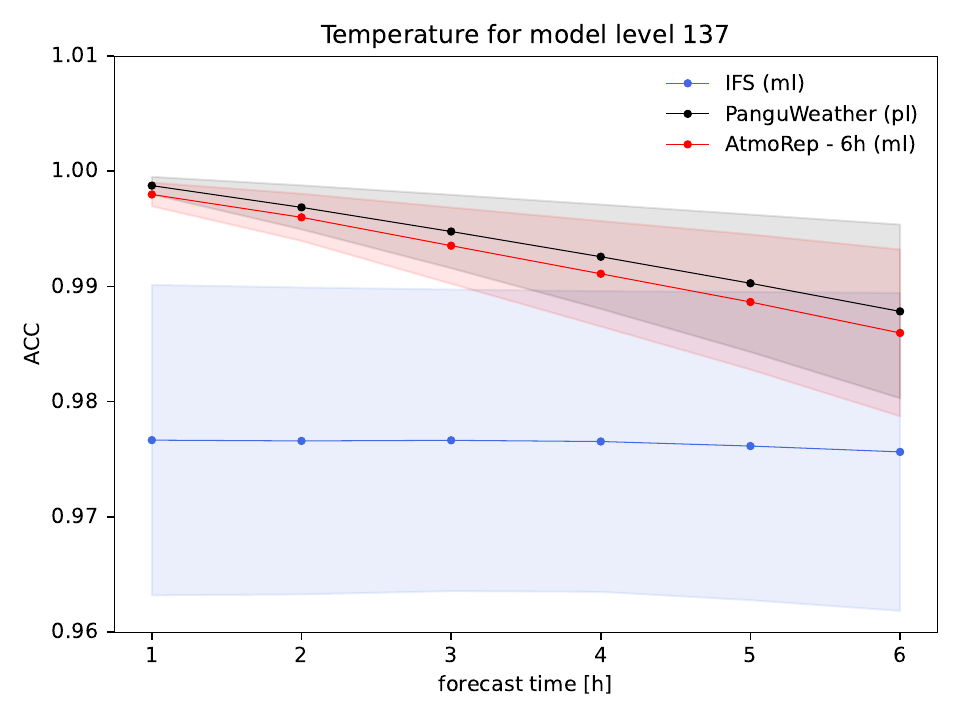}
 \hfill
 \includegraphics[width=0.19\textwidth]{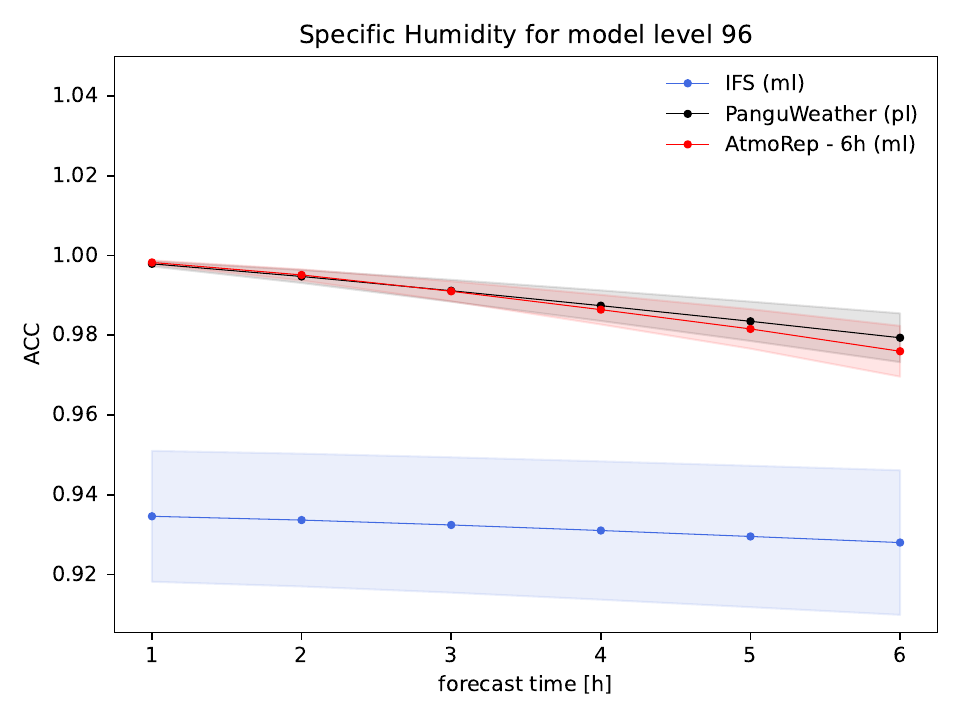}
 \includegraphics[width=0.19\textwidth]{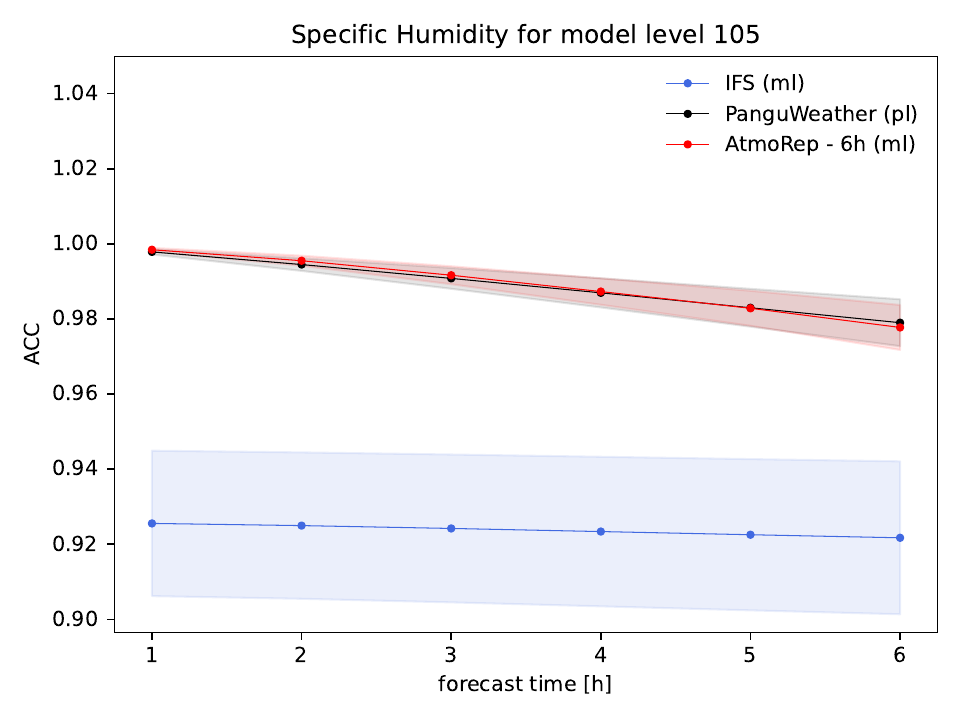}
 \includegraphics[width=0.19\textwidth]{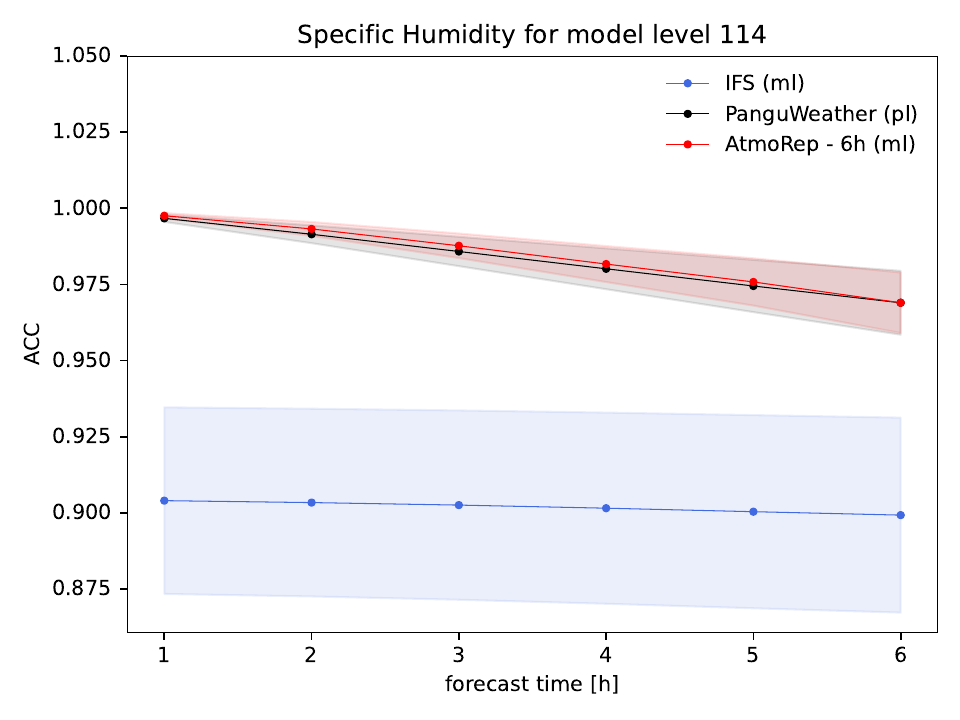}
 \includegraphics[width=0.19\textwidth]{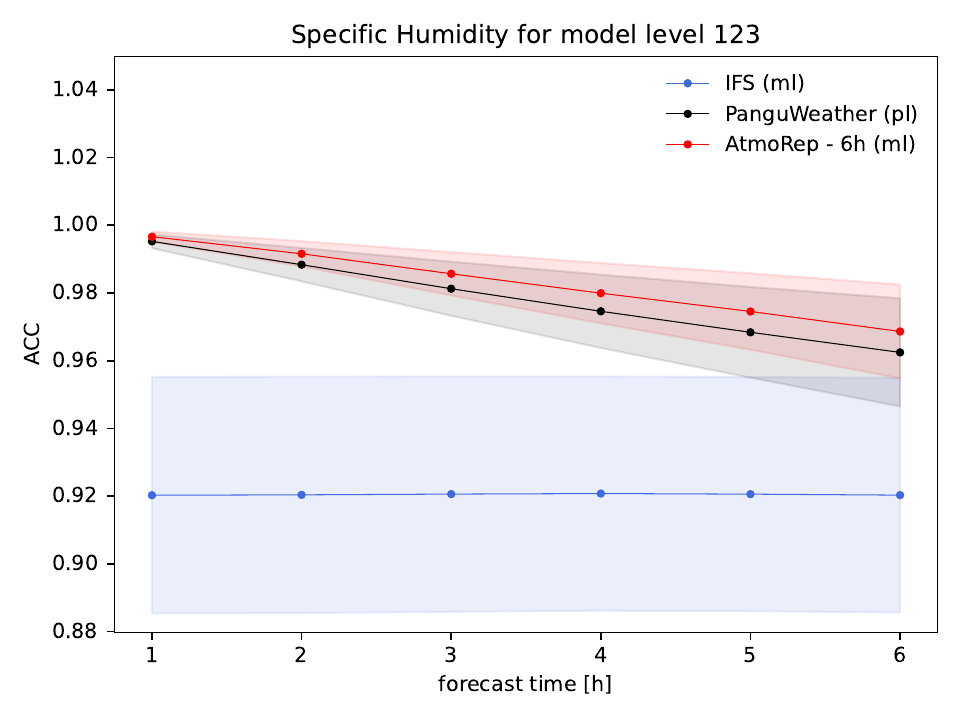}
 \includegraphics[width=0.19\textwidth]{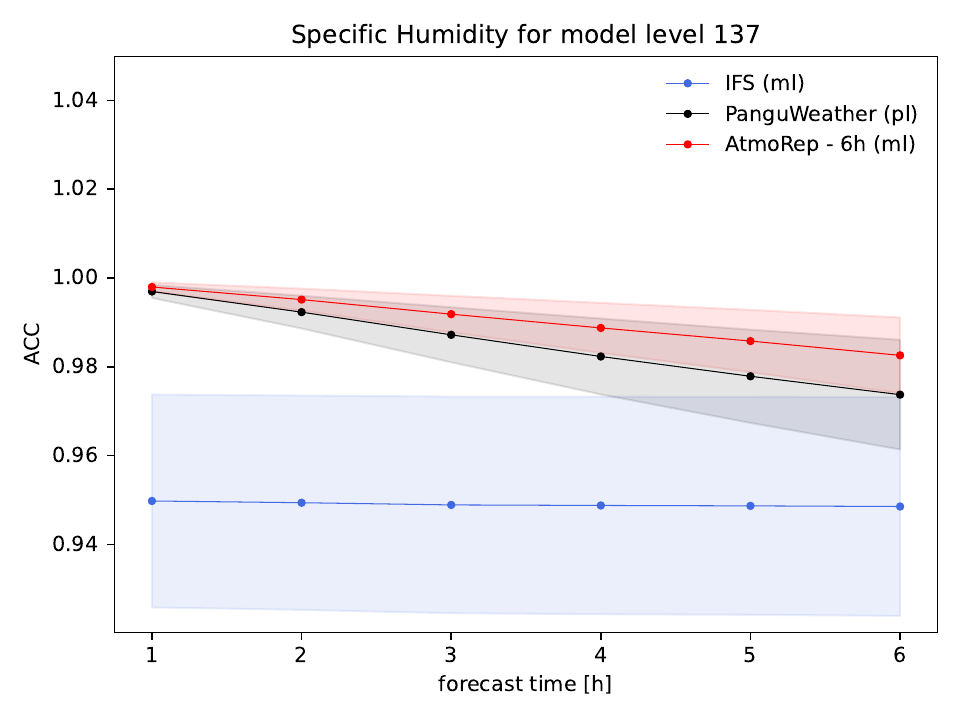}
 \hfill
\includegraphics[width=0.19\textwidth]{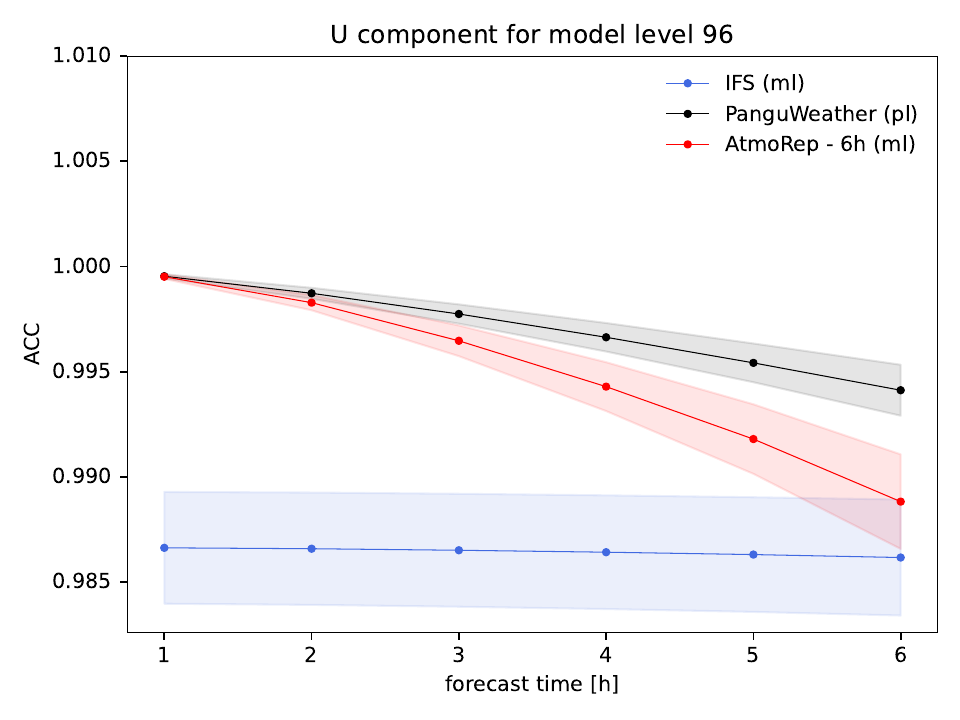}
\includegraphics[width=0.19\textwidth]{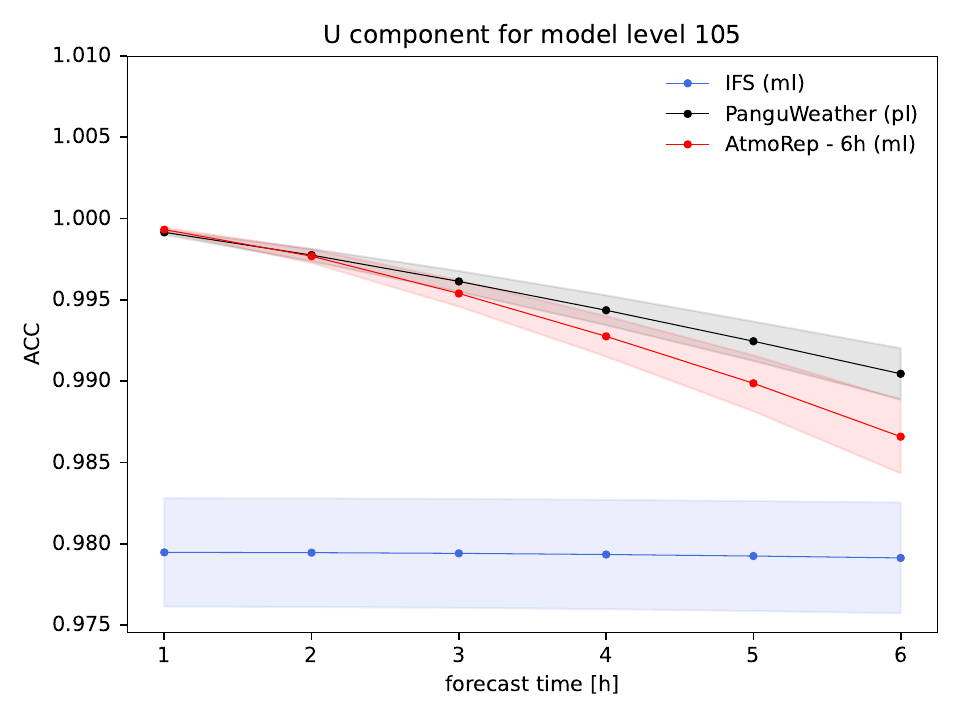}
 \includegraphics[width=0.19\textwidth]{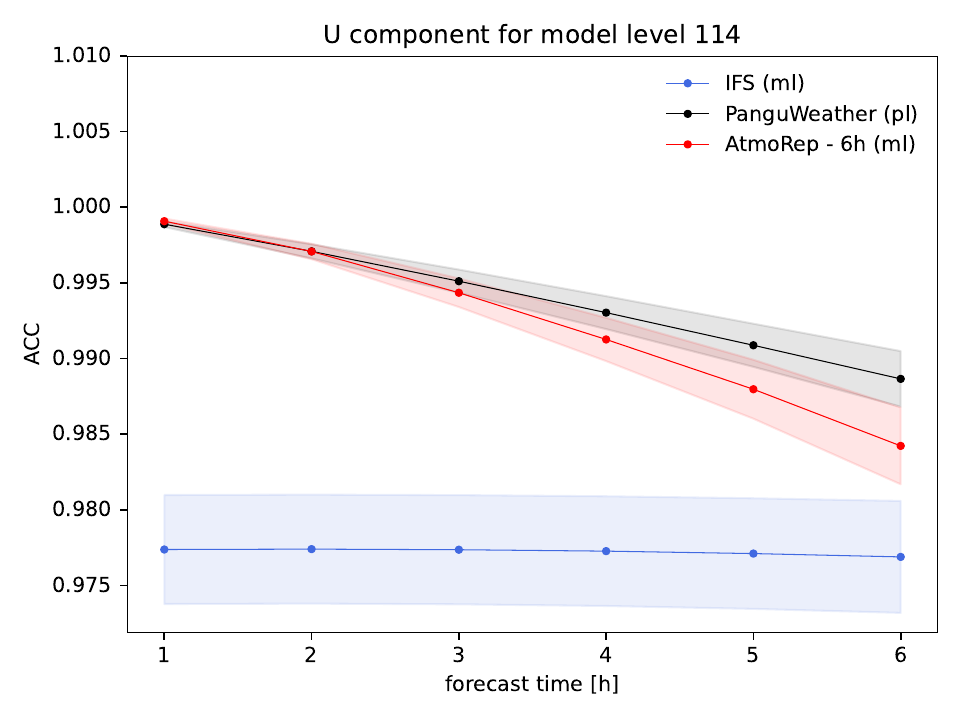}
 \includegraphics[width=0.19\textwidth]{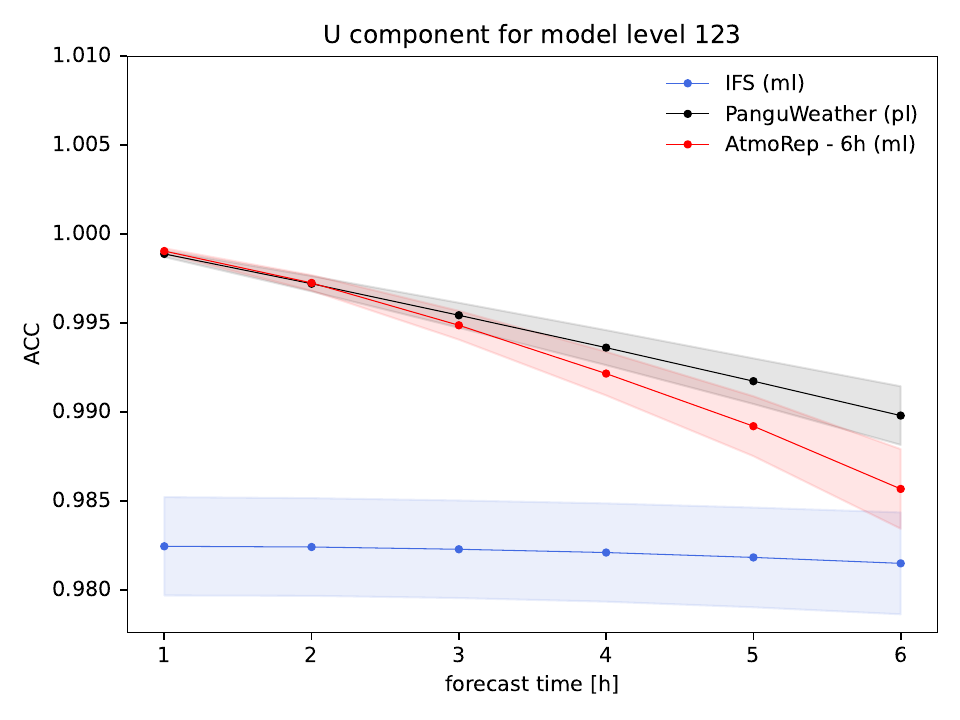}
 \includegraphics[width=0.19\textwidth]{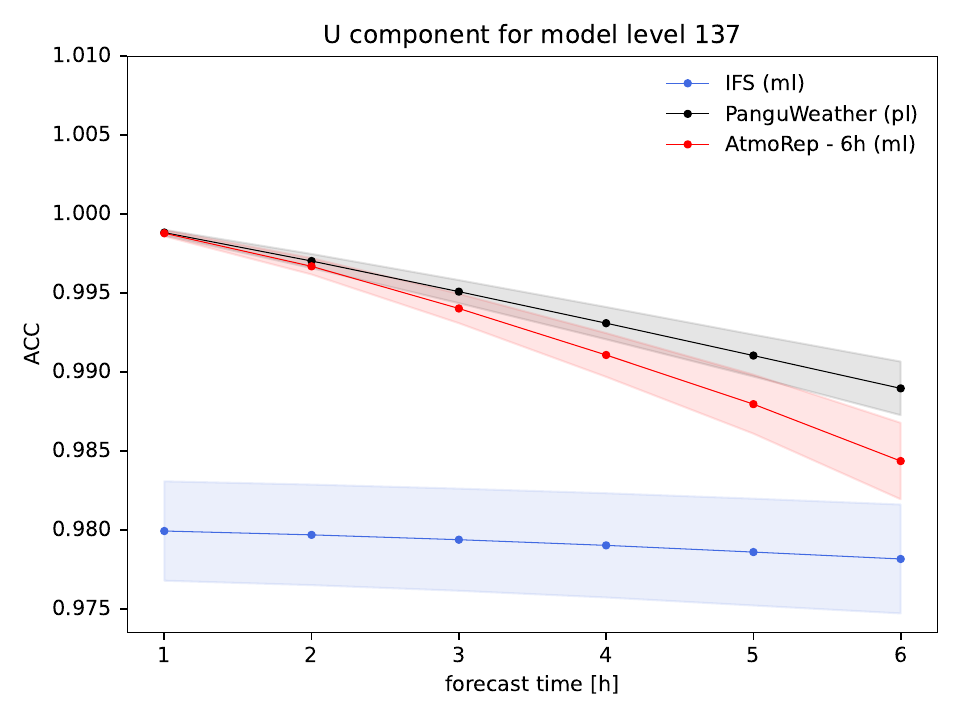}
 \hfill
 \includegraphics[width=0.19\textwidth]{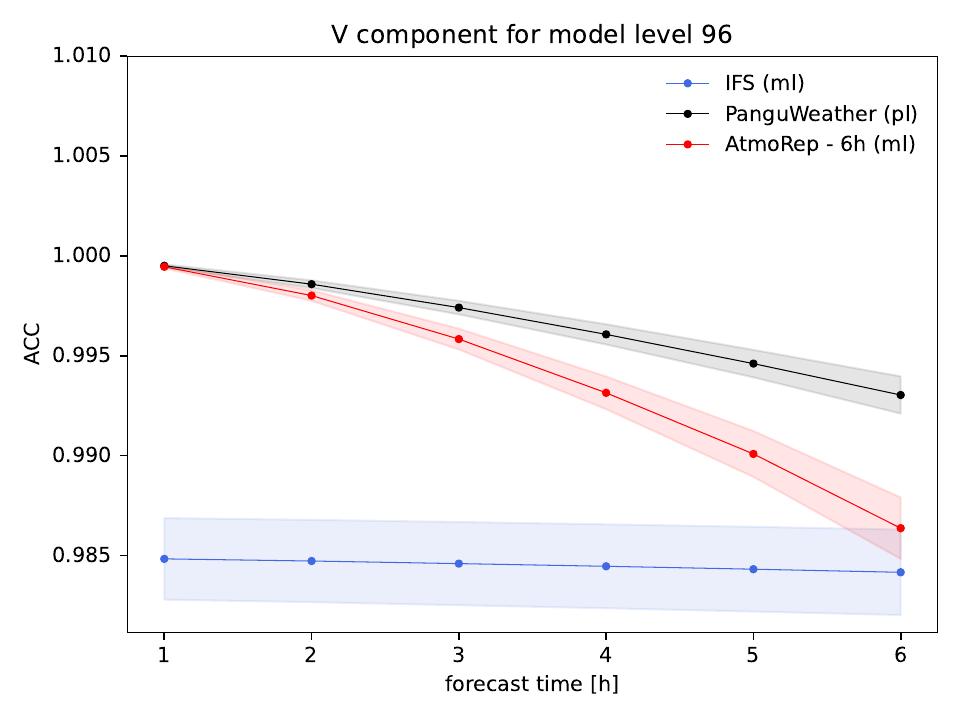}
 \includegraphics[width=0.19\textwidth]{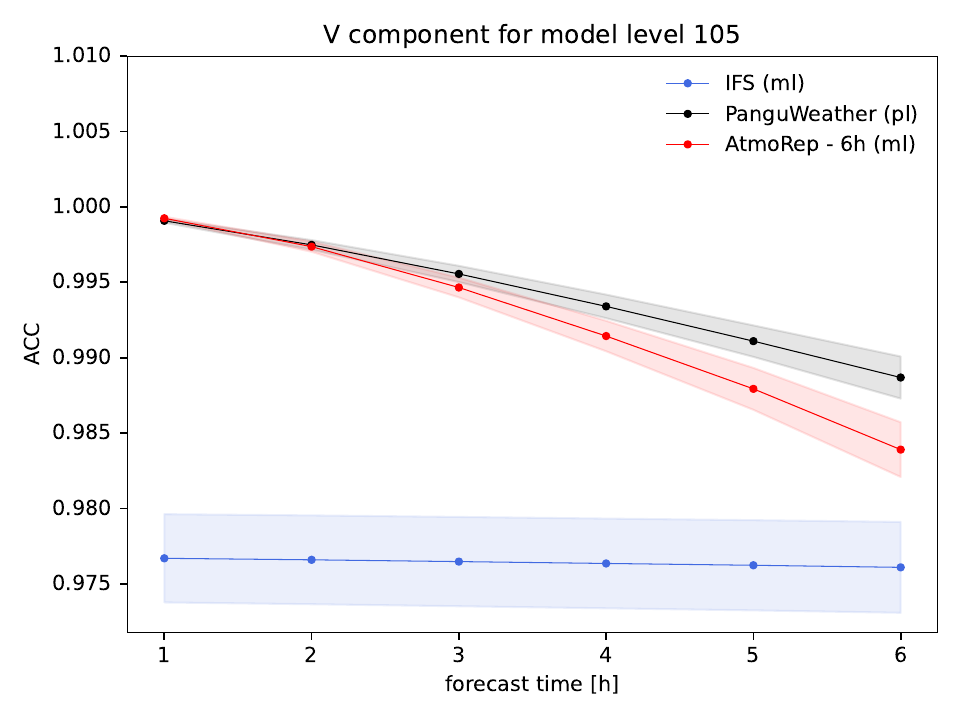}
 \includegraphics[width=0.19\textwidth]{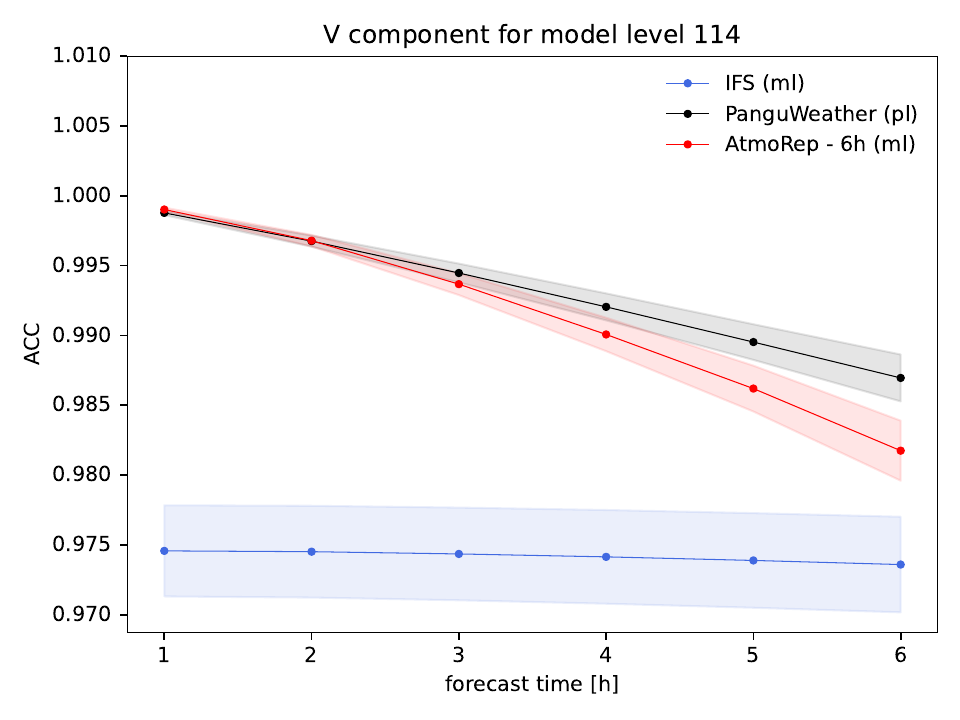}
 \includegraphics[width=0.19\textwidth]{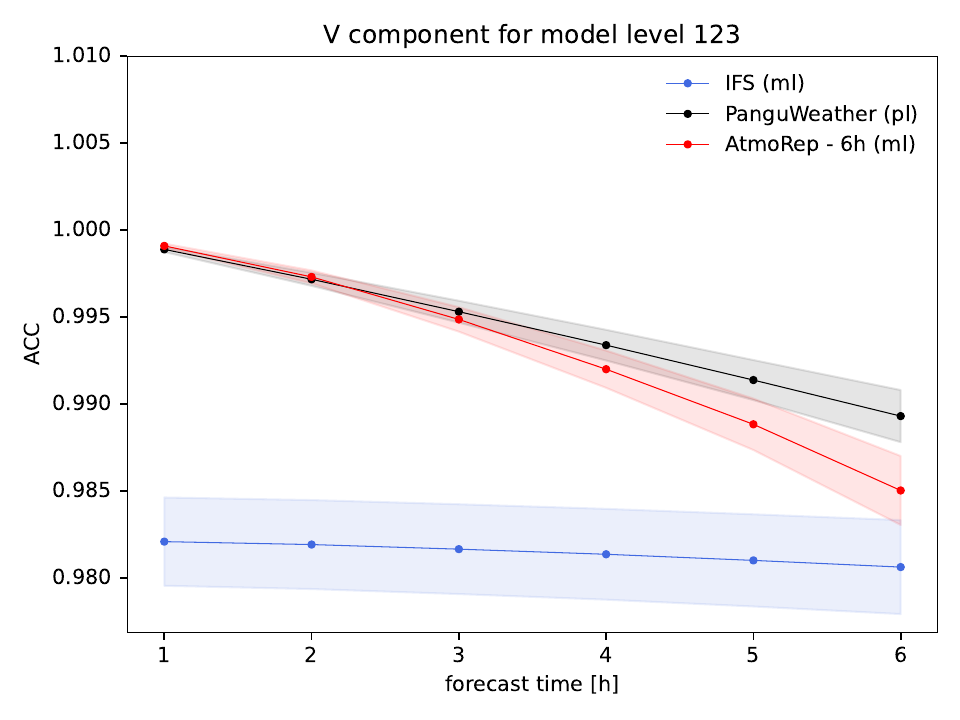}
 \includegraphics[width=0.19\textwidth]{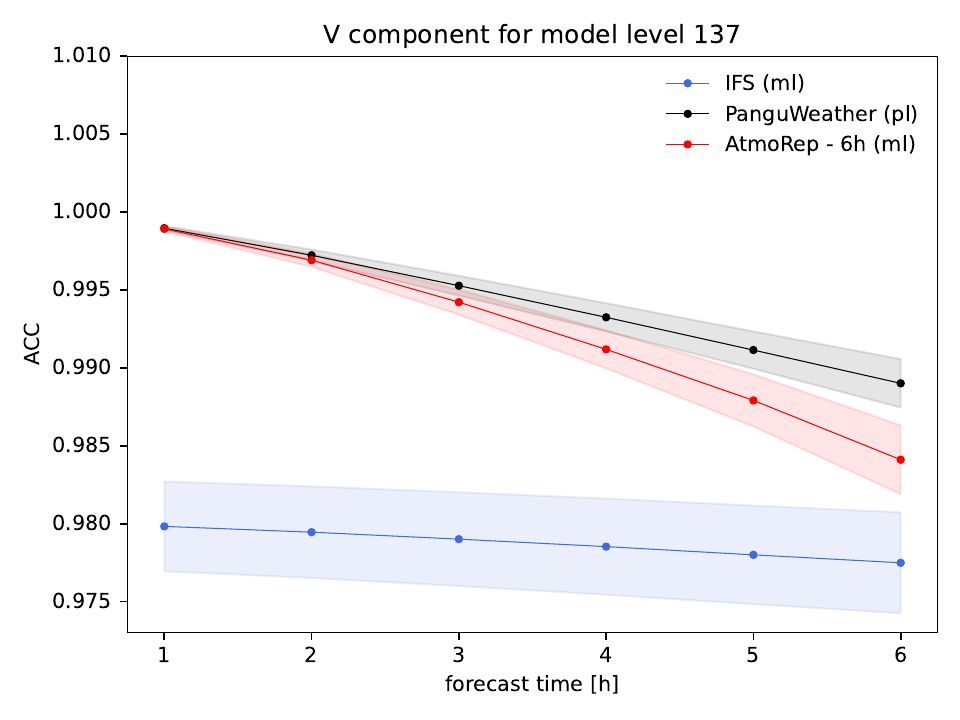}
  \hfill
  \\[10pt]
%
 \includegraphics[width=0.19\textwidth]{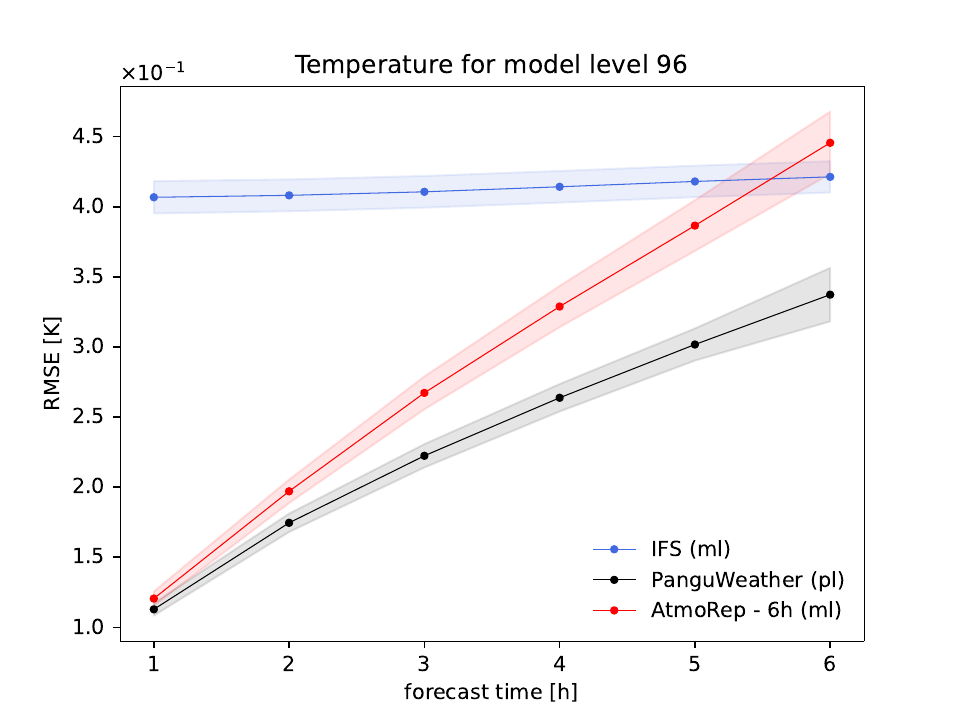}
 \includegraphics[width=0.19\textwidth]{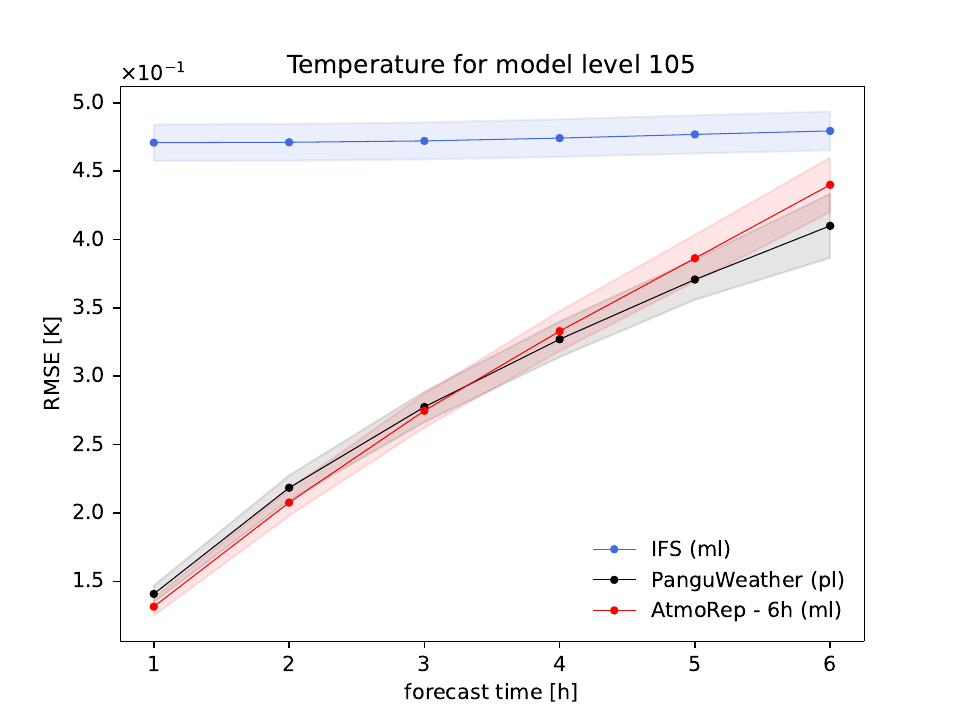}
 \includegraphics[width=0.19\textwidth]{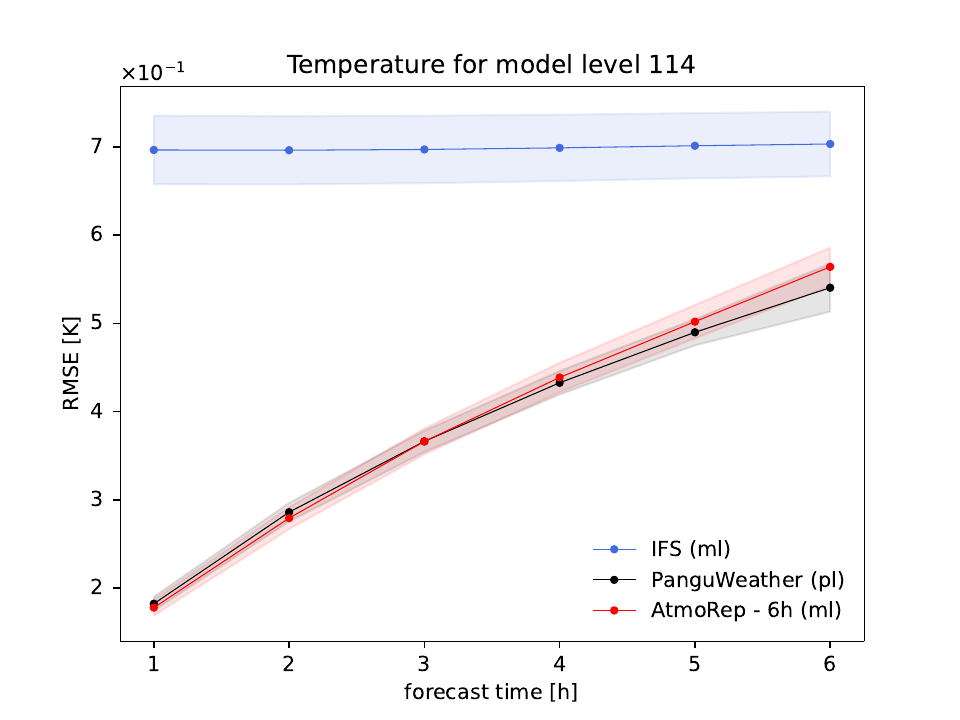}
 \includegraphics[width=0.19\textwidth]{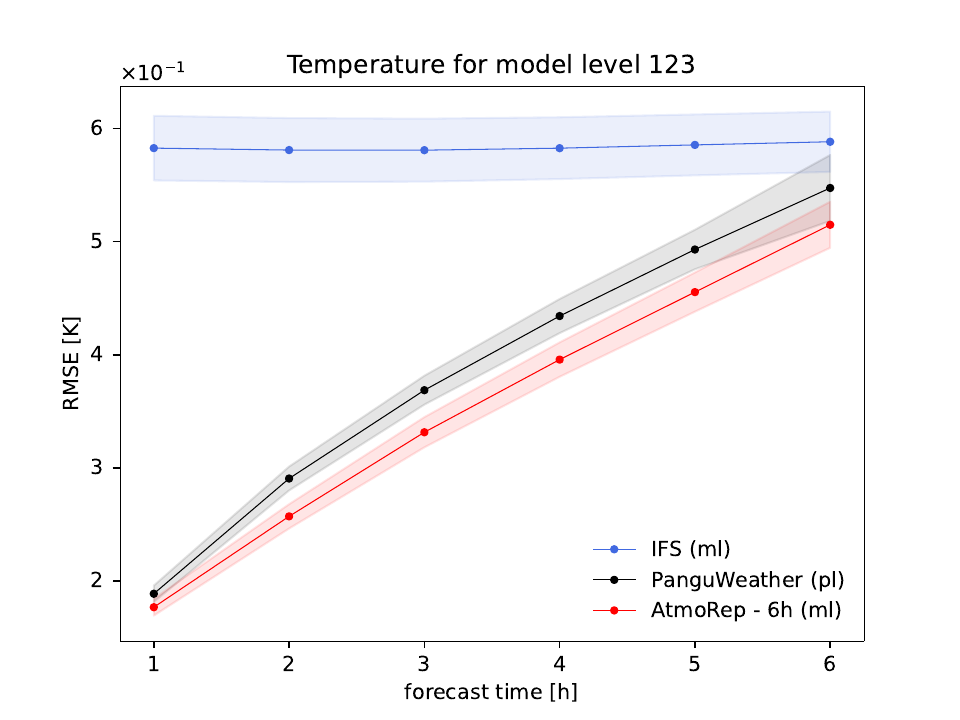}
 \includegraphics[width=0.19\textwidth]{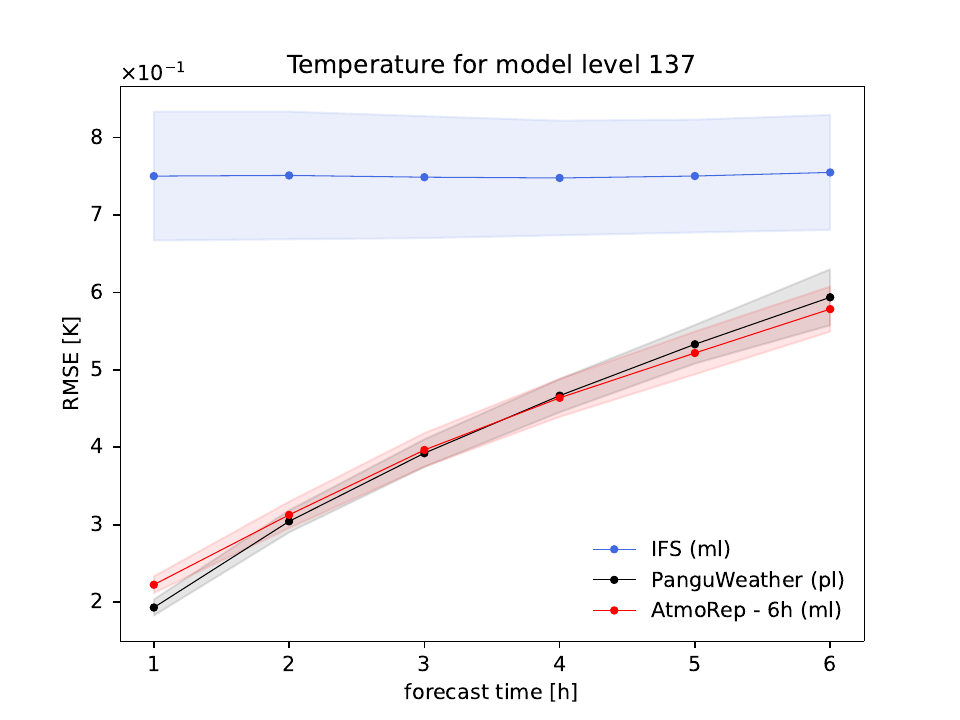}
 \hfill
 \includegraphics[width=0.19\textwidth]{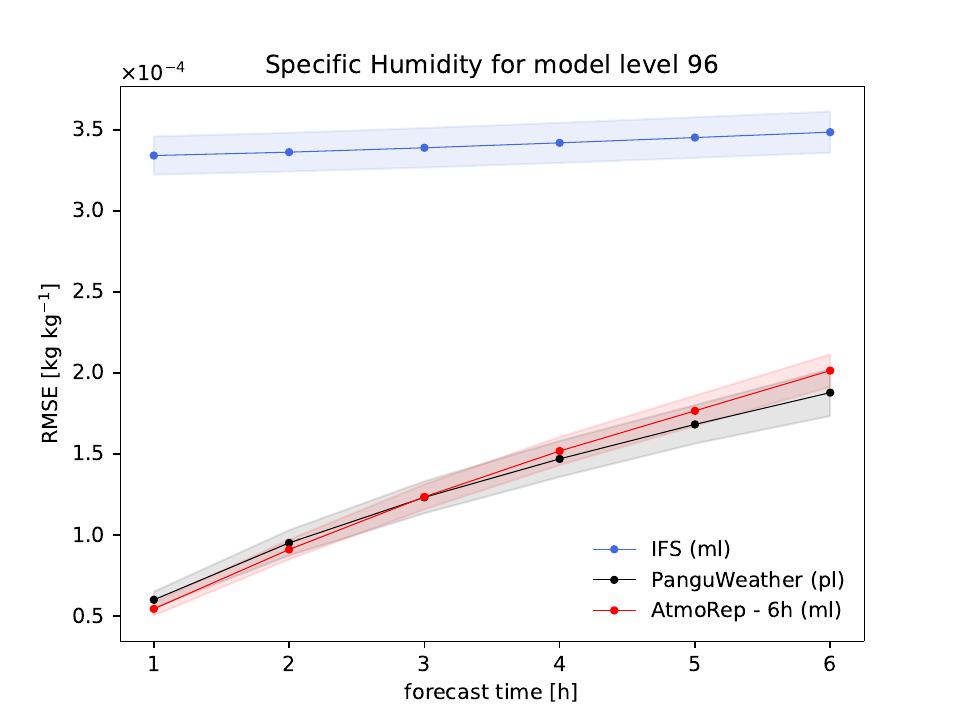}
 \includegraphics[width=0.19\textwidth]{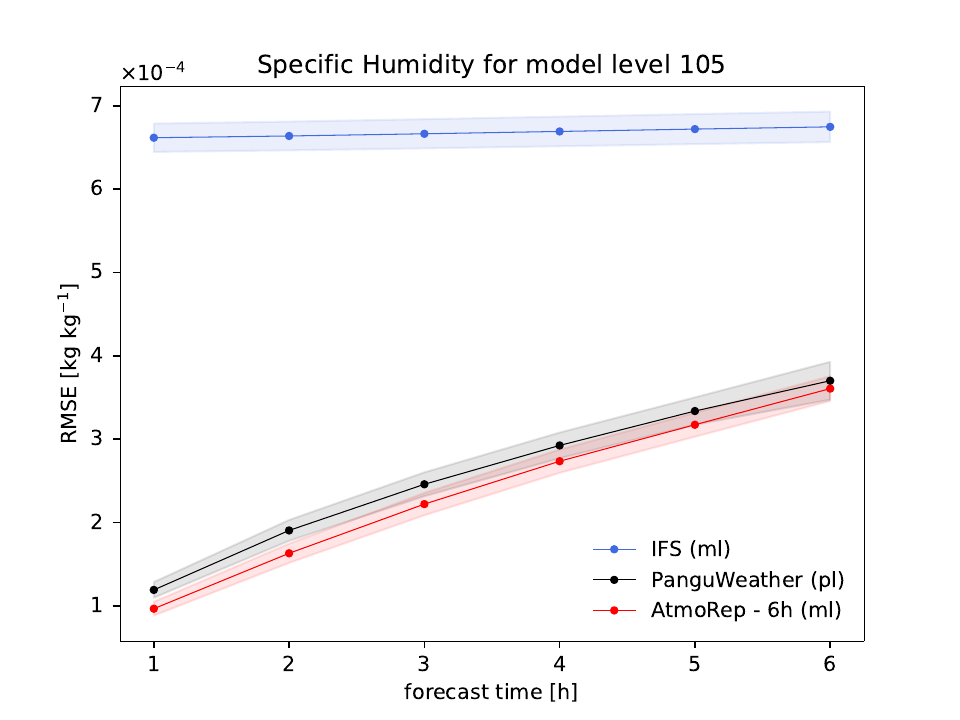}
\includegraphics[width=0.19\textwidth]{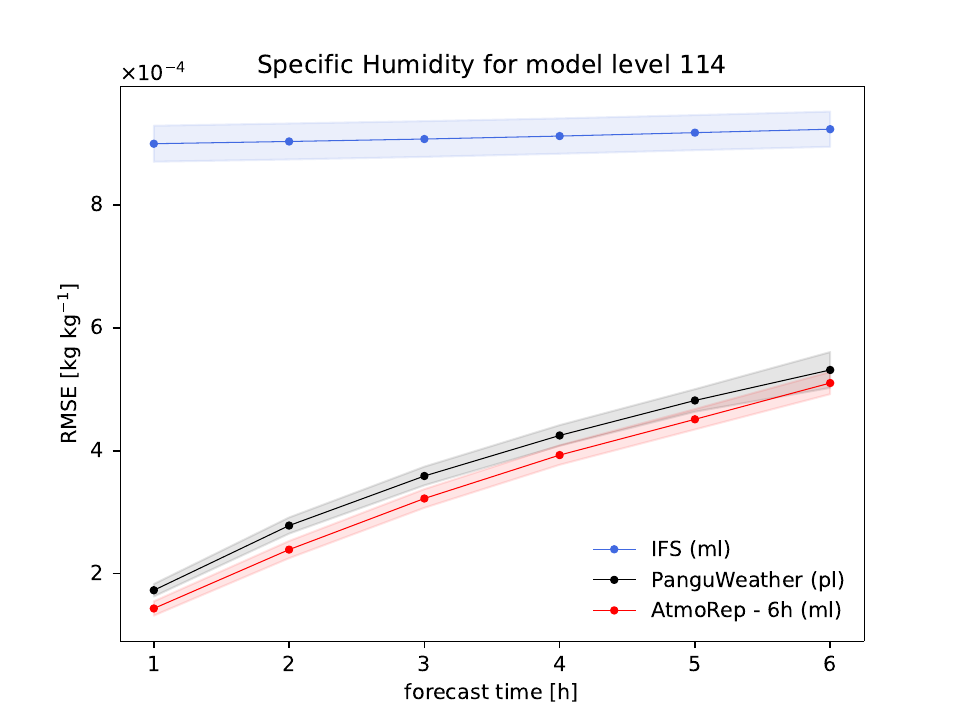}
 \includegraphics[width=0.19\textwidth]{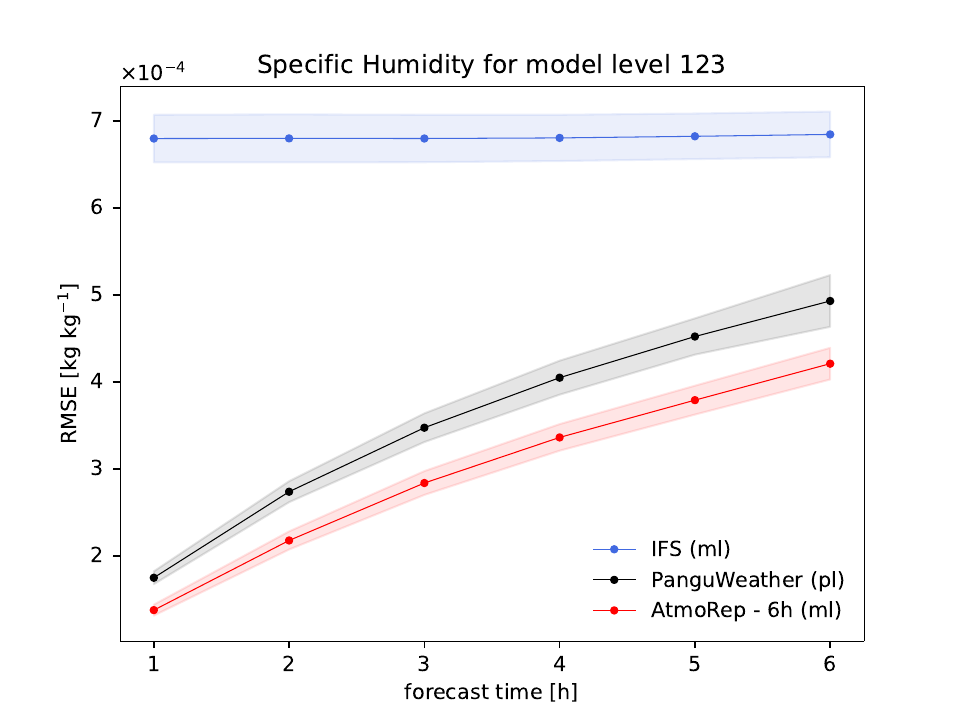}
 \includegraphics[width=0.19\textwidth]{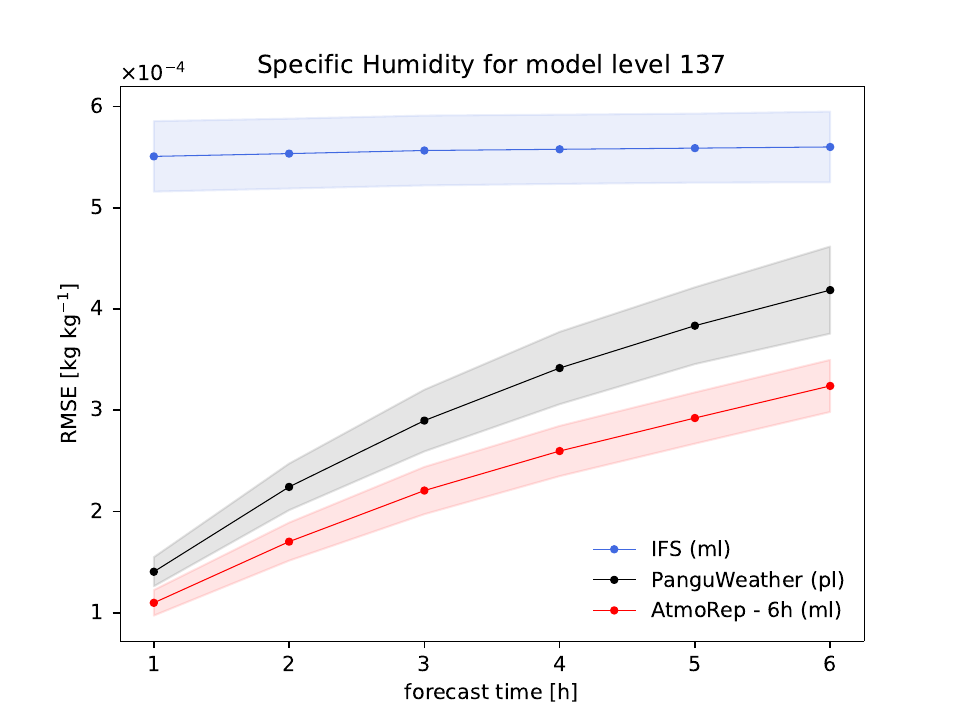}
 \hfill                                                                                     
 \includegraphics[width=0.19\textwidth]{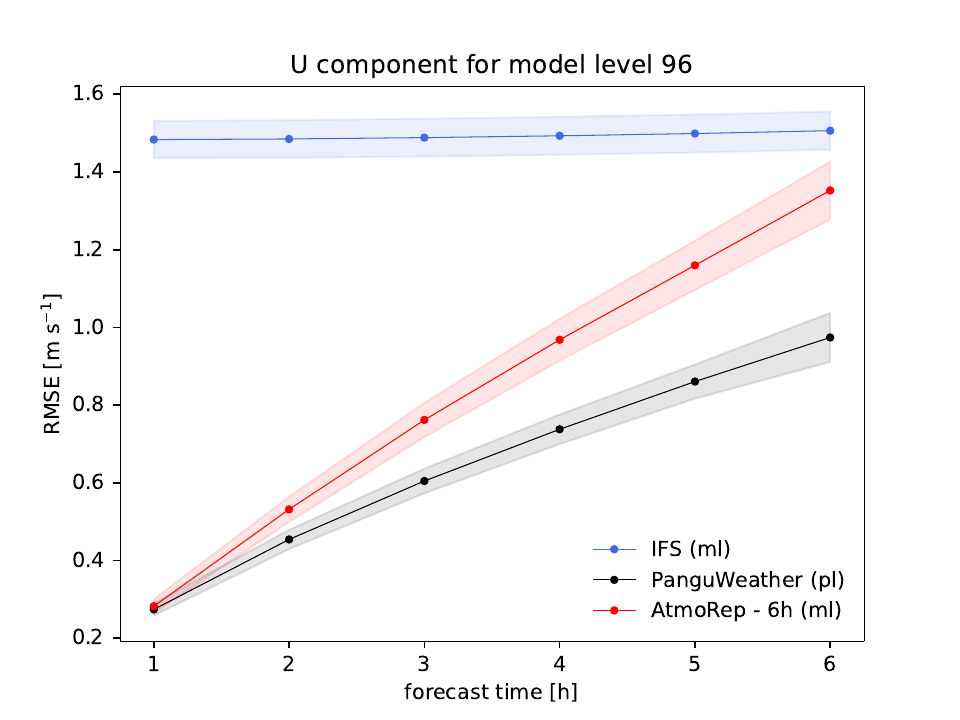}
 \includegraphics[width=0.19\textwidth]{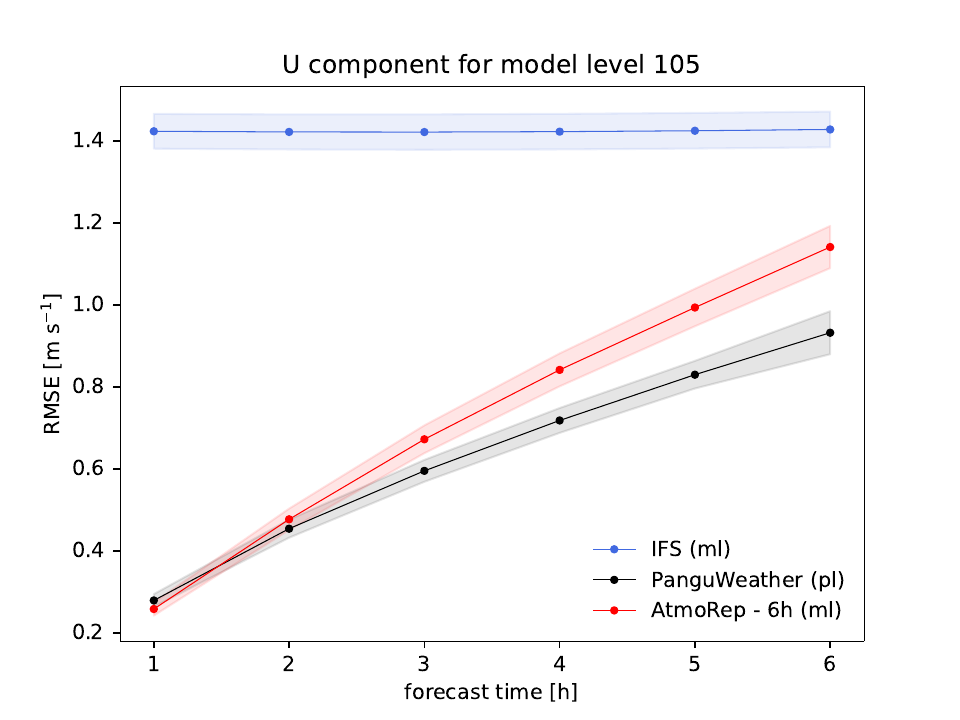}
 \includegraphics[width=0.19\textwidth]{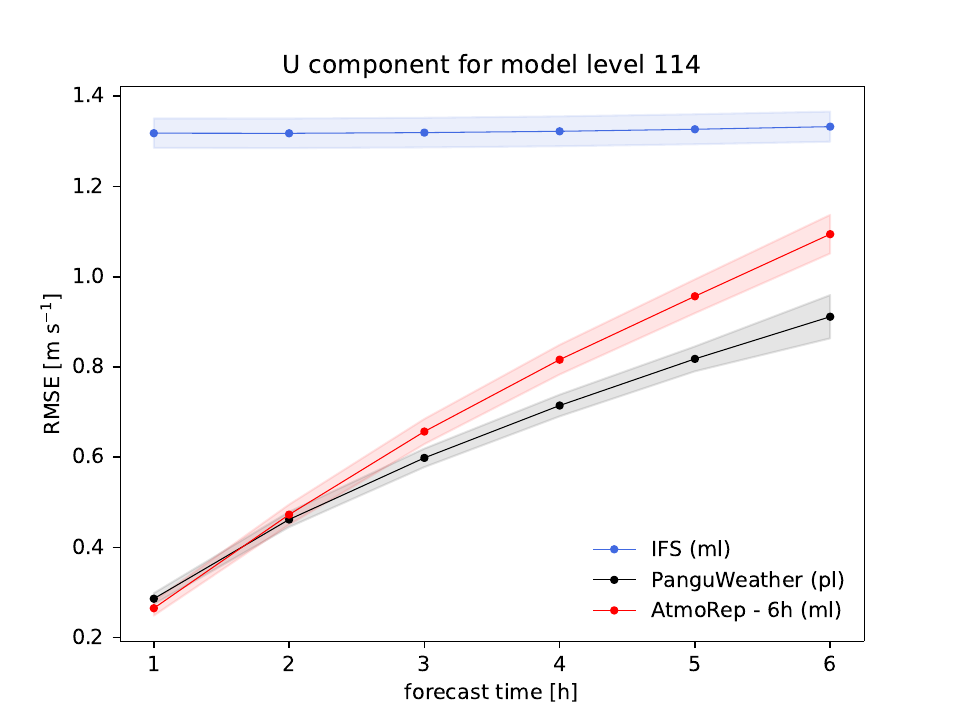}
 \includegraphics[width=0.19\textwidth]{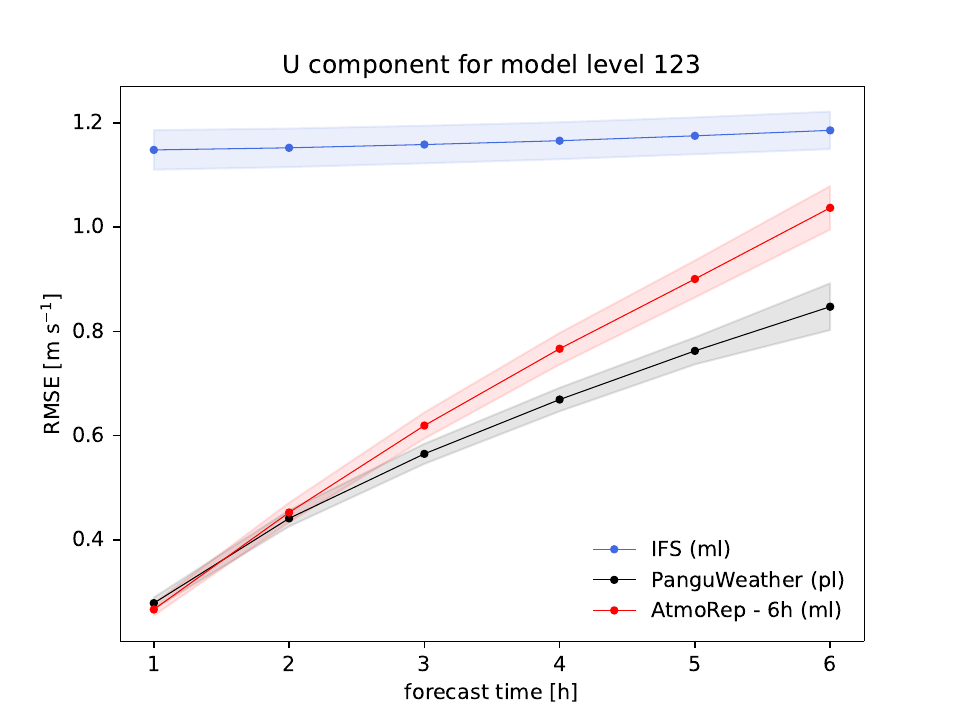}
 \includegraphics[width=0.19\textwidth]{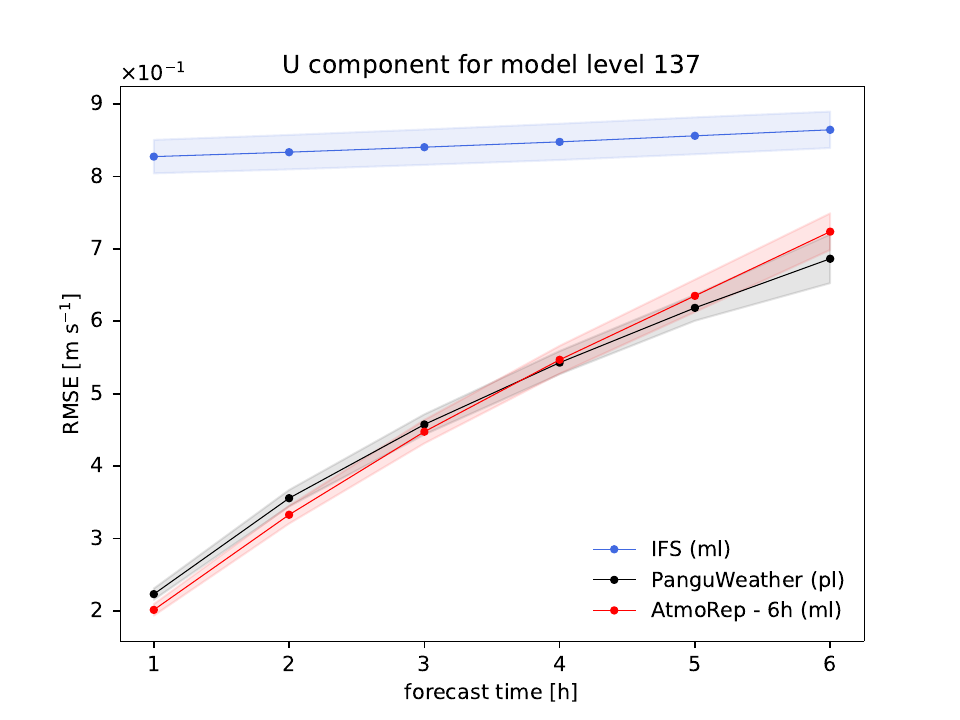}
 \hfill                                                                                    
 \includegraphics[width=0.19\textwidth]{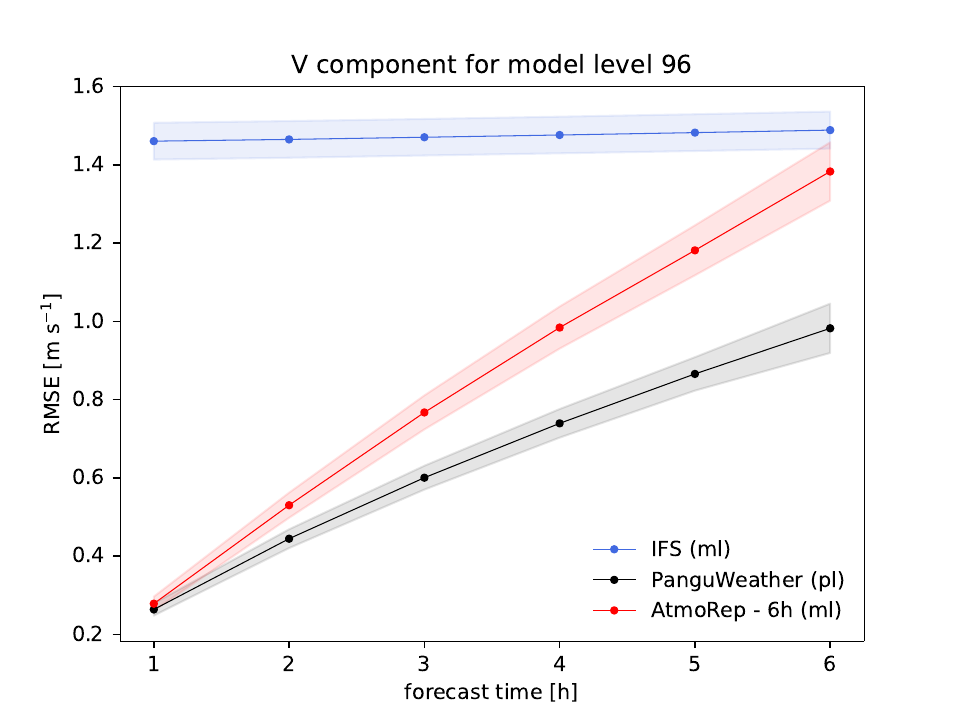}
 \includegraphics[width=0.19\textwidth]{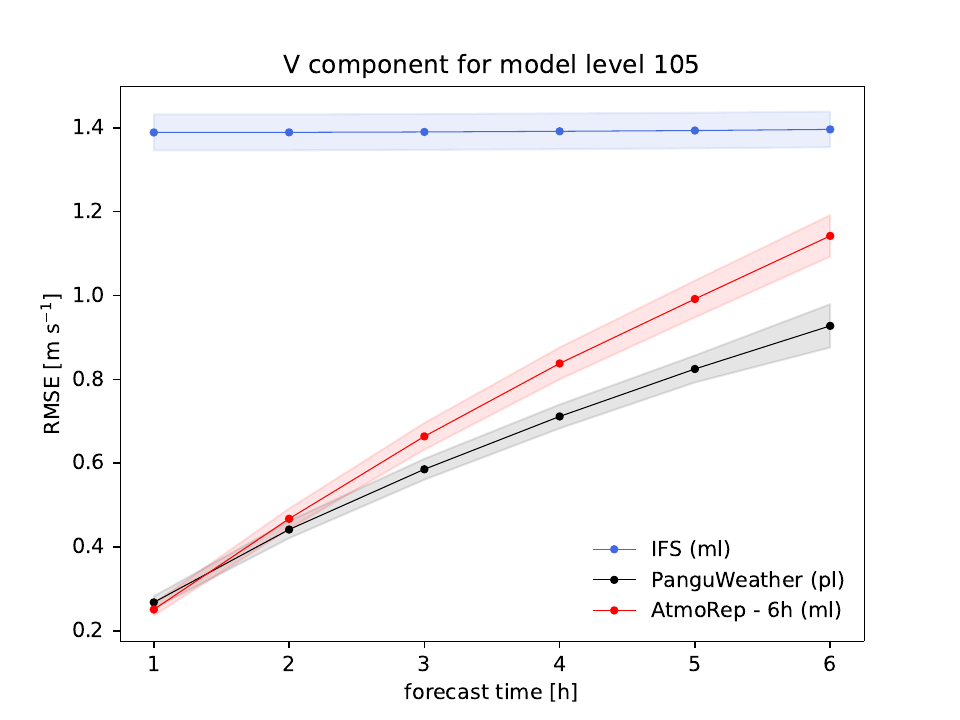}
 \includegraphics[width=0.19\textwidth]{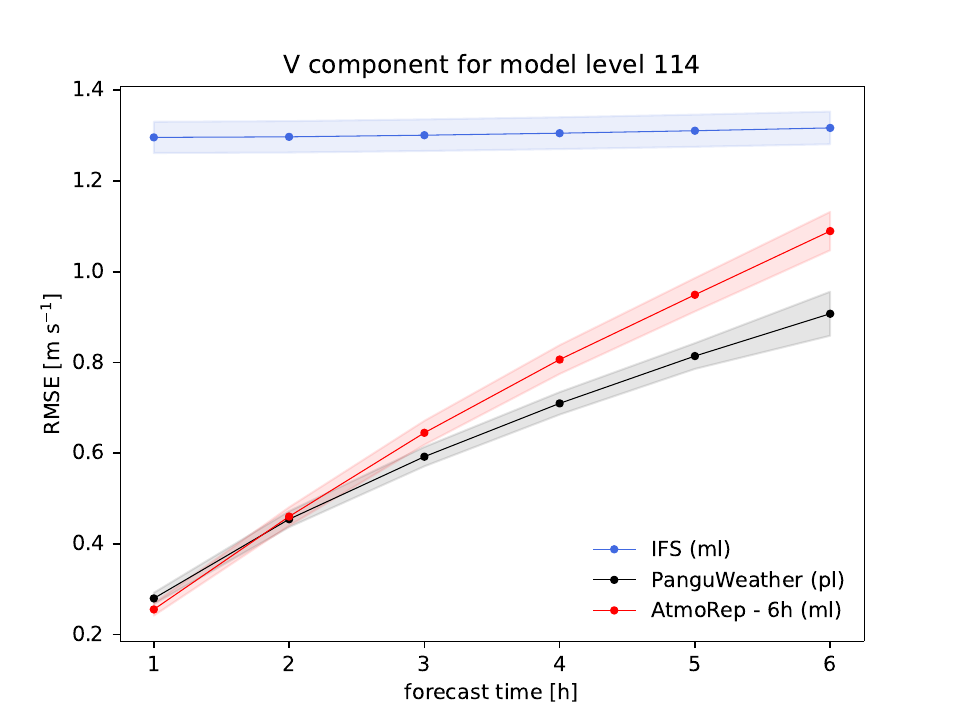}
 \includegraphics[width=0.19\textwidth]{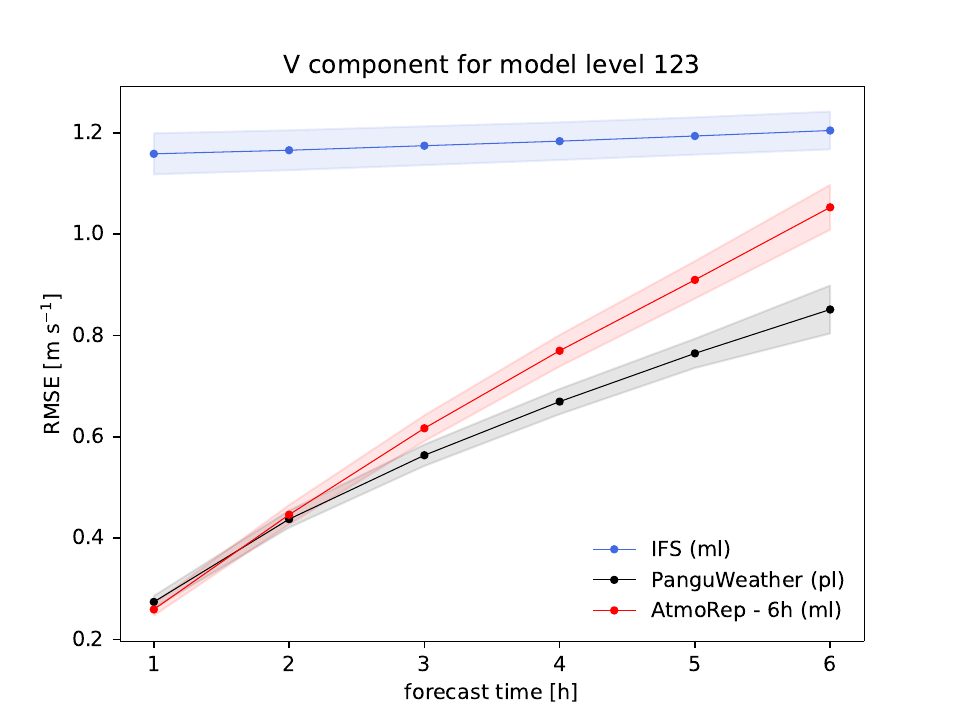}
 \includegraphics[width=0.19\textwidth]{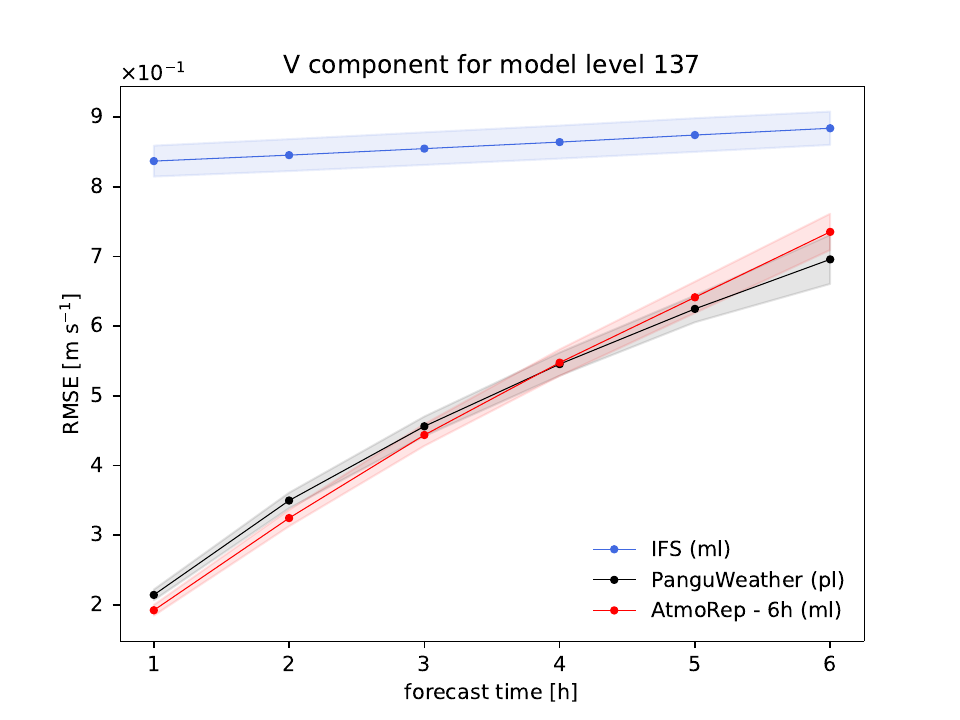}
 %
\caption{Detailed forecast evaluation compared to Pangu-Weather (black) and IFS (blue). The top four rows show ACC for several variables and levels and the bottom four rows RMSE. Also indicated is the standard deviation in each case (shaded areas). }
\label{fig:ext-fceval-RMSE}
\end{figure}

\begin{figure}[htpb!]
 \begin{center}
 \includegraphics[trim={1cm 0cm 1.0cm 0cm}, width=0.18\textwidth]{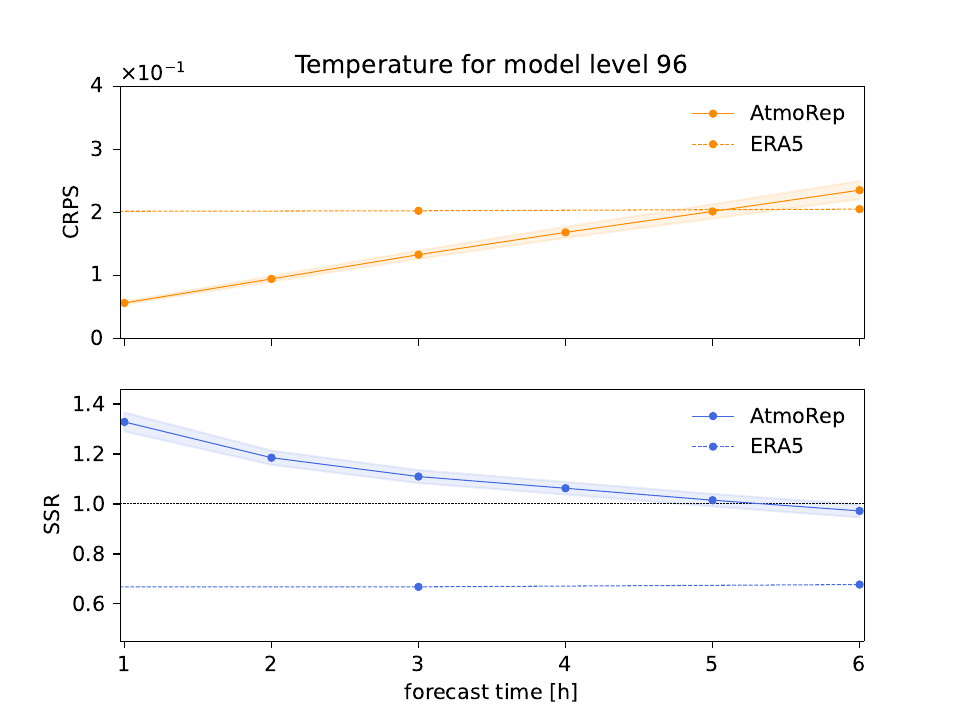}
 \includegraphics[trim={1cm 0cm 1.0cm 0cm}, width=0.18\textwidth]{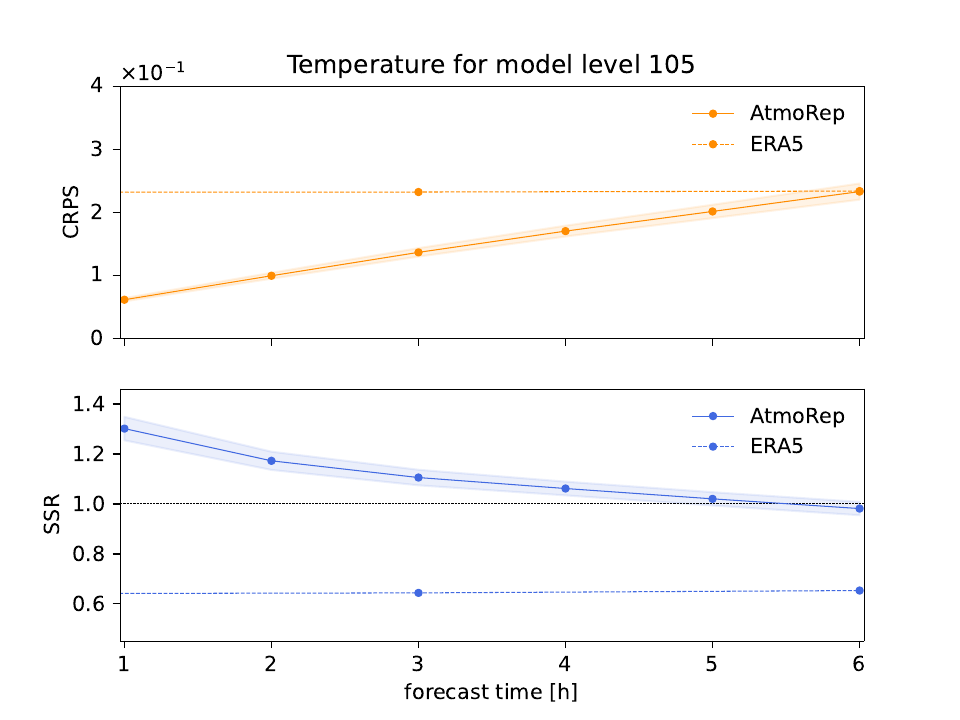}
 \includegraphics[trim={1cm 0cm 1.0cm 0cm}, width=0.18\textwidth]{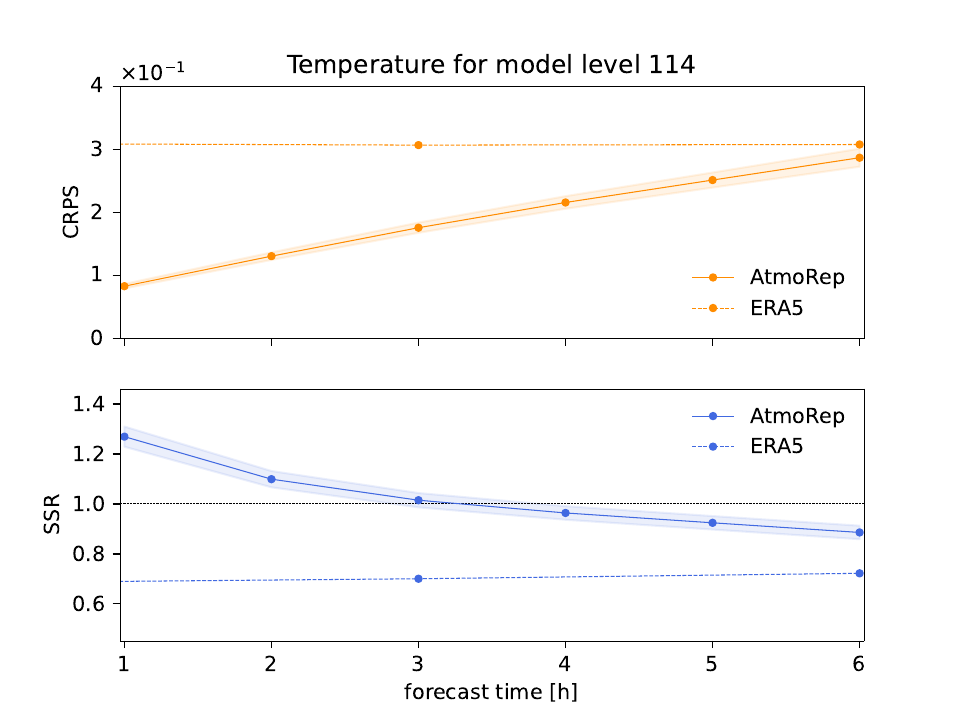}
 \includegraphics[trim={1cm 0cm 1.0cm 0cm}, width=0.18\textwidth]{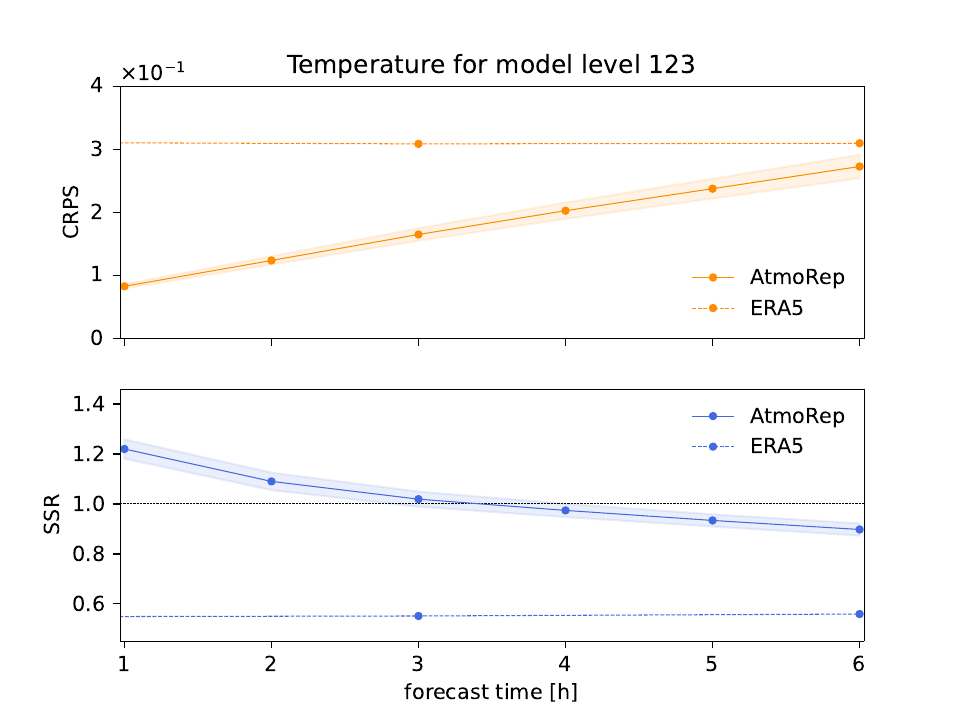}
 \includegraphics[trim={1cm 0cm 1.0cm 0cm}, width=0.18\textwidth]{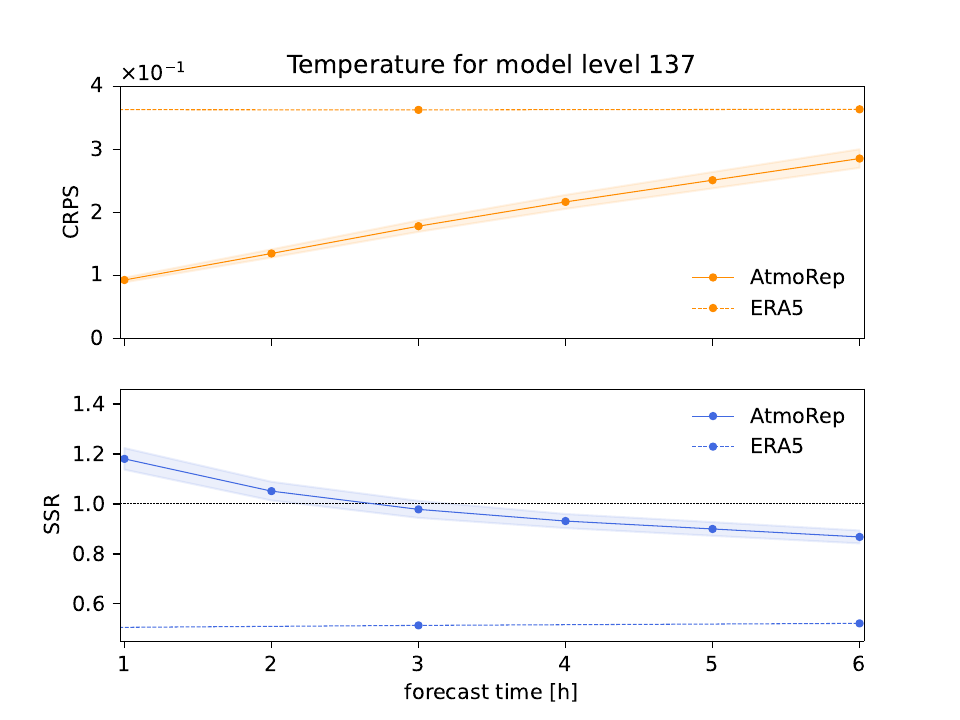}
 \hfill
 \includegraphics[trim={1cm 0cm 1.0cm 0cm}, width=0.18\textwidth]{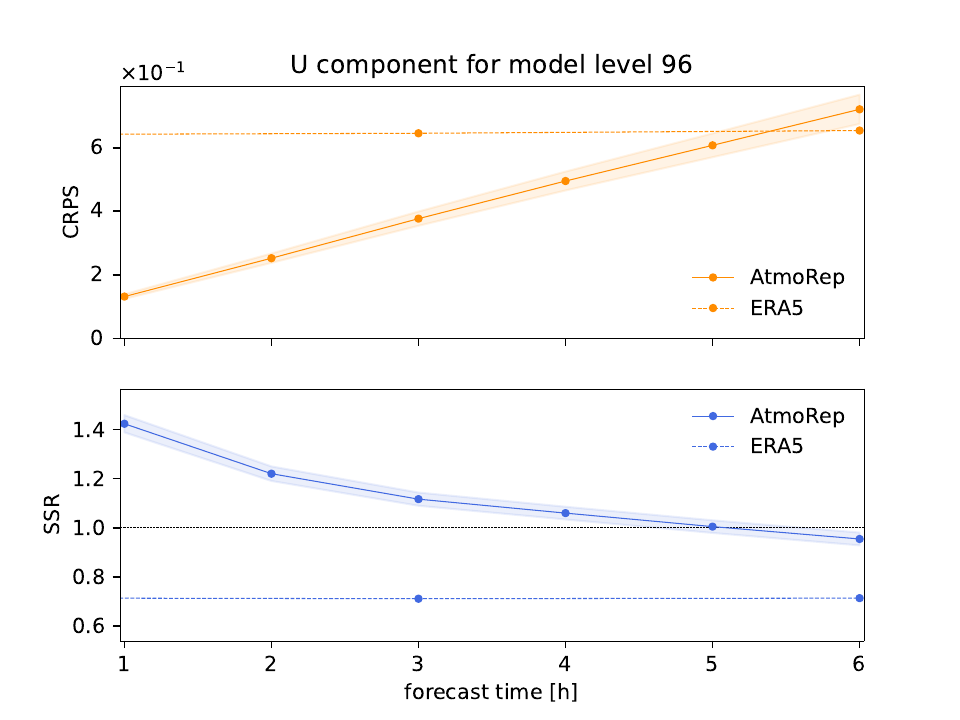}
 \includegraphics[trim={1cm 0cm 1.0cm 0cm}, width=0.18\textwidth]{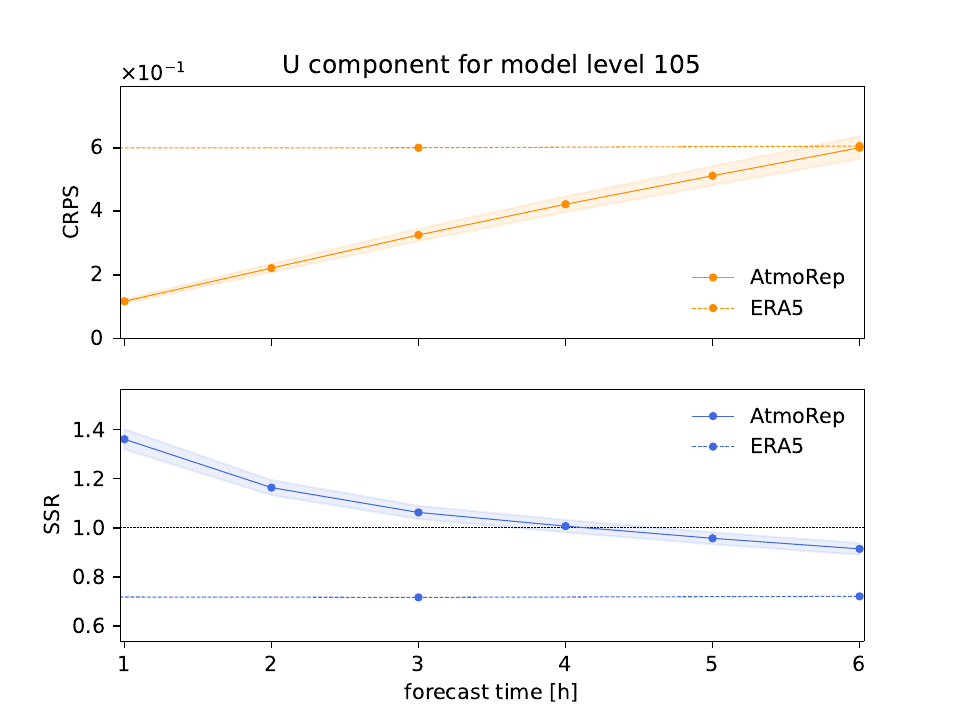}
 \includegraphics[trim={1cm 0cm 1.0cm 0cm}, width=0.18\textwidth]{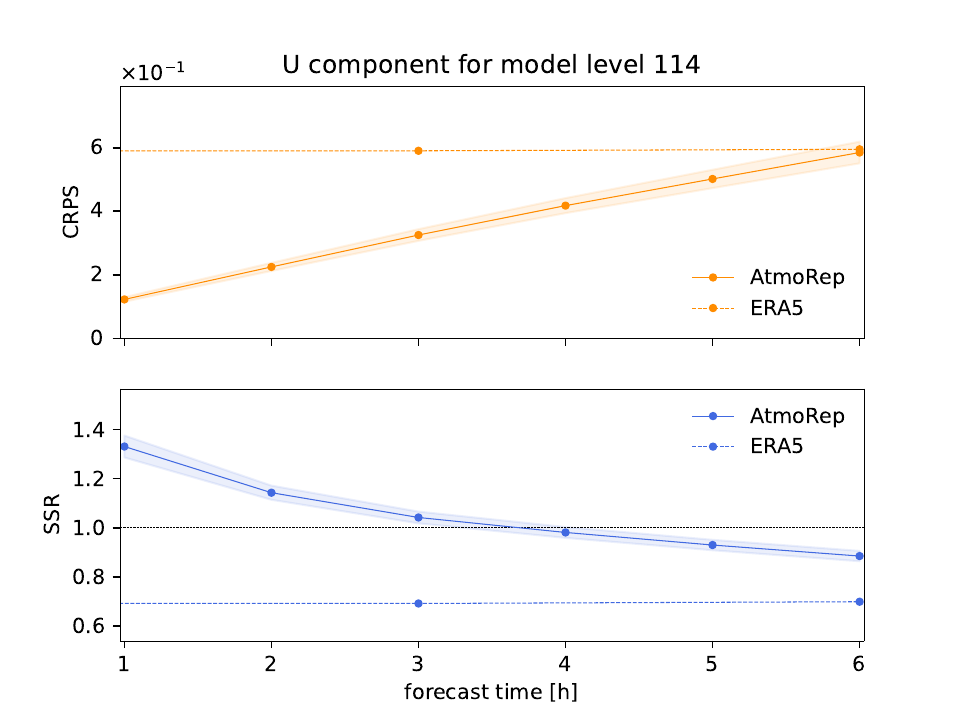}
 \includegraphics[trim={1cm 0cm 1.0cm 0cm}, width=0.18\textwidth]{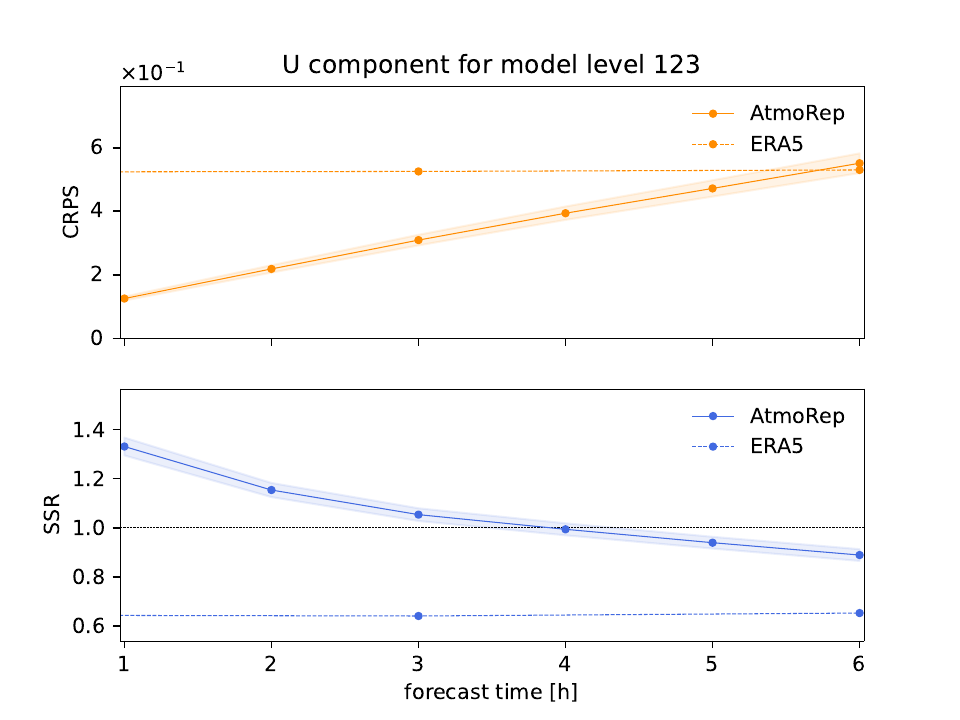}
 \includegraphics[trim={1cm 0cm 1.0cm 0cm}, width=0.18\textwidth]{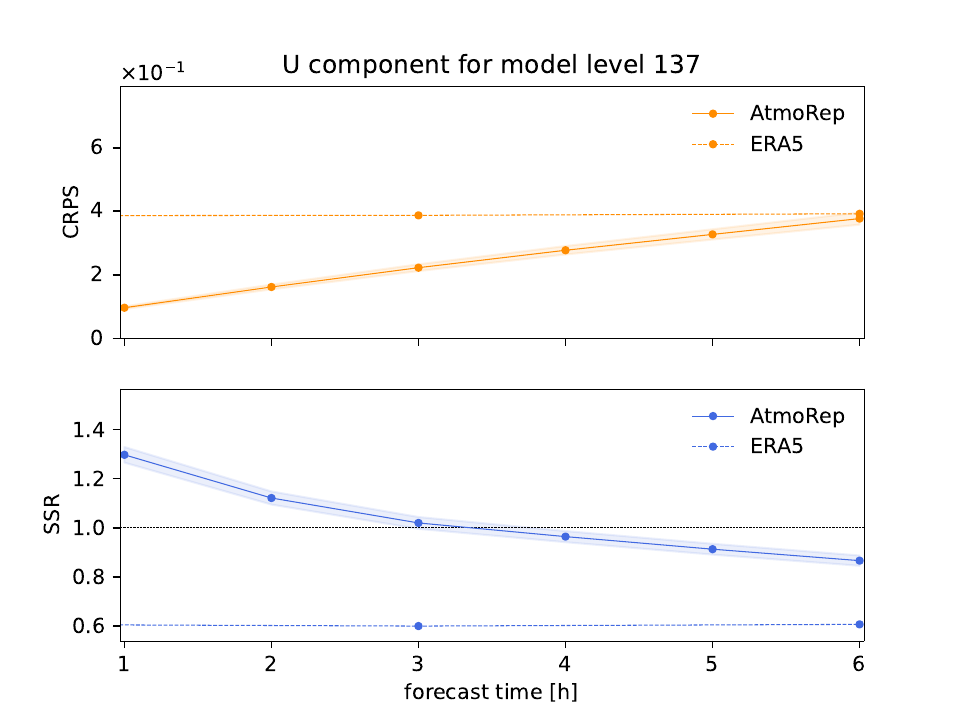}
 \hfill
 \includegraphics[trim={1cm 0cm 1.0cm 0cm}, width=0.18\textwidth]{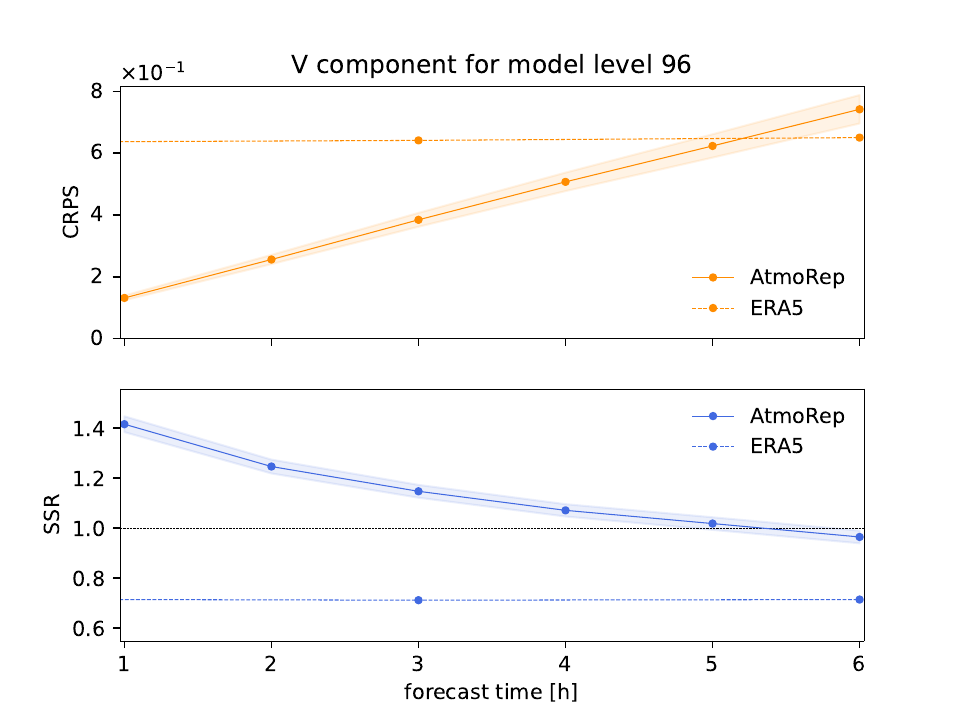}
 \includegraphics[trim={1cm 0cm 1.0cm 0cm}, width=0.18\textwidth]{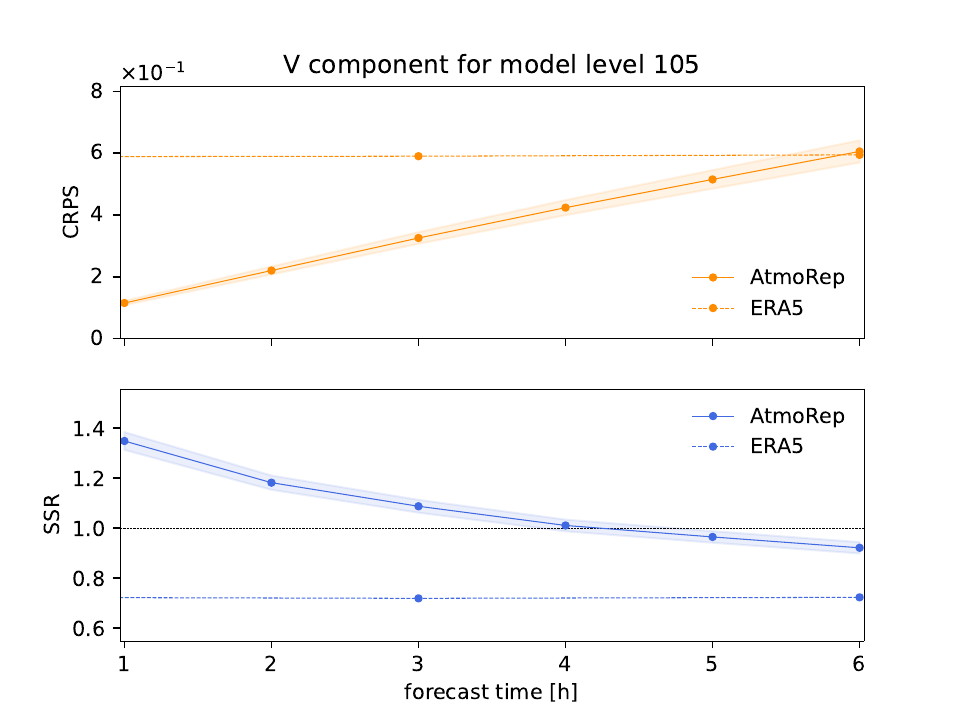}
 \includegraphics[trim={1cm 0cm 1.0cm 0cm}, width=0.18\textwidth]{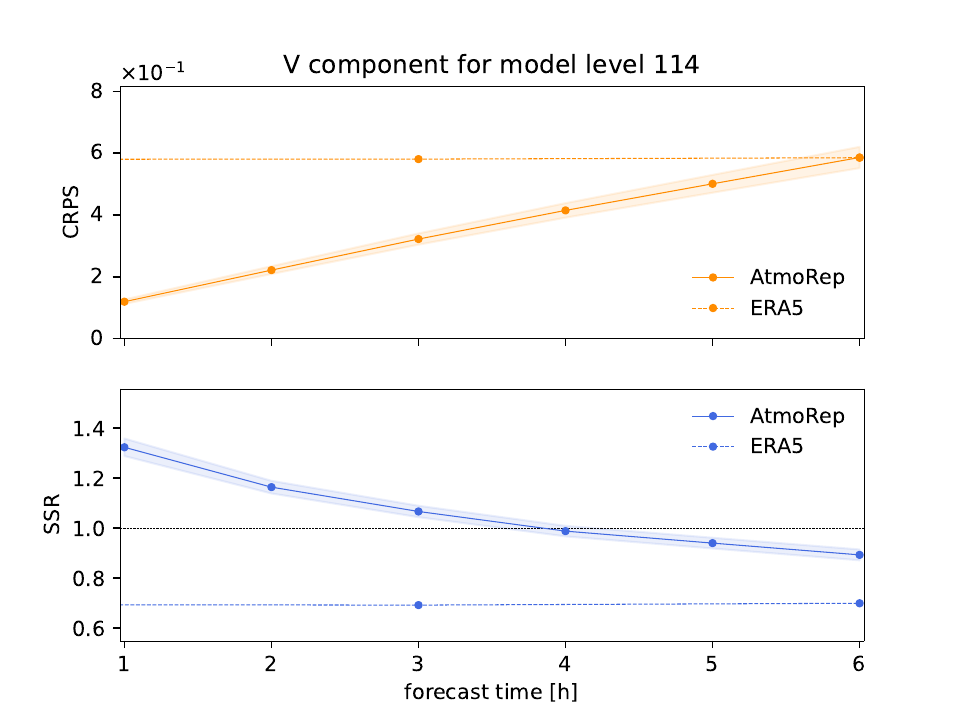}
 \includegraphics[trim={1cm 0cm 1.0cm 0cm}, width=0.18\textwidth]{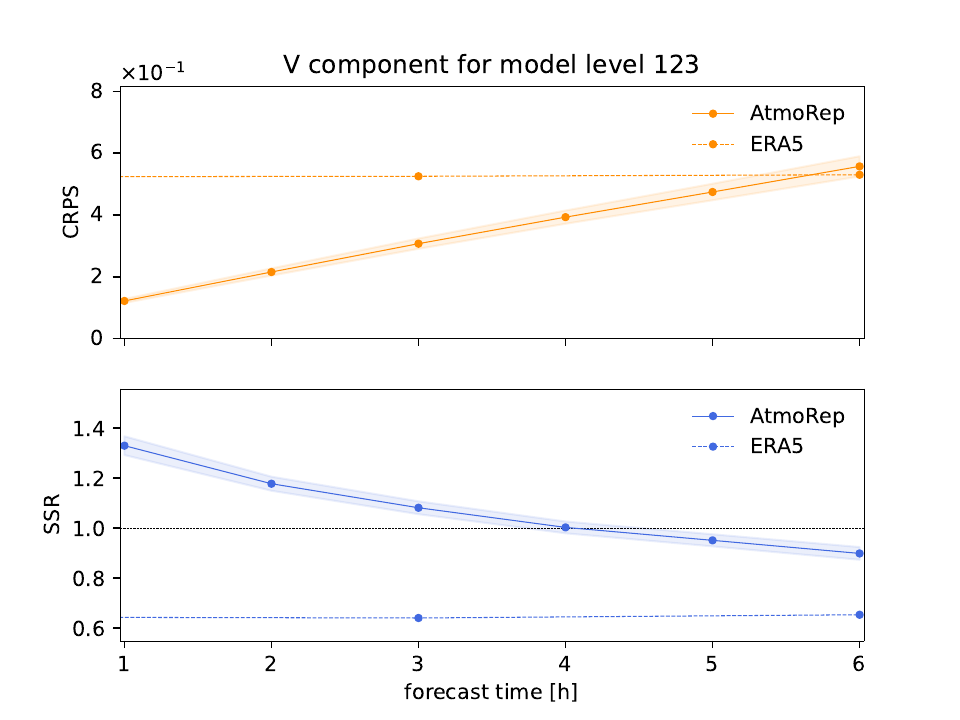}
 \includegraphics[trim={1cm 0cm 1.0cm 0cm}, width=0.18\textwidth]{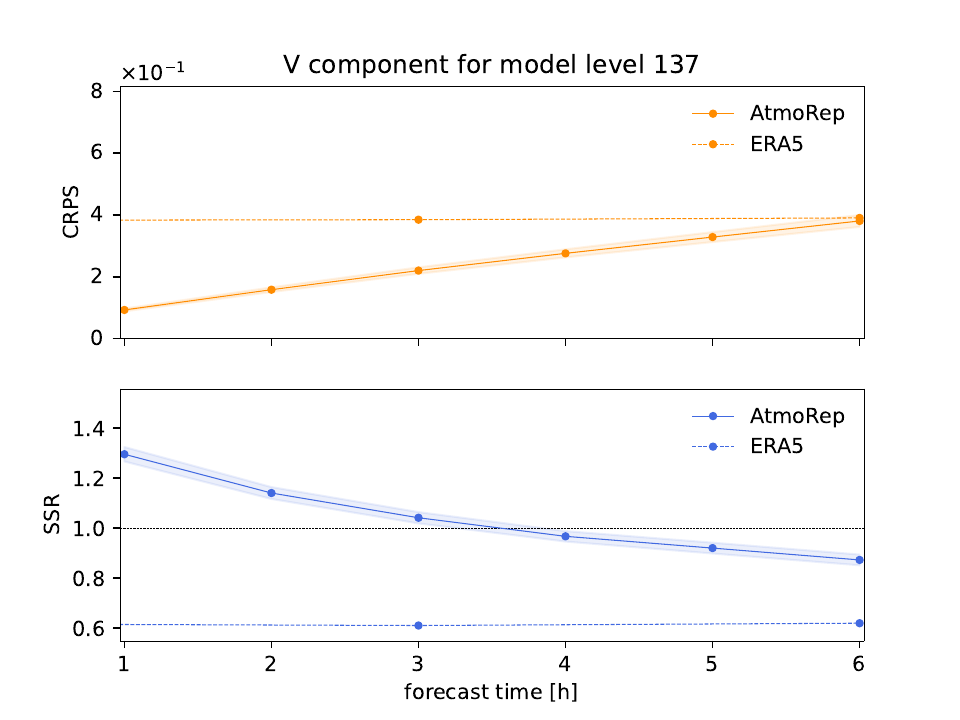}
 
    \caption{Evaluation of the AtmoRep ensemble using SSR (blue) the CRPS (red) for short-term forecasting. The dashed lines represent the respective metrics for the ERA5 forecast ensemble, which is available only with a time step of $3 \, \mathrm{h}$. The CRPS has been computed from the ensemble mean and spread assuming a Gaussian distribution. Shaded areas depict one standard deviation.}
    \label{fig:ext-fccrps}
 \end{center}
\end{figure}


\begin{figure}[htpb!]
    \centering
   \includegraphics[width=\textwidth]{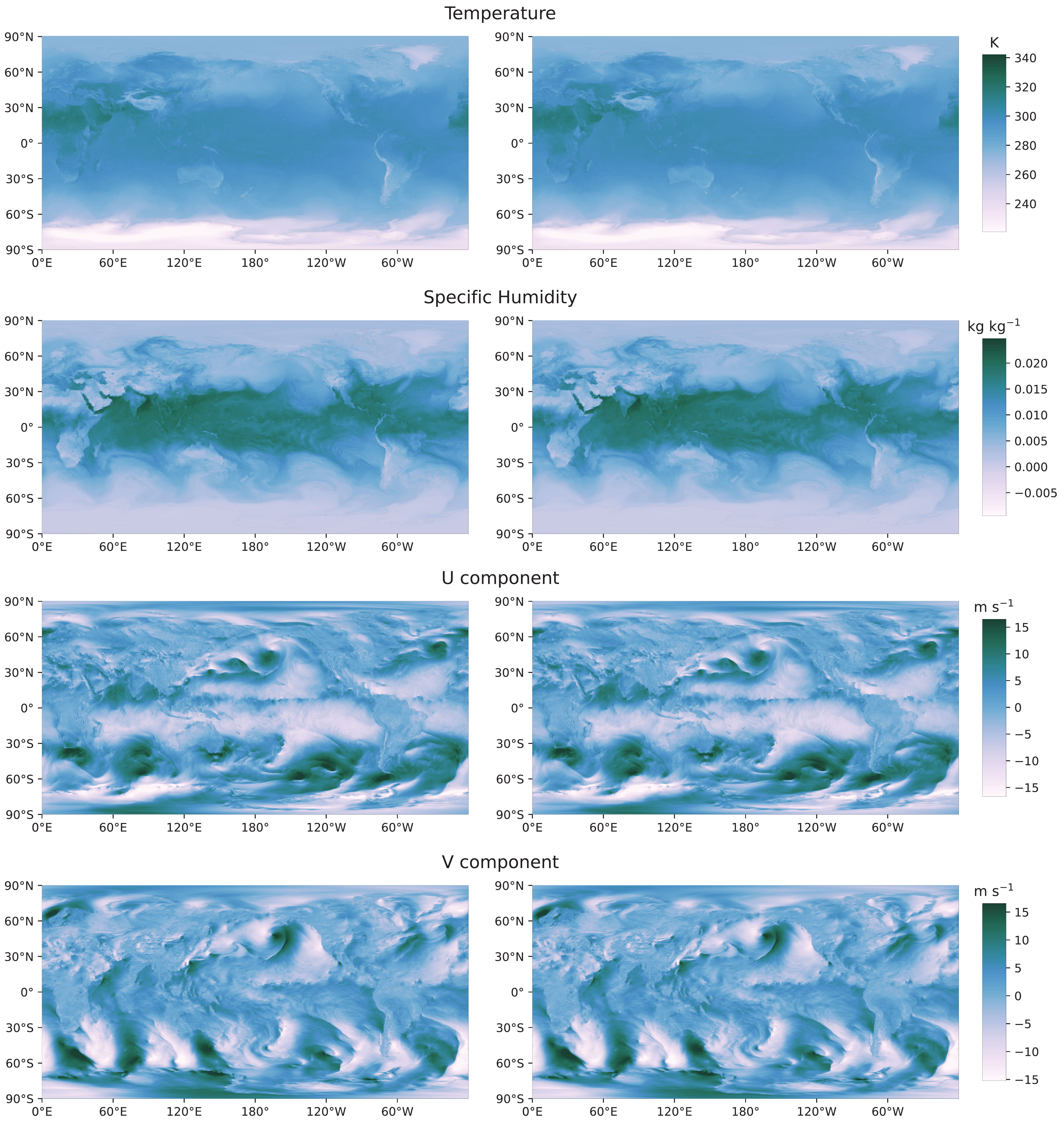}
    \caption{Example for $1\mathrm{h}$ short-term forecast for 15 June 2018 at 12:00 UTC for four variables at model level $137$. Left is the ERA5 ground truth and right the AtmoRep prediction in each case.}
    \label{fig:ext-fcmap}
\end{figure}

\newpage
\begin{figure}[htpb!]
    \centering
    \includegraphics[width=\textwidth]{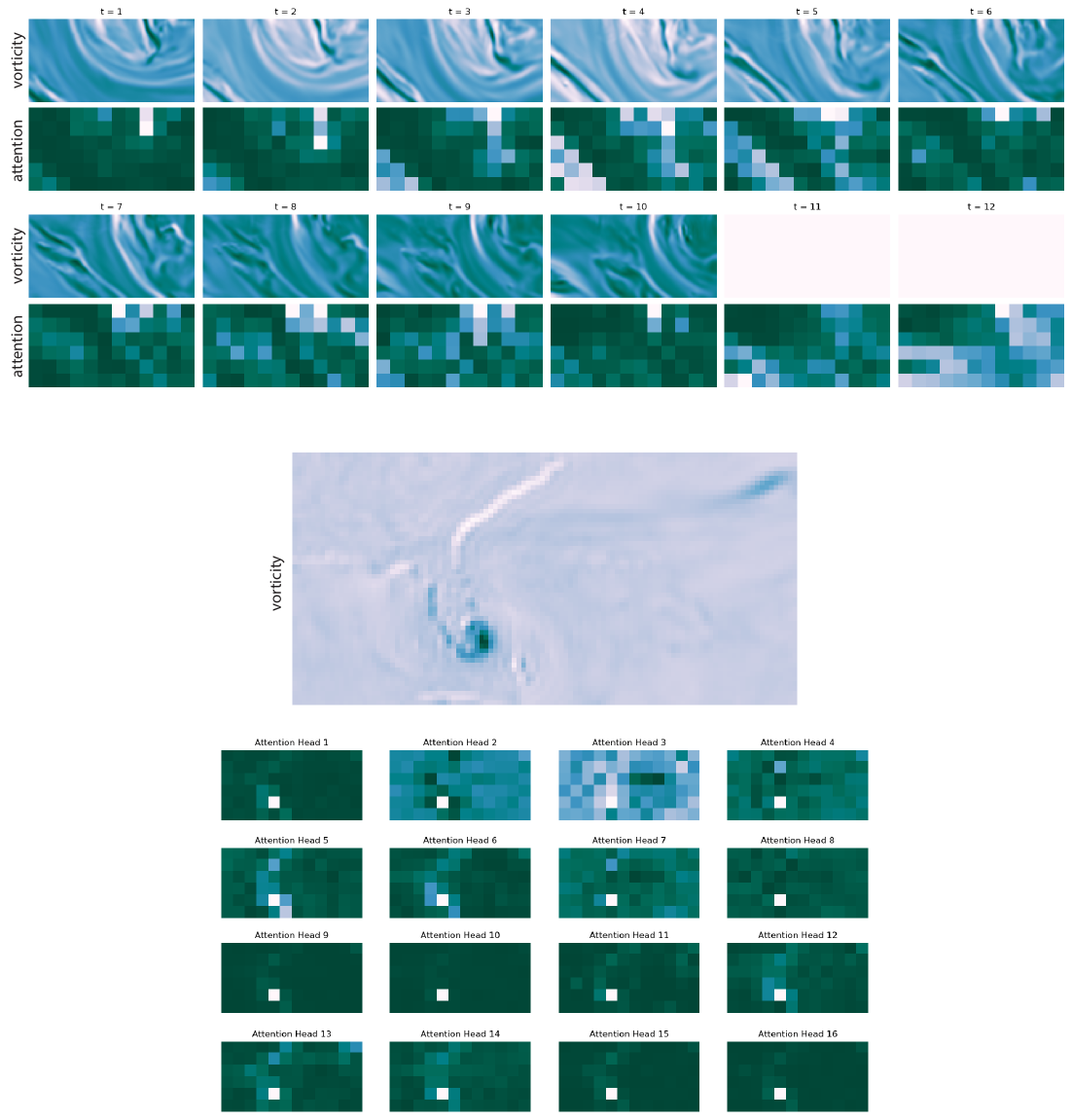}
    \caption{Attention maps provide a direct means to gain insight into what a trained network has learned and this has been used before both in natural language processing (e.g.~\cite{Vaswani2017}) and computer vision (e.g.~\cite{Caron2021}). 
    AtmoRep's attention maps are well aligned with physically relevant features, which we demonstrate with two examples. 
    To our knowledge, this has not been established before in Earth system science.
    The top figure shows attention maps of a large wave propagating through the atmosphere (vorticity, model level 96) for multiple time steps. 
      Shown is the average attention over all keys for a fixed head in the final decoder block in the model.
      The last two source images are masked since the model was evaluated with zero-shot forecasting.
    The bottom image shows attention for hurricane Katrina on August 26th, 2005 (vorticity, model level 137). The physical field is depicted at the top and the attention maps for the 16 attention heads underneath. 
    Shown is again the attention averaged over all keys and from the final decoder block of the model.
    Further details on the attention maps can be found in the supplementary material and more examples and the option to interactively explore these at~\url{https://www.atmorep.org/attention/}.}
    \label{fig:ext-attention}
\end{figure}

\end{appendices}

\clearpage
\newpage

\newpage

\noindent
{\huge Supplementary Material}

\section{Related work}

In the following, we put our work into a wider context with respect to the literature on atmospheric science and machine learning.

\subsection{Stochastic modeling of atmospheric dynamics}
\label{sec:related:stochastic_atmospheric_models}

Atmospheric dynamics are, in principle, governed by well-understood equations determined by the fundamental laws of classical physics, such as conservation of mass and energy and the laws of thermodynamics. 
However, due to the vast range of spatial and temporal scales involved, from tens of thousands of kilometers down to the scale of meters and below, it is impossible to resolve all atmospheric processes explicitly in numerical models. 
Furthermore, especially on smaller scales, we do not have the necessary data to constrain the initial conditions well enough and there are physical processes for which we do not have a complete understanding, for example for cloud lifecycles, aerosol formation, or the interaction with the biosphere~\cite{franzke_stochastic_2015}.
This closure problem is a major source for the forecast and projection uncertainties in current weather and climate models~\cite{Bauer2015}.

Already in 1976, the closure problem was a principle motivation for Hasselmann~\cite{hasselmann_stochastic_1976} to introduce his concept of stochastic modeling in climate science.
In his work, he described the long-term behavior of the atmosphere by a two-scale stochastic dynamical system with the long-term climate system driven by the ``integral response to continuous random excitation by short period `weather' disturbances.''~\cite{hasselmann_stochastic_1976}. 
This view has been refined in recent work and the observed power spectra of atmospheric variables are now understood as a result of cascade processes~\cite{franzke_stochastic_2015}.
This means that memory effects are relevant and climate should be modeled with non-Markovian processes~\cite{franzke_stochastic_2015}.


In numerical weather prediction, stochastic modeling was introduced by Palmer~\cite{palmer_nonlinear_2001} in 2001 but motivated by reasoning similar to the one by Hasselmann, i.e. that a stochastic representation of unresolved, small-scale physical processes is required to obtain the correct large scale behavior.
In particular, the abstract continuous dynamical system 
\begin{align}
  \label{eq:palmer_dynamical_continuous}
  \dot{\tilde{X}} = \tilde{F}\big[\tilde{X}\big]
\end{align}
of states $\tilde{X}$, which corresponds to the governing partial differential equations,
is numerically represented by
\begin{align}
  \dot{X} = F\big[X\big] + P[X,\alpha]
\end{align}
where $X$ is a finite dimensional representation of $\tilde{X}$, $F$ is the finite number of retained terms from the Galerkin projection of Eq.~\ref{eq:palmer_dynamical_continuous}, and $P[X,\alpha]$ corresponds to the residual of the projection~\cite[Sec. 2]{palmer_nonlinear_2001}.
Classically, $P[X,\alpha]$ is modeled by heuristic formulae, such as parametrizations.
Palmer argued, however, that a stochastic model of $P[X,\alpha]$ is required to obtain physically consistent long term dynamics. 
This led to the concept of stochastic parametrizations that play an important role in operational numerical weather prediction today~\cite{Bauer2015,Palmer2019}.
AtmoRep $p_{\theta}(y \vert x , \alpha)$ can be seen as a data-driven representation of $P[X,\alpha]$ that is learned from the processed observations in the ERA5 reanalysis and that includes the large-scale dynamics. 
However, with modifications, AtmoRep can also be used to only represent the small scale processes.

\subsection{Deep learning for weather forecasting}
  


The use of neural networks in Earth system science goes back to the early 2000s, e.g.~\cite{Chevallier1998,Krasnopolsky2002,Tolman2005}.
With the tremendous progress of deep learning methods beginning around 2010~\cite{LeCun2015}, efforts to exploit the methodology also in atmospheric science increased substantially around 2018.

Two earlier studies on the use of deep learning to Earth system modeling will be highlighted before briefly discussing recent works that have shown forecasting performance close to or on par with the best operational weather forecasting models. Ham et al.~\citep{ham2019deep} trained a CNN-based model and used transfer learning on simulation datasets and reanalysis data. They generated El Niño–Southern Oscillation projections with a lead time of up to one and a half years based on sea surface temperature and heat content anomaly maps, outperforming state-of-the-art dynamical prediction systems for lead times beyond six months.
In 2021, Ravuri et al. \citep{ravuri2021skilful} proposed a data-driven approach for probabilistic precipitation nowcasting with a lead time of up to two hours. Their deep generative model was trained on radar observations from the UK. The model's performance was comprehensively assessed using various verification metrics and subjective evaluations by operational forecasters. In comparison to earlier approaches based on CNNs, their generative model showed better nowcasting skill and much better performance with respect to capturing the local variability of precipitation. 

A significant breakthrough in AI-based weather forecasting was achieved by Pathak et al.~\citep{pathak2022fourcastnet}. Their model, FourCastNet, is based on adaptive Fourier neural operator (AFNO) with a vision transformer~\cite{Dosovitskiy2021} as backbone. FourCastNet delivers comparable forecasting results to the IFS model with a lead time of 3 days. 
Shortly after FourCastNet, Bi et al.~\citep{bi2023accurate} introduced Pangu-Weather, a 3D Swin-Trasnformer-based neural network.  The model was trained on 39 years of ERA5 reanalysis data and obtained better deterministic 7-day forecasting results than the operational IFS model, while time-to-solution is 10,000 times faster than for the IFS.
Concurrently to Pangu-Weather, Lam et al.~\citep{lam2022graphcast} introduced a graph neural network-based model, GraphCast, for weather forecasting. 
This model can generate forecasts with six-hour intervals for ten surface variables and six atmospheric variables on 37 vertical pressure levels. The training data spanned 39 years of historical weather data, again from ERA5. GraphCast performed on par and at times better than the IFS with respect to almost all forecasted fields~\cite[see also][]{ben-bouallegue_rise_2023}. 


The work by Chen et al.\citep{chen2023fuxi} presents the Fuxi neural network, designed to enhance the global ensemble weather forecasting system's capabilities in generating 15-day forecasts at a spatial resolution of 0.25. This deep-learning neural network utilizes a Swin Transformer-based model with 48 repeated blocks. The results indicate that Fuxi's performance is comparable to that of ECMWF's enhanced range model in the context of 15-day forecasting.

In a similar vein, Chen et al.~\citep{chen2023fengwu} introduced the FengWu deep learning neural network. This model utilizes model-specific encoder-decoder structures and a cross-modal fusion transformer. These innovations further enhance the forecasting capabilities, extending FengWu's deterministic skillful forecast lead time to 10.75 days. The results also demonstrate a superiority over GraphCast in predicting 80\% of the 880 reported predictands.

Gao et al.~\citep{gao2022earthformer} proposed the EarthFormer, which is a space-time transformer model for weather forecasting. 
It uses a cuboid attention mechanism that is adapted for space-time data. The EarthFormer model demonstrates strong performance in both forecasting sea surface temperature anomalies and precipitation nowcasting although it has not been used for high-resolution numerical weather prediction.

Another noteworthy recent contribution to the literature is ClimaX~\citep{nguyen2023climax} that developed a generalized deep learning model for weather and climate science through self-supervised learning. The pre-trained model was fine-tuned for various downstream tasks, in particular forecasting, climate projecting, and climate downscaling. 
The network used in the work is transformer-based and trained on the CMIP6 climate datasets with fine-tuning on ERA5 reanalysis.
While conceptually closest to AtmoRep in that a foundation-type model is developed, ClimaX also differs in fundamental aspects. 
For example, no zero-shot, intrinsic capabilites were demonstrated in~\citep{nguyen2023climax}.
The results obtained with AtmoRep are also superior to those of ClimaX although we do not yet demonstrate medium-range forecasting with roll-out.

\subsection{Representation learning and generative machine learning} 

AtmoRep builds on a substantial amount of work in the machine learning literature on large scale representation learning. 
The resulting models are sometimes referred to as foundation~\cite{Bommasani2022} or frontier models.

Representation learning~\cite{Bengio2013} is a machine learning methodology whose primary objective is not to obtain a model that is effective for a specific task but one that provides an effective encoding, or representation, of the data distribution.
Next to being of scientific interest, such an encoding can be used for a variety of applications, e.g. by fine-tuning or appending task-specific tail networks. 
While representation learning has a long history~\cite{Bengio2013}, it recently became central to many efforts in machine learning through the introduction of large language models~\cite{Devlin2019,Radford2018,Brown2020}. 
These are domain-specific but task-independent neural networks for natural language that can be specialized, for example, for translation, as chat bots, or for text auto-completion.

Large language models also popularized the use of self-supervised training protocols because a labelling of the very large training data sets would be impractical.
Instead, the pre-training objective, i.e. the one used to learn the task-independent representation, is defined based on the dataset itself.
A common approach is to mask part of the information and predict it based on unmasked ones, although alternatives are possible~\cite{Caron2021}.
For transformer-based large language models, masking is most commonly used in the form of masked token models~\cite{Devlin2019,Radford2018}. 
Since transformers are a highly generic architecture once a token has been defined~\cite{Dosovitskiy2021}, this approach has also been used in computer vision~\cite{He2022} and AtmoRep extended it to space-time neighborhoods.

Traditionally, representation learning required additional computations to make a pre-trained model applicable for a specific task.
Brown et al.~\cite{Brown2020} showed that sufficiently large models have skill for many tasks directly after pre-training and without task-specific refinement. 
This is referred to as zero-shot abilities in the literature~\cite{Brown2020}. 
An extension are few-shot abilities where a few examples are input to the model together with the task, again without update to the model weights, and this typically substantially improves performance.
Zero-shot abilities are enabled by pre-training on a large and diverse data set and through the pre-training task. 
For example, when pre-training uses a multi-lingual data set, then translation is a special case of the masked token model~\cite{Brown2020}. 
Interestingly, even interspersed foreign words, as are common in many everyday texts, are sufficient for zero-shot translation abilities~\cite{Chowdhery2022}.
A second key for the power of modern large language models is the size of their neural networks with the most advanced ones containing more than a trillion parameters today~\cite{OpenAI2023}. 
The network architecture plays thereby only a secondary role. 
Instead, the number of trainable parameters is the most significant factor determining the performance of the trained model (assuming sufficient computational resources for training and a sufficiently large training data set). 
This observation led to the introduction of scaling laws~\cite{Kaplan2020,Zhai2021} that allow to predict the effectiveness of a trained model a priori. 

Although large language models are powerful in the zero- and few-shot setting, they also have limitations, for example, in the generation of long coherent outputs, such as texts that are not repetitive and follow a coherent story line. 
This can, in principle, be addressed by sampling from the discrete probability distribution that is the output of large language models. 
How to obtain long-term coherent sequences remained, however, unclear for some time.
Fine-tuning with human feedback reinforcement learning~\cite{Ouyang2022,Christiano2023} provided substantial improvements in this respect.

Our mathematical formulation of AtmoRep follows closely those for generative AI models, e.g.~\cite{Radford2018,Ramesh2021-Dall-e,Luo2022}, which include large language models and generative image models.
These are commonly also formulated as joint or conditional probability distributions over known and unknown information. 
Interestingly, the theoretical approach for generative AI has strong similarities to the representation of unresolved processes in the stochastic weather and climate model formulations by Hasselmann and Palmer, cf. Sec.~\ref{sec:related:stochastic_atmospheric_models}.
AtmoRep, for the first time, makes this connection explicit and the training on observation-based data allows us to capture these processes since, at least in an integrated form, they are present in the observational record.
We consider this connection between generative AI models and stochastic atmospheric modeling of significant importance since it provides a physical grounding for our machine learning and its results.


\section{Detailed model description}
\label{sec:model:detail}

Below we provide detailed information on the model architecture and discuss design choices. The source code is available with the submission and should serve as a final reference.

\subsection{Model architecture}

The network architecture of AtmoRep is based on transformers~\cite{Vaswani2017} since these are known to scale well to very large model sizes, training data sets and to exploit compute hardware very efficiently. 
Architecturally, transformers differ from other neural networks by processing a sequence of inputs, known as tokens, and relating these to each other in the network through the so-called attention mechanism~\cite{Vaswani2017}. 
In AtmoRep, a token is a small space-time neighbourhood, which extends the conceptualization of tokens that has been introduced for the vision transformer~\cite{dosovitskiy_image_2021}. 
The entire input of AtmoRep is therefore a local spacetime hypercube subdivided into smaller regions representing the tokens (see Fig.~1 in Main).
The model levels are thereby treated as an independent dimension so that one has in total $n_v \times n_t \times n_{\theta} \times n_{\phi}$ tokens where $n_v$ is the number of vertical levels and $n_t$, $n_{\theta}$, $n_{\phi}$ are, respectively, the number of tokens in time, latitude, and longitude.
In contrast to related work~\cite{Pathak2022,bi2023accurate,Lam2022}, our model is spatially local but works on a significantly longer time span. 
Our primary motivation for the use of a local input domain is to learn general principles of atmospheric dynamics across different regions and to generalize better. 
It also alleviates memory pressure in the implementation of the neural network. With a local approach, care is, however, required to obtain globally coherent predictions.

AtmoRep uses a coupled stack of encoder-decoder transformers, which we call the Multiformer, an overview is provided in the Extended Data section in Fig.~1. 
The Multiformer consists of one transformer per physical field and the per-field encoders are coupled through the cross-attention mechanism that is classically used to couple the encoder and the decoder~\cite{Vaswani2017}.
The choice to represent each field with an individual transformer is motivated by the different physical mechanisms that drive the time evolution of the different fields and the resulting widely different signal characteristics. 
It allows, for example, to use a smaller internal embedding dimension and a larger token size for smoother fields, such as temperature. 
Having one transformer per field also leads to a modular design where fields can be combined flexibly to yield an overall Multiformer configuration suitable for the task at hand, as we demonstrated for downscaling. 
Another advantage is that it allows for a field-specific pre-training that speeds-up the overall training process substantially since the dense attention used in our Multiformer scales quadratically in the number of tokens. 

The transformer-encoders for the different physical fields are coupled through cross-attention to allow different fields to interact in the neural network. 
In particular, pre-training is performed with a fixed number of self-attention heads, as in a standard transformer-encoder. 
When a Multiformer is assembled, we preserve these heads and add a user-defined number of cross-attention heads per coupled field. 
Which fields are coupled is also user-specified, cf. Fig.~\ref{tab:sup-multiformer-params}. 
For example, one can allow for fields to only interact indirectly through a common coupled field, which reduces computational costs.
 
The per-field decoders use cross-attention with the encoder from the same field but it is not coupled to the other fields. 
In contrast to transformers as used in natural language processing, we do not use the output of the last layer of the encoder but a U-Net type coupling, see again the depiction in Extended Data Fig.~1.
For all attentions, we employ qk-layernorms. 
These are relevant for scaling to very large network sizes~\cite{Chowdhery2022,Dehghani2023scaling} and in a Multiformer they also of importance to stabilize the coupling of different fields with different characteristics.

\paragraph{Positional embeddings}


Vaswani et al.~\cite{Vaswani2017} developed a linear positional encoding based on trigonometric functions of different frequencies given by
\begin{align*}
    \mathrm{PE}_{2i}^{(k)} &= \sin( k / 10000^{2i/d_{\mathrm{model}}} )  
    \\
    \mathrm{PE}_{2i+1}^{(k)} &= \sin( k / 10000^{2i/d_{\mathrm{model}}})
\end{align*}
where $i$ is the the index along the embedding dimension, $k$ the linear, $1$-dimensional token position, and $d$ the embedding dimension of the model. 
We found that this embedding has limitations in that it creates aliasing when the number of tokens exceeds $\approx 100$ and it makes only inefficient use of the embedding space.
Furthermore, it is designed for a linear sequence of tokens, as one has in natural language processing, and does not respect the $4D$ structure of the token sequence in AtmoRep.
Therefore, we developed a modified hamonic positional embedding for AtmoRep's $4D$ token space that uses frequency modulation to encode all information.
It is given by
\begin{align*}
    \mathrm{PE}_{2i}^{(k)} &= \sin( i \, k_t  ) + \frac{1}{2} \sin(8 \, i \, k_{\theta})  \\[4pt]
    \mathrm{PE}_{2i+1}^{(k)} &= \cos( i \, k_v) + \frac{1}{2} \cos(8 \, i \, k_{\phi} )
\end{align*}
where the multi-index $k = (k_t, k_v, k_{\theta}, k_{\phi})$ provides the local, relative token position in time, vertical level, latitude and longitude, respectively. 
Note that the local positional encoding is complemented by the external information $\alpha$ that contains global time, level as well as latitude and longitude.

\subsection{Training and Loss}

The observational record provides the joint occurrence $(x,y)$ of atmospheric states through samples (or instantiations) from the instationary joint distribution $\tilde{p}( x, y ; \alpha)$, cf. Fig.~1, top-right, in Main.\footnote{For notational simplicity, we do not distinguish here between physical atmospheric states, denoted as $\tilde{x}$, $\tilde{y}$ in Main, as well as their discrete representations $x$, $y$.}
The parameter $\alpha$ controls again the instationarity of the distribution, e.g. its shift over time.
AtmoRep's training objective is the minimization of the distance measure
\begin{subequations}
\begin{align*}
  \mathcal{D}\big( \tilde{p}( y , x ; \alpha) , p_{\theta}( y , x ; \alpha) \big)
  &= \int_{\mathcal{X}_x} \int_{\mathcal{X}_y} d\big( \, \tilde{p}( y , x ; \alpha ) \, , \, p_{\theta}( y , x ; \alpha) \, \big) \, \mathrm{d}x \, \mathrm{d}y
  \\
  &= \int_{\mathcal{X}_x} \int_{\mathcal{X}_y} d\big( \, \tilde{p}( y , x ; \alpha ) \, , \, p_{\theta}( y , x ; \alpha) \big) \, \mathrm{d}x \, \mathrm{d}y
\end{align*}
for a suitable distance function $d( \cdot , \cdot )$, see below.
The joint distributions $\tilde{p}( y , x ; \alpha )$ and $p_{\theta}( y , x ; \alpha)$ factor as $\tilde{p}( y , x ; \alpha) = \tilde{p}( y \vert x ; \alpha ) \, p(x ; \alpha)$ and $p_{\theta}( y , x ; \alpha) = p_{\theta}( y | x ; \alpha) \, p(x ; \alpha)$ where $p_{\theta}( y | x ; \alpha)$ is the AtmoRep model.
Thus,
\begin{align*}
  \mathcal{D}\big( \tilde{p}( y , x ; \alpha ) , p_{\theta}( y , x ; \alpha) \big)
  &= \int_{\mathcal{X}_x} \int_{\mathcal{X}_y} d\big( \, \tilde{p}( y \vert x ; \alpha ) \, , \, p_{\theta}( y \vert x ; \alpha) \big) \, p(x ; \alpha ) \, \mathrm{d}x \, \mathrm{d}y
\end{align*}
where we assumed that $d( \cdot , \cdot )$ only acts on $y$ and hence $p(x ; \alpha)$ can be factored out (which is satisfied in our case, see again below).
The last equation equals
\begin{align*}
  \mathcal{D}\big( \tilde{p}( y , x ; \alpha) , p_{\theta}( y , x ; \alpha) \big)
  &= \mathbb{E}_{p(x ; \alpha)} \Bigg[ \int_{\mathcal{X}_y} d\big( \tilde{p}( y \vert x ; \alpha) , p_{\theta}( y \vert x ; \alpha) \big) \, \mathrm{d}y \Bigg] .
\end{align*}
The expected value over $p(x ; \alpha)$ can be estimated by the empirical mean over the large but finite observational record $\bar{\mathcal{X}}$ that consists of discrete samples $(\bar{y},\bar{x})$.
The available distribution $\tilde{p}( y , x ; \alpha)$ is thus a sequence of Dirac-deltas and we therefore have
\begin{align*}
  \mathcal{D}\big( \tilde{p}( y , x ; \alpha) , p_{\theta}( y , x ; \alpha) \big)
  \approx \frac{1}{N} \sum_{\bar{x} \in \bar{\mathcal{X}}} d\big( \bar{y} , p_{\theta}( y \vert \bar{x} ; \alpha) \big) .
\end{align*}
\end{subequations}
As an approximation for the observational record $\bar{\mathcal{X}}$, we employ in our work the ERA5 reanalysis and due to computational constraints we use a random subset of it in practice, i.e. we use a Monte Carlo estimate of the last equation in the actual computations (as is standard when stochastic gradient descent or one of its variants are employed for training).

\begin{figure}
  \centering
  \includegraphics[width=0.8\textwidth]{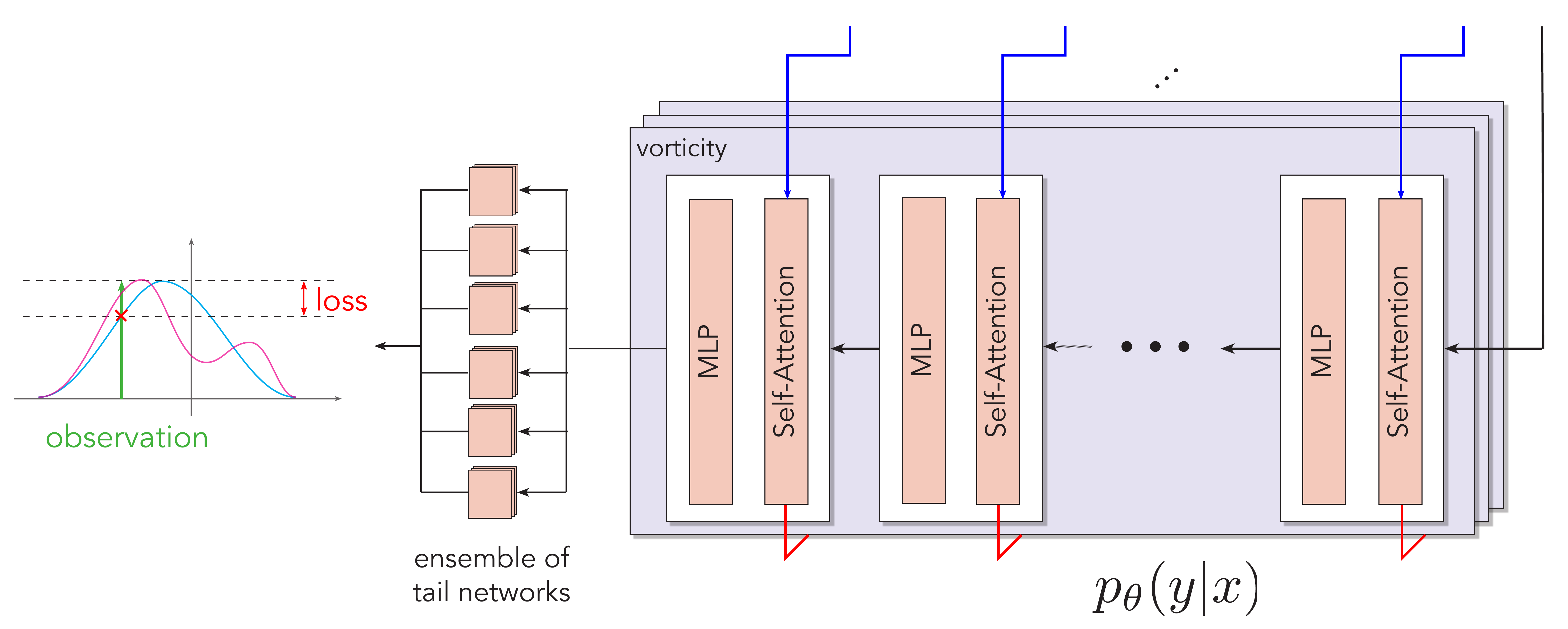}
  \caption{Conceptual depiction of ensemble prediction and associated statistical loss. The magenta curve on the left represents the non-parametric distribution that is represented by the ensemble, the cyan curve the first two statistical moments that are used in the loss computation and which corresponds to a Gaussian. The loss is given by the difference between Gaussian evaluated at the observation and $1$, which is the value of the Gaussian at its mean. Hence, when the ensemble mean coincides with the observation then the loss vanishes.}
  \label{fig:ensemble_loss}
\end{figure}

As discussed in Methods, AtmoRep employs an ensemble to provide an non-parametric representation of the conditional probability distribution over the output state $y$ and a novel statistical loss function for its training.
These are motivated by the cross-entropy loss, which is the de facto standard for the training of discrete probability distributions~\cite{Murphy2022}.
In the discrete case, an explicit representation of the probabilities over valid outputs is possible (even when the number of classes is very large, as for large language model, e.g. $32,768$).
In the continuous, regression case, however, this is not possible. 
One remedy is to work with a parametric probability distribution with a finite number of parameters that then can be learned. 
Inspired the success of ensemble methods for numerical weather prediction~\cite{Palmer2006,Palmer2019}, we instead generalize the discrete case using an ensemble. 
It provides a sample from an non-parametric distribution and therefore no assumptions on the true distribution are required.
As discussed in Methods, the ensemble is realized by a set of predictions heads and it is therefore computationally inexpensive.

To compute the loss based on the ensemble prediction $\hat{y}$, we currently consider its first two statistical moments, i.e. mean $\mu$ and standard deviation $\sigma$. 
The corresponding Gaussian $G_{\mu,\sigma}$ has a value of $1$ at its mean when normalized accordingly. 
With a single observation $\bar{y}$ given by a per grid point value for a single physical field, a loss function is
\begin{align}
  \label{eq:statistical_loss}
  d_s\big( \bar{y} , \hat{y} \big) 
  = \Big\vert 1 - \int \delta_x(y) \, G_{\mu,\sigma}(y) \, dy \Big\vert^2
  = \Big\vert 1 - G_{\mu,\sigma}(y) \Big\vert^2 .
\end{align} 

\begin{figure}
  \includegraphics[width=0.49\textwidth]{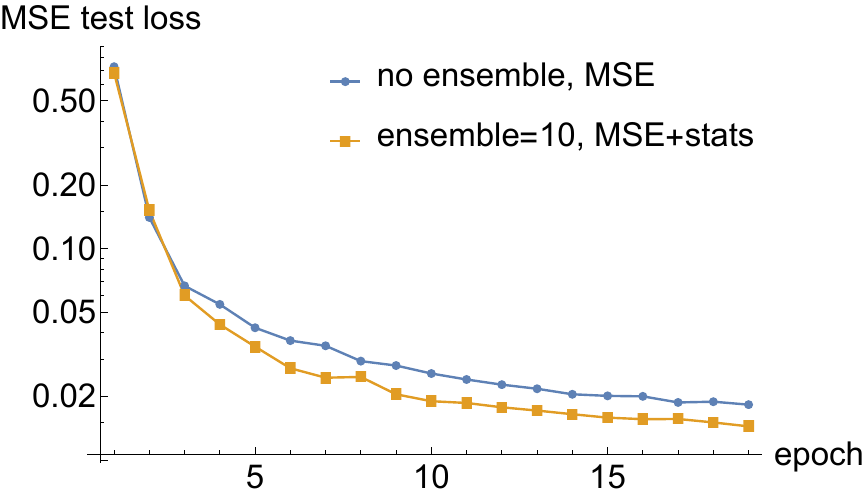}
  \includegraphics[width=0.49\textwidth]{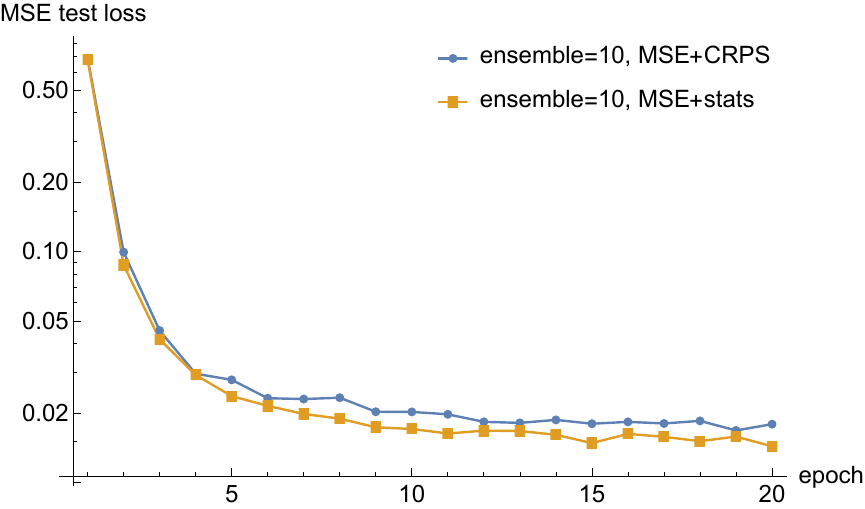}
  \caption{Evaluation of statistical loss; results are shown for a single field model vorticity with vertical level ml$=137$. Note that MSE is always used for comparison independent of the loss function. \emph{Left:} \emph{Right:} Comparison between our ensemble loss and CRPS. Despite a close similarity in the mathematical formulation, our statistical loss performs better in terms of the training dynamics.}
  \label{fig:statistical_loss}
\end{figure}

The above loss $d_s( \, \cdot , \cdot )$ is very similar to the Gaussian continuous ranked probability function (CRPS), a proper scoring rule; in particular, it can be seen as probability density function form of the CRPS which is formulated in terms of cumulative distribution function, cf.~\cite{Rasp2018}.
However, in our experiments, Eq.~\ref{eq:statistical_loss} performed significantly better as loss function than CRPS, see Fig.~\ref{fig:statistical_loss}.
The figure also shows that our statistical ensemble loss leads to a lower test-set MSE than training with MSE directly. 

As already discussed in Methods, the full loss function per grid point per physical field is
\begin{align}
  d \big( \bar{y} , \hat{y} \big) 
  = \sum_k \vert \bar{y} - \hat{y}_k \vert^2 + d_s\big( y , \hat{y} ) + \sqrt{ \sigma } .
\end{align}
where the $\hat{y}_k$ are the individual ensemble members.
The loss is computed independently for each predicted grid point and for each physical field. 
We thereby use a field-specific weighting but a uniform weight across the vertical levels. The weights are given in Table~\ref{tab:sup-field-token-params}.

The states $x$,$y$ we use in the training of AtmoRep are based on the tokens that are the input to the transformer-based Multiformer and which are given by tiles of a larger space-time neighborhood, e.g. $12 \times 5 \times 6 \times 12$ tokens for a $36 \, \mathrm{h} \times 5 \, \textrm{vertical levels} \times 1800 \, \mathrm{km} \times 3600 \, \mathrm{km}$ neighborhood for each physical field.
In particular,  we randomly designate a subset of tokens as $y$ and the remainder of tokens as $x$, see Fig.~1 in Main.
The $y$- or target tokens are probabilistically masked or distorted.
For masking, we set them to $0$, which is for all fields a physically valid value due to the data normalization. 
For distortions, we either add noise with a magnitude that is consistent with the standard deviation of the values in the token or we coarsen it by a factor of $2$ or $4$. 
The choice which tokens are masked and which are distorted is made fully randomly by sampling in each case from a uniform distribution over a linear indexing of the tokens.
The user-specified masking ratio, cf. Table~\ref{tab:sup-field-token-params}, is thereby a maximum ratio and the actual one used per training example is also sampled randomly.
The loss is computed over all tokens in $y$, which in general includes unmasked tokens due to the random sampling.
The masking is thereby performed independently for different physical fields and different levels. 
Our training strategy with masking and distorting tokens is an analogue of the masked token models used in natural language processing, e.g.~\cite{Devlin2019,Brown2020}, which recently have also been adopted to computer vision, e.g.~\cite{He2022}.
The distortions are inspired by work in natural language processing~\cite{Devlin2019} that used random word permutations to encourage the learning of a robust and probabilistic representation.
We employ our distortions towards the same end.

Processing the entire ERA5 training set in multiple epochs was beyond the scope of the compute resources available to us. 
We therefore formulate the training as a Monte Carlo approximation where in each epoch we processed a randomly sampled sub-set of the full data set.
The purpose of the epochs used in our work were therefore to have a manageable number of training periods that can be used to adjust the learning rate and to evaluate the test set for monitoring progress during training.
We chose the amount of data for each epoch hence accordingly so that the training would complete within a small number of hours.


\subsection{Implementation}

The AtmoRep model was implemented with PyTorch~\cite{pytorch}.
Data parallel training was used extensively throughout the project. We employed PyTorch's DDP library for it.
The final model configuration was chosen so that a complete Multiformer with six fields can be placed on one large compute node with four A100 GPUs. 
Different fields thereby reside on different GPUs with temperature and total precipitation on one since they use a smaller embedding dimension and also process a smaller number of tokens.

We implemented a custom hierarchical sampling of the per-batch data. 
This was critical since random access to small parts of a large data set is highly inefficient on standard file systems when implemented naively.
A good randomization per batch is, however, required for effective stochastic gradient descent and to obtain an unbiased estimator with AtmoRep.
The formulation of training as a Monte Carlo method allowed us thereby to perform the batch assembly per parallel task fully independent from other tasks so that only gradients needed to be communicated in the data parallel training. 
Concretely, we sampled on each data parallel task first a small number of year-month pairs with the number being controlled by the available CPU RAM. For each time slice in these, a user-specified number of local neighborhoods was sampled. 
The data chunks corresponding to the year-month pairs were loaded from disk and the spatial sampling from these was performed on the fly in multiple parallel data loaders.
The variable number of local neighborhoods per time slice allowed us to hide latencies in disk I/O.



\subsection{Model configuration}
\label{sec:model-config}

  
The results presented in the main text were obtained with three different configurations of the AtmoRep Multiformer  (Table~\ref{tab:sup-multiformer-params}). 
All of these configurations build on the same pre-trained single-field models, which were used with the configuration given in Table~\ref{tab:sup-field-token-params}.
These parameters were not changed for the Multiformers. 
Relevant parameters can be summarized as follows:
\begin{itemize}
    \item Physical fields: velocity u, v (or vorticity and divergence), vertical velocity, temperature, specific humidity, total precipitation
    \item Model levels: 96, 105, 114, 123, 137
    \item Resolution: $0.25^{\circ}$ equi-angular grid ($721 \times 1440$ grid points)
    \item Training period: 1980 -- 2017, test period: 2018
    \item Neighbourhood: 12 x 6 x 12 tokens with 3 x 9 x 9 grid points for pre-training
    \item UNet-like encoder-decoder architecture with 10 transformer layers in each branch
    \item Self-attention: 16 heads in the encoder, 8 in the decoder
    \item Cross-attention (same field): 8 heads in the decoder
    \item Cross attention (inter-field): 2 heads per field in the encoder (in addition to self-attention heads)
    \item Ensemble tail networks: 16 linear layers per field for ensemble generation
    \item Learning rate: $10^{-5}$ - $2\cdot 10^{-5}$, five epochs warm-up, then exponential decay to $2\cdot 10^{-5}$
    \item Dropout rate: 0.05
    \item Optimizer: AdamW (weight decay: 0.05)
    \item Masked token model with distortions; up to 90~\% of modified tokens are masked, up to 20~\% get noise added, and up to 5~\% are smoothed
    \item Multiformer with five fields: 3.5 billion parameters, parallelized across the 4 GPUs available in one node
    \item Training on up to 32 nodes with 4 $\times$ A100 GPUs on JUWELS-BOOSTER at the Jülich Supercomputing Center
    \item Training time: 4.5 hours per epoch, where an epoch is per node given by 2 spatial samples per time step for all time steps in 1 randomly sampled months.
    \item Inference time: $\approx 10 s$ for a global 3-hour forecast on 1 node
\end{itemize}
Due to the large model size, no hyper-parameter optimization was possible. 
The selected parameters were those that showed the best performance and trade-off in preliminary experiments with smaller scale models and training runs.

\begin{table}[]
    \centering
    \begin{tabular}{l|c|c|c|c|c|c|c}
         field name & short &  \#tokens & tok. size & norm. & mask & weight & embed. dim. \\
         \hline
         zonal wind  & u & 12, 6, 12  & 3, 9, 9 & local & 0.7 & 0.3 & 2048 \\[3pt] 
         meridional wind  & v & 12, 6, 12  & 3, 9, 9 & local & 0.7 & 0.3 & 2048 \\[3pt]
         vertical wind & w & 12, 6, 12  & 3, 9, 9 & global  & 0.65 & 0.025 & 1024 \\[3pt]
         vorticity & vor & 12, 6, 12  & 3, 9, 9 & global & 0.7  & 0.25 & 2048 \\[3pt] 
         divergence & div & 12, 6, 12  & 3, 9, 9 & global & 0.7 & 0.25 & 2048 \\[3pt] 
         (dry) temperature & T & 12, 2, 4  & 3, 27, 27 & local & 0.85 & 0.2 & 1024 \\[3pt] 
         specific humidity & q & 12, 6, 12  & 3, 9, 9 & local  & 0.85 & 0.15 & 2048 \\[3pt]
         total precipitation & precip & 12, 6, 12  & 3, 9, 9 & global & 0.5 & 0.025 & 1536\\ 
         \hline
    \end{tabular}
    \smallskip
    \caption{AtmoRep configurations for the pre-training of individual fields. The same parameters were used in the Multiformer model. Tokens are listed in the order time, latitude, and longitude. The masking ratio refers to the maximum fraction of masked or distorted tokens in the BERT-style training. Note that some parameters, such as the masking ratio, have been adjusted during training and we report here the final values.}
    \label{tab:sup-field-token-params}
\end{table}


\begin{table}[]
    \centering
    \begin{tabular}{l|c|c}
         configuration name &  fields &  application  \\
         \hline
         multi6-uv &  u $\leftarrow$ v, w, T, q   &  nowcasting, \\
                    &  v $\leftarrow$ u, w, T, q   &  interpolation \\
                    &  w $\leftarrow$ u, v, T      &   \\
                    &  T $\leftarrow$ u, v, w, q   &   \\
                    &  q $\leftarrow$ u, v, w, T, precip   &   \\
                    &  precip $\leftarrow$ u, v, w, q   &   \\[3pt]
                    \hline
         multi6-vd  & vor $\leftarrow$ div, w, T   &  counterfactuals, \\ 
                         & div $\leftarrow$ vor, w, T   &  IFS model correction, \\
                         & w $\leftarrow$ vor, div, T   &  extrapolation, \\
                         & T $\leftarrow$ vor, div, q   &  bias correction \\
                         & q $\leftarrow$ vor, div, w, T   &   \\
                         & precip $\leftarrow$ vor, div, w, q   &   \\[3pt]
                      \hline
         multi3-uv  &   u $\leftarrow$ v, T   &  downscaling \\
                     &   v $\leftarrow$ u, T   &    \\
                     &   T $\leftarrow$ u, v   &    \\[3pt]
        \hline
    \end{tabular}
    \smallskip
    \caption{AtmoRep Multiformer configurations used in this work. Left arrows indicate cross attentions, i.e. which variables impact on the respective target variable.}
    \label{tab:sup-multiformer-params}
\end{table}

A summary of the model configurations used for the different application together with an overview of the datasets and the metrics used in the analysis is reported in Table~\ref{tab:downstream_config}.

\begin{table}[]
\begin{tabular}{ | C{6.0em} | C{5.75em}| C{5.0em} | C{8.0em}| C{5.25em} | C{5em}|   } 
  \hline
  \textbf{application} & \textbf{model}  & \textbf{mode} & \textbf{domain} & \textbf{other datasets} & \textbf{metrics}  \\[5pt]
   \hline
  nowcasting & multi6-uv, non- and fine-tuned & 6h-forecast & 2018 preds. at 0 and 12 UTC, global forecasts & IFS, Pangu-Weather & RMSE, ACC, CRPS, SSR \\[5pt]
   \hline
 counter-factuals & multi6-vd & 3h-forecast & global, June 2018, local random sampling & -  & histogram differences \\[5pt]
  \hline
 temporal interpolation & multi6-vd & temporal interpolation & global, 2018, local random sampling & linear interpol. & RMSE \\[5pt]
  \hline
 extrapolation to 2022 & multi6-vd & 3h-forecast & global, 2018, local random sampling & - & histogram differences \\[5pt]
  \hline
 model correction & multi6-vd & random masking & global, Feb. 2022, local random sampling  & IFS & histogram differences \\[5pt]
  \hline
 downscaling & multi3-uv with downscaling tail  & identity, hourly & $[27.375^{\circ}, 70.375^{\circ}]$ $[-12.625^{\circ}, 37.125^{\circ}]$ 2017  &  COSMO REA6, GAN (Stengel et al.) & RMSE, power spectrum \\[5pt]
  \hline
 bias correction & multi6-vd (12x6x6), fine-tuned & identity, hourly & $[44.5^{\circ}, 57.75^{\circ}]$ $[3.5^{\circ}, 16.75^{\circ}]$ 2019 & RADKLIM & RMSE, FBI, PSS, EST \\[5pt]
  \hline
\end{tabular}
\smallskip 
\caption{Overview of the model configurations and the datasets used for each application. The third column reports the mode used for AtmoRep's masked token model. The fourth column also reports how the model output was generated; in particular, local random sampling means that we sampled data as during pre-training. The metrics in the fourth column are defined in Sec.~\ref{sup:metrics}. }
\label{tab:downstream_config}
\end{table}

\subsection{Attention maps}

Attention maps have been used before in natural language processing and computer vision to understand what a transformer neural network has learned and to visualize and exemplify their generalization abilities, e.g.~\cite{Devlin2019,Serrano2019,Abnar2020,Caron2021}.
An advantage compared to other introspection methods is that these are immediately available when evaluating a model without further computations.
To the best of our knowledge, attention maps have so far not been studied for large scale transformer models in Earth system science. 

An attention map $A_h \in R^{N \times N}$ is the matrix derived from multiplying queries $Q_h \in \mathbb{R}^{N \times E_h}$ by keys $K_h \in \mathbb{R}^{N \times E_h}$ within an attention block, i.e.
\begin{align}
  \label{eq:attention:map}
  A_h = Q_h K_h^T .
\end{align}
where $N$ is the number of tokens and $E_h$ the per-head embedding dimension.
The factors are thereby given by
\begin{align}
  \label{eq:attention:per_head_projections}
  Q_h = W_Q^h \, T \qquad K_h = W_K^h \, T \qquad V_h = W_V^h \, T
\end{align}
with $T \in \mathbb{R}^{N \times E}$ being the matrix formed by stacking the tokens $t \in R^{E}$ as rows.
$W_Q^h$, $W_K^h$, $W_V^h$ are learnable, head-specific projection matrices for queries, keys, and values, respectively.
In cross-attention, a different token matrix $T$ is used for queries and for keys and values.
In the transformer, the per-head attention map $A_h$ is employed as
\[ 
T_h' = \text{softmax}\left(\frac{A_h}{\sqrt{dk}}\right) V_h = \text{softmax}\left(\frac{Q_h K_h^T}{\sqrt{E}}\right) V_h 
\]
where $T_h'$ is the matrix of updated tokens that, after passing through the MLP of the attention block, is processed by the subsequent blocks in the transformer.
By the last equation, an attention map hence provides a direct measure of the token mixing within the block, i.e. how much information from the $j^{\textrm{th}}$ token is used to update the $i^{\textrm{th}}$ one.
$A_h$ is thereby determined by the scalar product, i.e. a standard, linear measure for the similarity of two vectors from linear algebra. 

To reveal space-time structures in AtmoRep's attention maps, we arrange the $N$ tokens again in their space-time form, e.g. the $12 \times 5 \times 6 \times 12$ used during pre-training, so that $A \in \mathbb{R}^{(n_t \times n_v \times n_{\theta} \times n_{\phi}) \times (n_t \times n_v \times n_{\theta} \times n_{\phi})}$. 
We can then generate spatio-temporal plots by either fixing one token that is attended to, i.e. fixing an index in one of the ``legs'' $n_t \times n_v \times n_{\theta} \times n_{\phi}$, or by averaging over one of them. 
In either case, this enables us to visualize $n_t \times n_v$ latitude-longitude plots that are directly interpretable. 
In Fig.~10 in the Extended Data section, for the top plot we also averaged over the different attention heads in the last layer of the decoder, while in the bottom plot we show the different heads for a fixed time step and level.

%

\subsection{Pre-training results}

Training was performed on JUWELS-BOOSTER at the Jülich Supercomputing Centre. 
The per-field transformers were pre-trained on 4 nodes each for multiple weeks with a per-node batch size of $6$ (i.e. with an effective data-parallel batch size of $96$).
For the multiformers, we employed $32$ nodes with a per-node batch size of $2$ (i.e. with an effective data-parallel batch size of $64$). 
The batch size was in all cases controlled by the available GPU memory.

\begin{figure}
  \centering
  \includegraphics[trim=0 0 0cm 20cm,clip,width=0.7\textwidth]{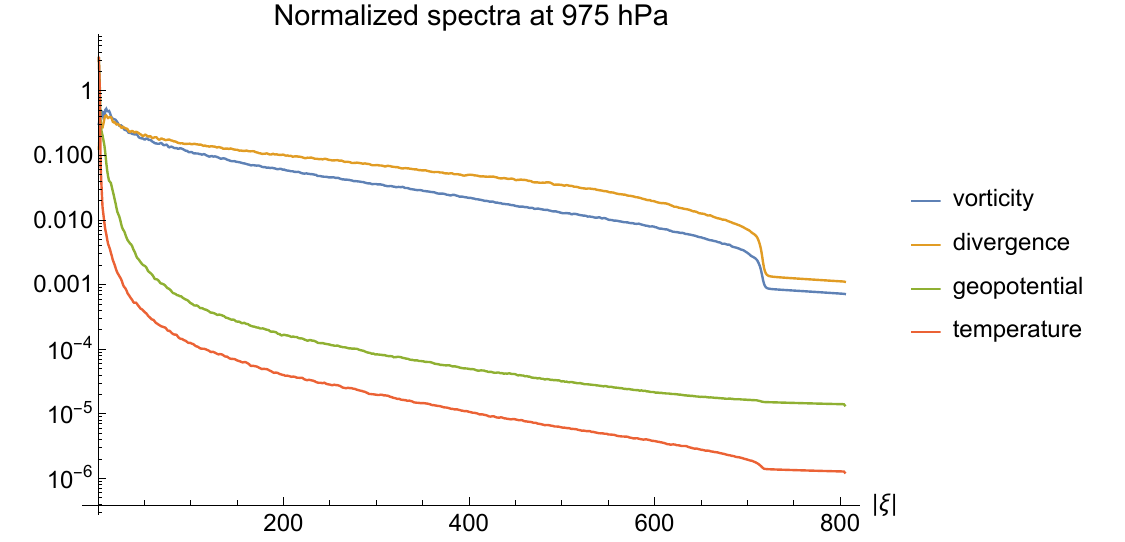}
  \caption{2D radial spectrum for different fields on the $721 \times 1440$ ERA5 grid. }
  \label{fig:data:spectra_levels}
\end{figure}



\begin{figure}[t]
    \centering
    \includegraphics[width=0.7\textwidth]{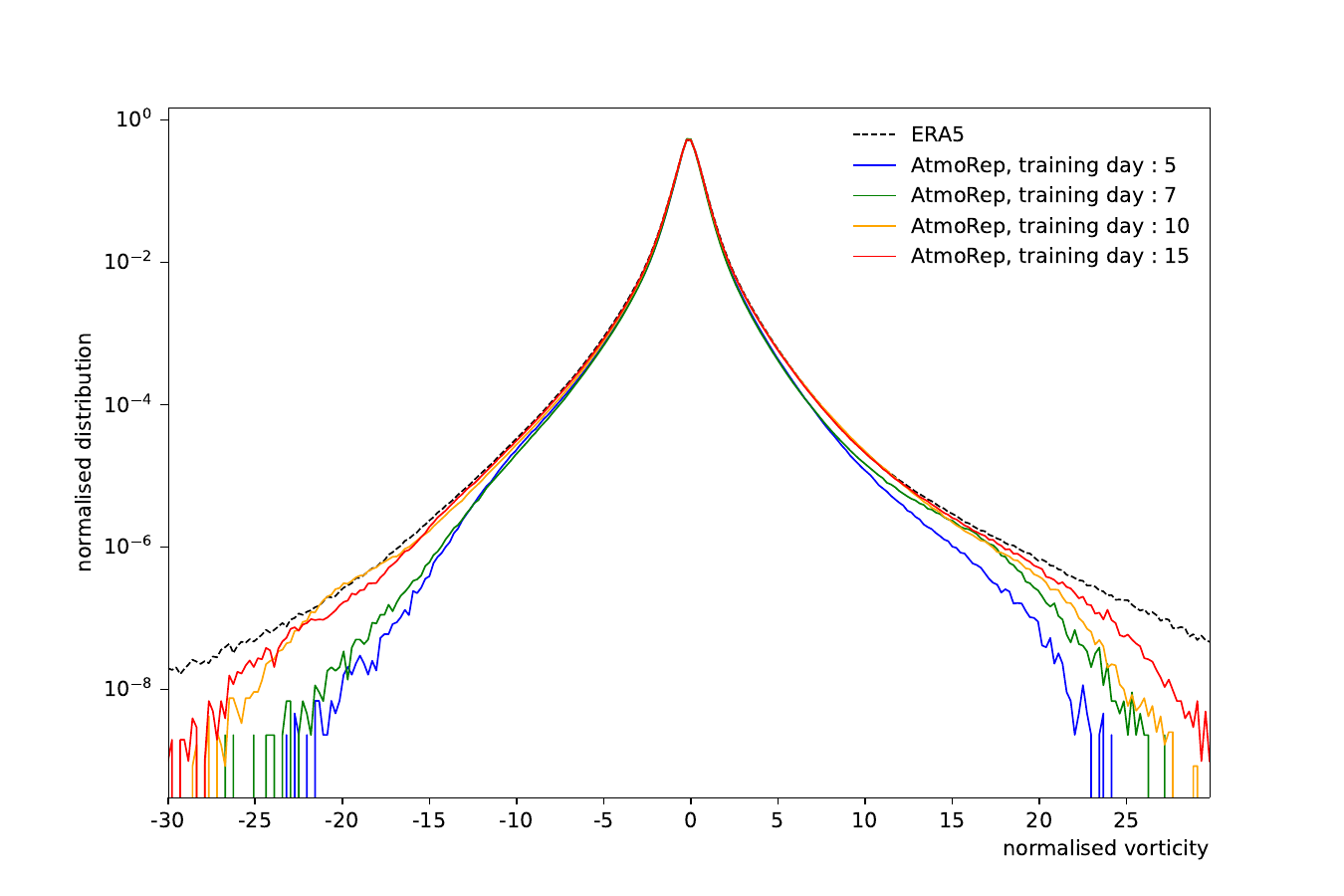}
    \caption{Comparison between the normalized vorticity distributions in ERA5 and AtmoRep as a function of the training epoch.}
    \label{fig:ext-epochs}
\end{figure}

\begin{figure}[t]
    \centering
    \includegraphics[width=0.8\textwidth]{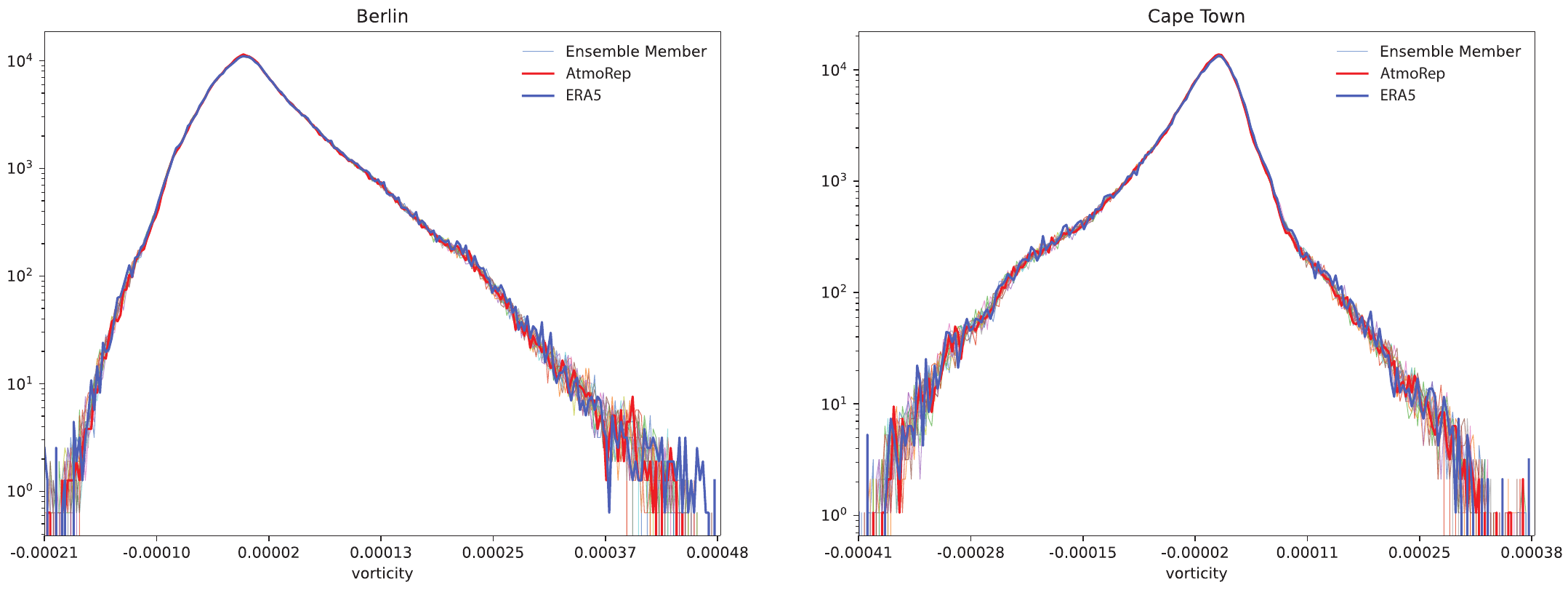}
    \caption{Histograms for vorticity at model level $137$ for Berlin and Cape Town after pre-training obtained with the pre-training task and by using a large number of random samples from the training data set.}
    \label{fig:pretraining_histograms_cities}
\end{figure}

To monitor progress during pre-training, we used the validation loss for the masked token model training task.
This provides, however, only limited insight into the generalization abilities of a model.
We therefore complemented it by the zero-shot forecasting performance.
We also regularly evaluated error plots and similar metrics to monitor the progress of the training.
To evaluate the suitability of AtmoRep as a statistical model, we also extensively used histograms. Two examples are shown in Fig.~\ref{fig:ext-epochs} and Fig.~\ref{fig:pretraining_histograms_cities}.
Fig.~\ref{fig:ext-epochs} shows the global histogram for vorticity at model level $137$ for a selected number of training epochs. 
In early training, deficiencies are in particular present in the tails of the distribution but these improve with more training.
Fig.~\ref{fig:pretraining_histograms_cities} shows histograms for two selected but representative point locations. 
The highly non-Gaussian nature of the distributions is well captured by our model, even in the tails.

For all models, we increased the masking rate during pre-training to make the training task more difficult and this led to a measurable improvement of the zero-shot generalization abilities of AtmoRep. 

We observed substantially different behavior for different fields during pre-training. 
For example, for the velocity components the statistical loss was sufficient and no MSE term was needed whereas no convergence was obtained for vorticity without it.
For divergence, obtaining convergence was difficult even with both MSE and statistical loss.
Using a model pre-trained for vorticity yielded substantially better results although the loss remained higher than for vorticity.
We also observed that we obtained sub-optimal predictions with visually apparent artifacts for temperature, despite a very small MSE loss. 
Using a larger token size helped to alleviate the problem, likely since small tokens are essentially constant and result in an uninformative attention computation.  
We believe that most of the above observations can be understood through the widely different frequency spectra of the different physical fields, cf. Fig.~\ref{fig:data:spectra_levels}, but we leave a more systematic study to future work.
The ability to adapt parameters and computational protocols to the properties of the physical fields is, in our opinion, a significant advantage of the Multiformer architecture.

We did not observe convergence for any of the models during pre-training and the loss was continuously decreasing. 
The pre-training was terminated when the available computational resources were exhausted.


\subsection{Design Choices}

Below, we discuss some of the design choices we made for AtmoRep as well possible alternatives.

\begin{itemize}
  \item \emph{Use of dense attention:} Dense attention is computationally expensive since its computational costs scale quadratically with the number of tokens. This problem is exaggerated by the four-dimensional domain AtmoRep works on. One alternative to dense attention is axial attention, which has been used before for problems in Earth system science e.g. in~\cite{Sonderby2020}. We also implemented it in AtmoRep but it lead to substantially worse results for the intrinsic zero-shot abilities. We believe that sparse attention, as used for example in large language models~\cite{Brown2020}, might provide an alternative that improves computational costs without negatively impacting the skill.
  \item \emph{Training with forecasting instead of a bidirectional masked token model:} The masked token model in our work is inspired by~\cite{Devlin2019}.
  An alternative is to train with a forecasting task, which would preserve causality. Preliminary experiments led to worse performance but we believe that also forecasting is a suitable pre-training task when properly tuned. (The difference is analogous to BERT-type pre-training~\cite{Devlin2019} and next-word-prediction pre-training~\cite{Brown2020} in natural language processing; both provide comparable performance.)
  \item \emph{Masking value:} In our work, we mask tokens with $0$. Due to the data normalization, this is leads to the masked tokens being physically valid, at least in a statistical sense. For the training, a statistically valid token is, in our opinion, desirable, e.g. because it facilitates the learning of a robust representations as needed for model correction. It also leads, however, to some downsides. For instances, the reconstruction of a field only from other fields or only from a given external condition $\alpha$ is not possible in this case, cf. Sec.~\ref{sec:future}.
  \item \emph{Statistical loss formulation:} The choice of using the difference to an unnormalized Gaussian is unorthodox from the point of view of probability theory where, e.g. the area between the observation and the mean to a normalized Gaussian would be a more natural choice. However, in our experiments the current loss performed better than alternatives. We leave a theoretical investigation of this to future work.
  \item \emph{Number of statistical moments in loss computation:} We currently employ only the first two statistical moments in the loss computation. This is not equivalent to assuming a Gaussian distribution but means that we only control the first two moments of the output (somewhat similar to a weak formulation in finite elements). Using the first two moments worked well in all of our experiments and  we believe that one reason for this is that these are considered independently per grid point and hence arbitrarily complex inter-grid point distributions are possible. However, with a sufficient number of ensemble members one could also consider higher order moments, e.g. curtosis.
  \item \emph{Velocity versus Vorticity and Divergence:} The wind velocity vector field can either be specified through its u-v components or through two potentials, which are equivalent to vorticity and divergence. 
      We therefore trained single-field models for both variants and also assembled them into two Multiformer configurations (see Table~\ref{tab:sup-multiformer-params}). Vorticity had in our experience thereby the best performance and was in particular better than the velocity components. 
      However, divergence had the worst performance from the $4$ fields so that we did not reach a conclusion if either the velocity components or vorticity and divergence are better suited.
      Note, however, that a direct comparison of, e.g., loss values for the velocity components on the one side and vorticity and divergence on the other is not possible since they are related by differential operators. Instead, a prediction for vorticity and divergence needs to be converted to velocity space, or vice versa, for a fair comparison. 
      In Fig.~\ref{fig:ext-temp-interpol} we present preliminary results for an alternative way to compare the two configurations. Specifically, there we consider the performance of the two large multiformer configurations for temporal interpolation on a common field, in this case specific humidity. 
      There we see a significant benefit for the vorticity/divergence Multiformer, which is also consistent with other preliminary results. A more detailed investigation would, however, be required to obtain a more complete understanding.
  \end{itemize}

\section{Detailed Experimental Protocols and Further Results}

Below we provide further details on experimental protocols and evaluation. 

\subsection{Definition of Metrics}
\label{sup:metrics}

The most common metric used by multiple applications to evaluate model skill is the root mean square error (RMSE),  defined as
\begin{equation}
    \text{RMSE}= \frac{1}{N}\sum_{n = 1}^{N} \sqrt{\frac{1}{W \cdot H}  \sum_{w = 1}^{W} \sum_{h = 1}^{H} \Lambda_{w, h}  \cdot (x_{w, h}^n - \tilde{x}_{w, h}^n)^{2}} .
    \label{eq:RMSE}
\end{equation}
with $\Lambda_{w, h}$ either being the identity or 
\begin{equation}
 \Lambda_{w, h} =  W \cdot \frac{\cos(\alpha_{w, h})}{\sum_{w'=1}^{W} \cos(\alpha_{w', h})}
\end{equation}
for latitude-weighted Root Mean Square Error~\cite{chen2023fengwu}.
Here, $\tilde{x}$ denotes a prediction and and $x$ its ground truth (target) value. 
$N$ is a sequence length, $w$ and $h$ are the latitude and longitude indices of each grid point in the given region,  and $\alpha_{w, h}$ is the latitude value at point $(w, h)$.

The (latitude-weighted) anomaly correlation coefficient (ACC) is defined as
\begin{equation*}
    \text{ACC}= \frac{1}{N}\sum_{n = 1}^{N} \frac{ \sum_{w = 1}^{W} \sum_{h = 1}^{H} \Lambda_{w, h} (x_{w, h}^n - C_{w, h} )(\tilde{x}_{w, h}^n - C_{w, h})}{\sqrt{ \sum_{w = 1}^{W} \sum_{h = 1}^{H} \Lambda_{w, h} (x_{w, h}^n - C_{w, h})^{2} \cdot  \sum_{w = 1}^{W} \sum_{h = 1}^{H} \Lambda_{w, h} (\tilde{x}_{w, h}^n - C_{w, h})^{2}}} .
\end{equation*}
%

A classical measure for ensemble skill is the continuous ranked probability score (CRPS). 
It is mathematically defined as
\begin{equation}
\mathrm{CRPS} = \int_{-\infty}^{+\infty} (F(x) - 1_{\{x \geq y\}})^{2} dx  
\end{equation}
and represents a quadratic measure of the difference between the forecast cumulative distribution function, $F(x)$, and an empirical observation represented by a delta function. 

The (latitude-weighted) spread-skill ratio (SSR) represents a measure of the correlation between the ensemble spread and the prediction error. 
It is the ratio between the ensemble spread and the RMSE, with the ensemble spread calculated as
\begin{equation*}
\mathrm{SSR} = \sqrt{\frac{ \sum_{w = 1}^{W} \sum_{h = 1}^{H} \Lambda_{w, h} \cdot \sigma^2_{w, h}}{W \cdot H}} 
\end{equation*}
and the RMSE as in Eq.~\ref{eq:RMSE}. Here $\sigma^2_{w,h}$ is the variance of the ensemble spread.
For a well-calibrated ensemble, a larger error on the prediction corresponds to a larger ensemble spread. 
The SSR is usually between $0$ and $1$ with values close to zero indicating an undersampling of the ensemble distribution while values above $1$ indicate that the ensemble spread is larger than the error on the prediction. 

To evaluate the bias-corrected  precipitation forecasts, we made use of common metrics for dichotomous events \cite{haiden_intercomparison_2012}.
 To separate dry and rainy events, a threshold of $0.1\,\mathrm{mm}/3\,\mathrm{h}$ for the precipitation rates was applied. Letting $a$, $b$, $c$ and $d$ denote the number of hits, false alarms, misses and correct negatives of the $2 \times 2$ contingency table, the equitable threat score (ETS), the Pierce skill score (PSS) and the frequency bias (FBI) are given by
\begin{align*}
\text{ETS} = \frac{1}{N}\sum_{n = 1}^{N}\frac{a_n - a_n^{ref}}{a_n-a_n^{ref}+b_n+c_n} &\text{~~~with~~~} a_n^{ref} = \frac{(a_n+b_n)(a_n+c_n)}{a_n+b_n+c_n+d_n} 
\\[6pt]
\text{PSS} = &  \frac{1}{N}\sum_{n = 1}^{N} \frac{a_n d_n - b_n c_n}{(a_n+c_n)(b_n+d_n)} 
\\[6pt]
\text{FBI} =&  \frac{1}{N}\sum_{n = 1}^{N} \frac{a_n+b_n}{a_n+c_n} .
\end{align*}
The ETS thereby constitutes a skill score formulation for which the performance of a random forecast in terms of the treat score serves as reference. The Peirce skill score is based on the proportion of correct events and takes an unbiased random forecast as reference ($a_n^{ref}$). By contrast, the FBI is not an accuracy measure itself, since it evaluates the marginal event frequency of the forecast and ground truth data. However, the FBI provides information on systematical biases of a forecasting system with $\mathrm{FBI} > 1$ ($\mathrm{FBI} < 1$) indicating over-confidence (under-confidence).

\subsection{Nowcasting}

The fine-tuning for the $6\,\mathrm{h}$ forecasting was performed for $4$ days, i.e. for approximately $8$ epochs where each of them used data from $32 \times 1$ months (distributed across the 32 nodes used in training).
The training task was a modified masked token model with always all tokens in the last two rows in the temporal dimension masked (and no randomness). Since each token has a width of $3\,\mathrm{h}$ this corresponds to training for $6\,\mathrm{h}$ lead time.
CRPS for specific humidity has not been reported since the Gaussian approximation for the ensemble distribution does not hold.
Other details can be found in the Methods section.

\subsection{Temporal interpolation}

\begin{figure}
    \centering
  \includegraphics[width=0.7\textwidth]{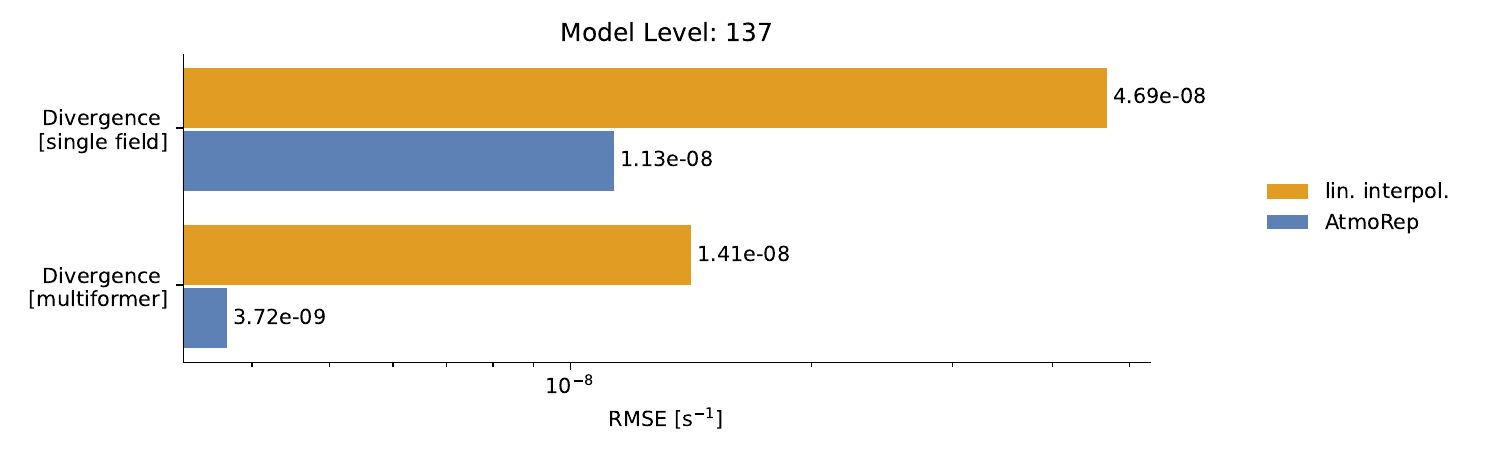}

   \vspace{0.2 cm}
   
  \includegraphics[trim={0cm 0cm 0cm 1cm},clip,width=0.7\textwidth]{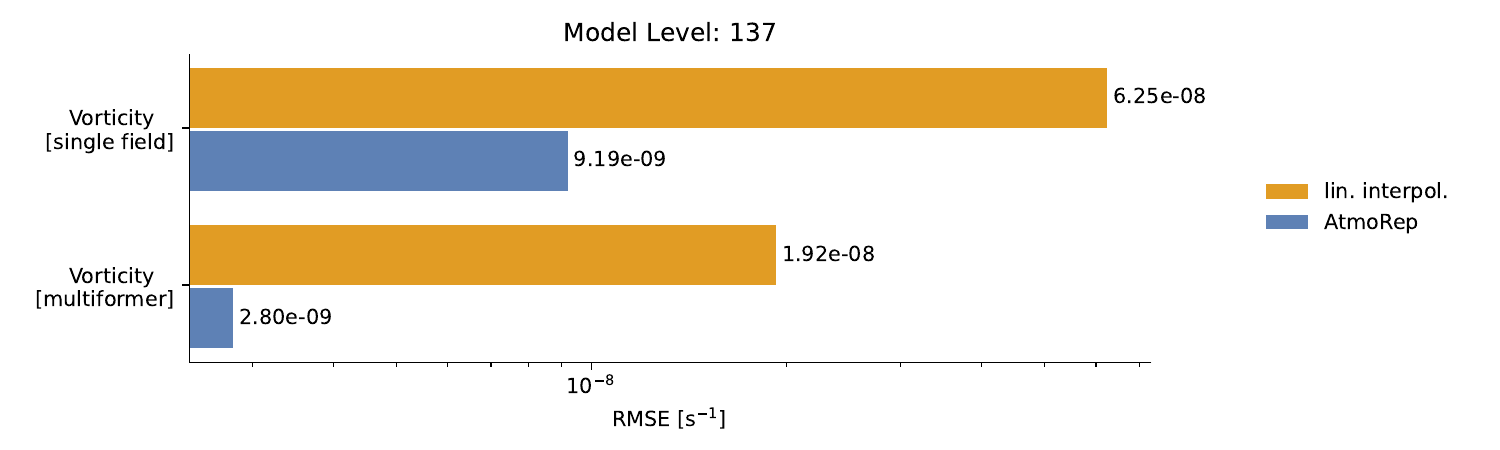}
  
  \vspace{0.2 cm}

 \includegraphics[width=0.7\textwidth]{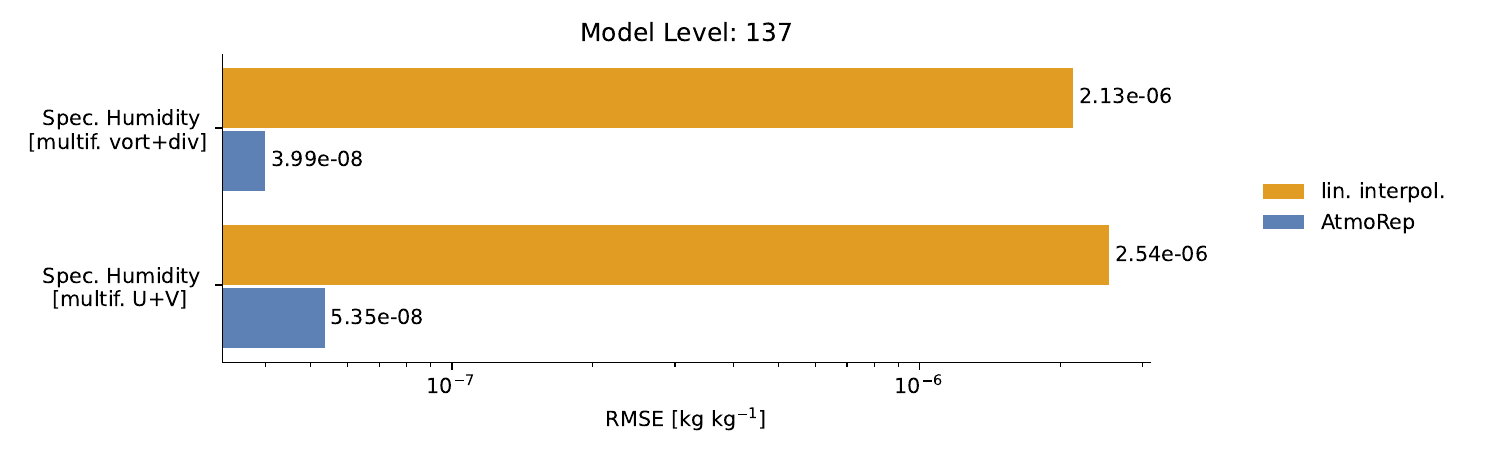}
    \caption{Average RMSE of temporal interpolation at model level 137. \textit{Top panel}: Results for vorticity and divergence at model level 137 with the respective single-field model and for the fields in the Multiformer (multi6-vd).
    As reference we also show the average RMSE obtained with linear interpolation.  
    \textit{Bottom panel:} Results for specific humidity for the two large multiformer configurations with wind components (multi6-uv) and with vorticity and divergence (multi6-vd). 
    }
    \label{fig:ext-temp-interpol}
\end{figure}

For temporal interpolation, we report some additional results in Fig.~\ref{fig:ext-temp-interpol}.
In particular, there we provide a comparison between the single-field models and results obtained with the two different $6$-field Multiformers.
The results show the clear advantages of the coupled models. 
By evaluating the skill for temporal interpolation for specific humidity, which is a common field in both large Multiformers, we can also compare the effect of either working with the wind components or with vorticity and divergence, i.e. with the different equivalent ways to represent the wind vector field.
A substantially lower RMSE can be observed for the configuration with vorticity and divergence. 
Experiments for other applications are, however, needed to establish a clear result.
The differences in the linear interpolation in the bottom plot in Fig.~\ref{fig:ext-temp-interpol}e can be attributed to differences in the random sampling for the experiment.

\subsection{Counterfactuals}
\label{sec:experimental:counterfactuals}

For the counterfactuals, we report results for vorticity since the ERA5 distribution has a relatively simple shape (in contrast, e.g., for temperature and specific humidity) and a clear shift between the early and late ERA5 years exists.
Furthermore, vorticity is rapidly changing in time so that a $3\,\mathrm{h}$ short-term forecasts is more likely to remove imprints from the initial conditions, e.g. compared to temperature that is changing much more slowly. 
Further details on the experimental setup can be found in Methods and in Fig.~5 in the Extended Data section.

%

\subsection{Extrapolation to 2022}

The experimental setup was very similar to that of the counterfactuals in Sec.~\ref{sec:experimental:counterfactuals}, see also Fig.~5 in the Extended Data section.
In particular, we used a large, random set of initial conditions from 2017 and evaluated short-term forecasts with these once with the correct external conditions $\alpha$ and once with modified $\hat{\alpha}$ where $\hat{\alpha}_y = 2022$. 
To avoid that results are biased by differences between ERA5 and model predictions, the reference distribution for $2022$ was also obtained through short-term forecasting with AtmoRep.

\subsection{Model correction}

ECMWF's operational Integrated Forecasting System (IFS) is a state-of-the-art numerical weather prediction model. 
A version from 2016 with a spectral resolution of T$639$ was used for the production of the ERA5 reanalysis. 
The operational IFS, in contrast, works with T$1220$, which corresponds to an approximately $4$-times finer resolution. 
The additional fine-scale spatial details in IFS data are not fully representable on the $721 \times 1440$ grid used for ERA5 but one still has a substantially higher effective signal resolution compared to ERA5, see the zoom-ins in Fig.~4 in the Extended Data section. 
Furthermore, the current IFS differs also in other respects (e.g. the parametrizations used) from those from 2016 so that next to the resolution also the overall data distribution is different.

With data from the IFS distribution as input to AtmoRep, it is a priori not clear that the model will produce meaningful predictions.  
While ``blow-up'' as in conventional models is not typical for AI-based ones, these can show other artefcats, e.g. regular patterns.
No such artifacts are observed in AtmoRep's predictions with IFS data as input. 
Furthermore, our model is able to partially preserve the higher frequency content, despite never having seen data with such a high resolution during training. 
The robustness of AtmoRep is broadly consistent with what has been observed for other large scale representation models, e.g. large language models, and a sufficiently large and diverse training dataset seems key to achieve these properties. 
Since AtmoRep is trained to produce data from the ERA5 distribution as output, it will do so also for IFS input. 
This leads to the distributional shifts towards ERA5 that are presented in Fig.~2 in Main and Fig.~5 in the Extended Data section.
We thereby used data from a single month only, namely February, since averaging over months led to a less robust shift in the background distribution.
For computing the distributions, we used AtmoRep with the masked token model while for the spectra in Fig.~1 in Main global forecasts where computed analogous to the nowcasting experiments. 
Spectra have been computed with the \texttt{spharm} python package.

\subsection{Downscaling}

For downscaling, we used the $3$-field multiformer configuration with the two wind velocity components and temperature as base model. 
As discussed in Main, the target field was $2\,\mathrm{m}$ temperature from the COSMO REA6 reanalysis.
As downscaling network we employed a transformer with six blocks and an embedding dimension of $2048$, i.e. twice the size as for temperature in the pre-trained model.
The larger embedding dimension was motivated by the much larger number of grid points in the COSMO REA6 target data set for each token (whose spatial extent was preserved). 
A linear layer was used to map the token for temperature in the AtmoRep core model to the larger embedding dimensions in the downscaling network. 
In the ensemble tail, we employed only $4$ members since each of them had a large number of parameters.
The wind components contributed to the downscaling only through the inter-field cross-attention in the encoder.
The overall network had about $1.8$ billion parameters and it was trained end-to-end for the training task, although fine-tuning of, e.g., only the decoder and the downscaling network is an interesting direction for future work.
The task was the direct prediction of the downscaled field, fixed at the center of the COSMO REA6 domain and with an extent determined by the $6 \times 12$ spatial tokens used during pre-training.
The loss was the same as during pre-training.
A noteworthy observation during training was that a small batch size was critical to obtain convergence.
Also, results improved considerably when we changed from a global to local data normalization for the $2\,\mathrm{m}$ temperature target field.

We used the recent work by Stengel et al.~\cite{stengel_adversarial_2020} as baseline for our downscaling experiments. 
We retrained the GAN used in this work for our domain and data, using only ERA5 temperature at model level $137$ as input. 
In contrast to Stengel et al., we also performed a single downscaling step since the super-resolution factor was smaller than in the original work. 


\subsection{Bias correction}

The RADKLIM dataset contains missing values. These are coherent regions at the boundary of the domain without any observations but also individual grid points within the domain at times. 
We treated these values by masking them for the loss computation. 
Despite this masking, the obtained predictions were spatially and temporally coherent, see Fig.~4 in Main.

For training we used an MSE loss together with a loss that provided a stronger emphasis on large values to better captures the tails of the highly skewed data distribution.
In particular, instead of squaring the per ensemble member difference between prediction and truth and the difference in the statistical loss, we took these values to the sixth power.
The training task was the direct prediction of the RADKLIM data from the ERA5 data. 
The $6$-field vorticity and divergence multiformer was used as base model and the entire model was fine-tuned. 

Data for the year 2018 was used as validation set and those from 2019 for test. 
This separation was necessary for bias correction since the loss used for training and evaluation scores were substantially different. 





\section{Additional experiments}

\subsection{Effect of using different domain sizes than during training}

As described in Main and Methods, AtmoRep's Multiformer is a highly flexible neural network architecture that allows one to change the number of tokens after training and through this to adapt the domain size as well as the number of vertical levels.
In Table ~\ref{tab:sup-multiformer-domain-size} we show the effect of changing the number of tokens in the input along different dimensions away from the $5$ vertical levels and $12 \times 6 \times 12$ tokens in time, latitude and longitude used during pre-training.
The results demonstrate that reducing the domain size leads to only a limited degradation in the performance, in particular along the vertical dimension.
A larger number of tokens, however, incurs a more substantial loss in skill.
Note that small amounts of fine-tuning allow one to substantially reduce the degradation with little additional computations.

\begin{table}
\begin{tabular}{c|c|c|c|c}
         field    &  $12 \times 6 \times 12$ tokens & $12 \times 6 \times 6$ tokens & $12 \times 6 \times 15$ tokens & $\textrm{ml}=[123, 137]$ \\
         \hline
         mean     &   0.166 &   0.210  &   0.375 &    0.166 \\[3pt]
           vor    &   0.091 &   0.125  &   0.300 &    0.096 \\[3pt]
           div    &   0.284 &   0.363  &   0.531 &    0.267 \\[3pt]
           w      &   0.260 &   0.326  &   0.568 &    0.184 \\[3pt]
           T      &   0.035 &   0.064  &   0.097 &    0.055 \\[3pt]
           q      &   0.034 &   0.040  &   0.170 &    0.054 \\[3pt]
           precip &   0.291 &   0.340  &   0.583 &    0.351 \\[3pt]
           \hline
    \end{tabular}
    \smallskip
    \caption{MSE test loss for $5$-field Multiformer with vorticity and divergence when used with inputs with a different number of tokens than during training. The first column is the reference that matches the training configuration followed by fewer and more tokens in longitude, respectively. In these three cases all five vertical levels were used.
    The last column shows results with $12 \times 6 \times 12$ tokens but only the lowest two vertical levels as input. }
    \label{tab:sup-multiformer-domain-size}
\end{table}

\subsection{Effect of coupling individually pre-trained transformers}

In Table~\ref{tab:sup-multiformer-coupling} we show the test error for the extended masked token model task used during training for individual per-field transformers and compare it to the performance when the fields are evaluated together in a Multiformer.
The results demonstrate that the model combines the information from the different fields effectively in a Multiformer and that this leads to a substantial improvement of the results.
Similar results are obtained for forecasting.

\begin{table}[]
\centering
\begin{tabular}[t]{l|c|c}
         field    &  Multiformer & single field   \\
         \hline
           vor    &    0.020 &    0.024 \\[3pt]
           div    &    0.074 &    0.076 \\[3pt]
           w      &    0.053 &    0.057 \\[3pt]
           T      &    0.020 &    0.029 \\[3pt]
           q      &    0.009 &    0.010 \\[3pt]
           precip &    0.116 &    0.142 \\[3pt]
           \hline
    \end{tabular}
    \smallskip
    \caption{MSE test loss for the coupled Multiformer configuration (multi6-vd) compared to the values attained with single-field models, all evaluated with the masked token task used in pre-training.}
    \label{tab:sup-multiformer-coupling}
\end{table}








\section{Directions for future work}
\label{sec:future}

There are several other potential applications of AtmoRep, some of which might be possible with the existing model and without task-specific training, i.e. they would be further intrinsic capabilities.
Others will require some, or even substantial fine-tuning and/or model extensions. 
To illustrate the generality of AtmoReps concept, we discuss a selected set of potential applications and their realization with AtmoRep below. 
Possible extensions of the AtmoRep model are also presented.

\paragraph{Medium range forecasting}

Several recent studies focused on medium-range weather forecasting and demonstrated that sufficiently large deep learning models can produce weather forecasts with skill comparable to state-of-the-art conventional models when trained on ERA5~\cite{Pathak2022,bi2023accurate,Lam2022,chen_fengwu_2023,chen_fuxi_2023}. 
While AtmoRep does not specifically target medium-range weather forecasting, the model can be easily extended to it.
More specifically, forecasting can be formulated as an auto-regressive problem
\begin{equation}
   p\big(x_N \vert x_0 \big) = \prod_{t_i = 1}^N p\big( x_{t_i} \vert x_{t_{i-1}} \big)
\end{equation}
where the $t_i$ form a sequence of discrete time steps.
Hence, by representing $p\big( x_{t_i} \vert x_{t_{i-1}} \big)$ with our numerical stochastic model $p_{\theta}( y \vert x , \alpha)$, i.e. $p\big( x_{t_i} \vert x_{t_{i-1}} \big) = p_{\theta}( x_{t_i} \vert x_{t_{i-1}} , \alpha)$, we can realize forecasting by iterating it.
This is known as roll-out in the machine learning literature.

For medium-range weather forecasting, AtmoRep's intrinsic ensemble provides an interesting avenue to develop an ensemble forecasting system.
This is critical to achieve skillful and practically applicable medium-range forecasts~\cite{Bauer2015,Palmer2019} although none of the existing large-scale AI-based forecasting models currently supports this natively.
A central question for ensemble forecasting is how to achieve coherent long-term preditions with a practical number of ensemble members and with a physical spread~\cite{Leutbecher2008}. 
In natural language processing, a similar challenge has been met using reinforcement learning~\cite{Ouyang2022,Christiano2023} to generate long, coherent model output.
The principal applicability of reinforcement learning for forecasting has already been demonstrated in~\cite{chen_fengwu_2023} and we therefore believe that reinforcement learning-based ensemble forecasting methods provide a promising direction for further exploration..

\paragraph{Error estimates through the ensemble}

\begin{figure}
  \centering
  \includegraphics[width=0.32\textwidth]{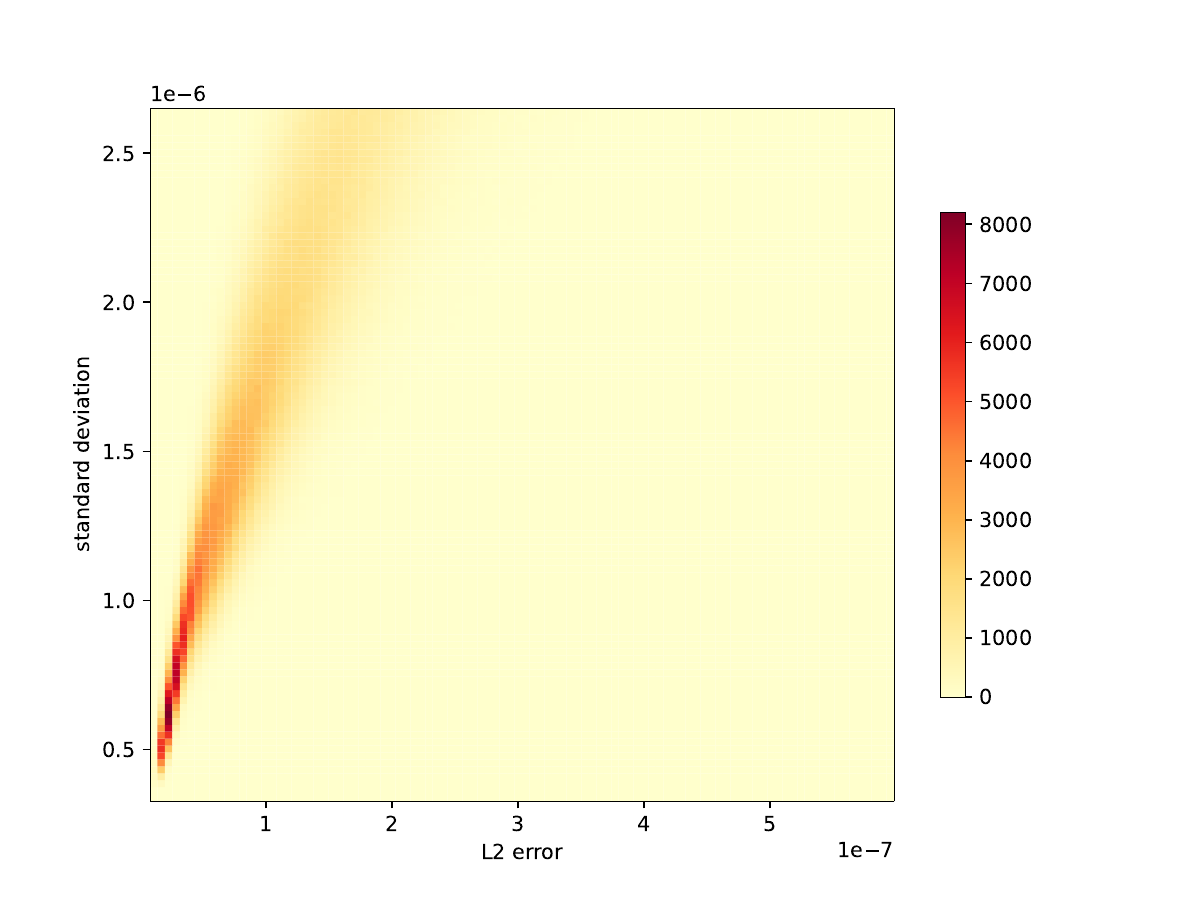}
  \includegraphics[width=0.32\textwidth]{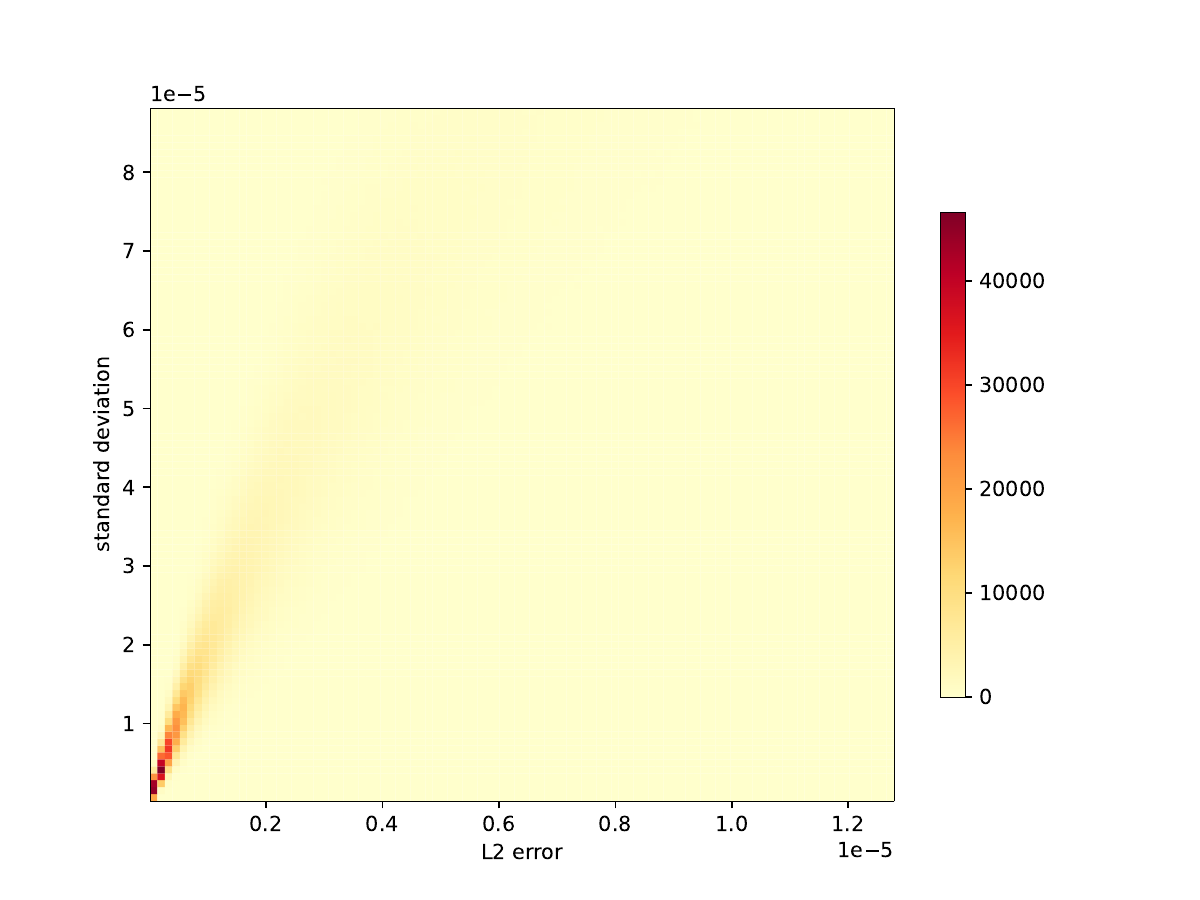}
  \includegraphics[width=0.32\textwidth]{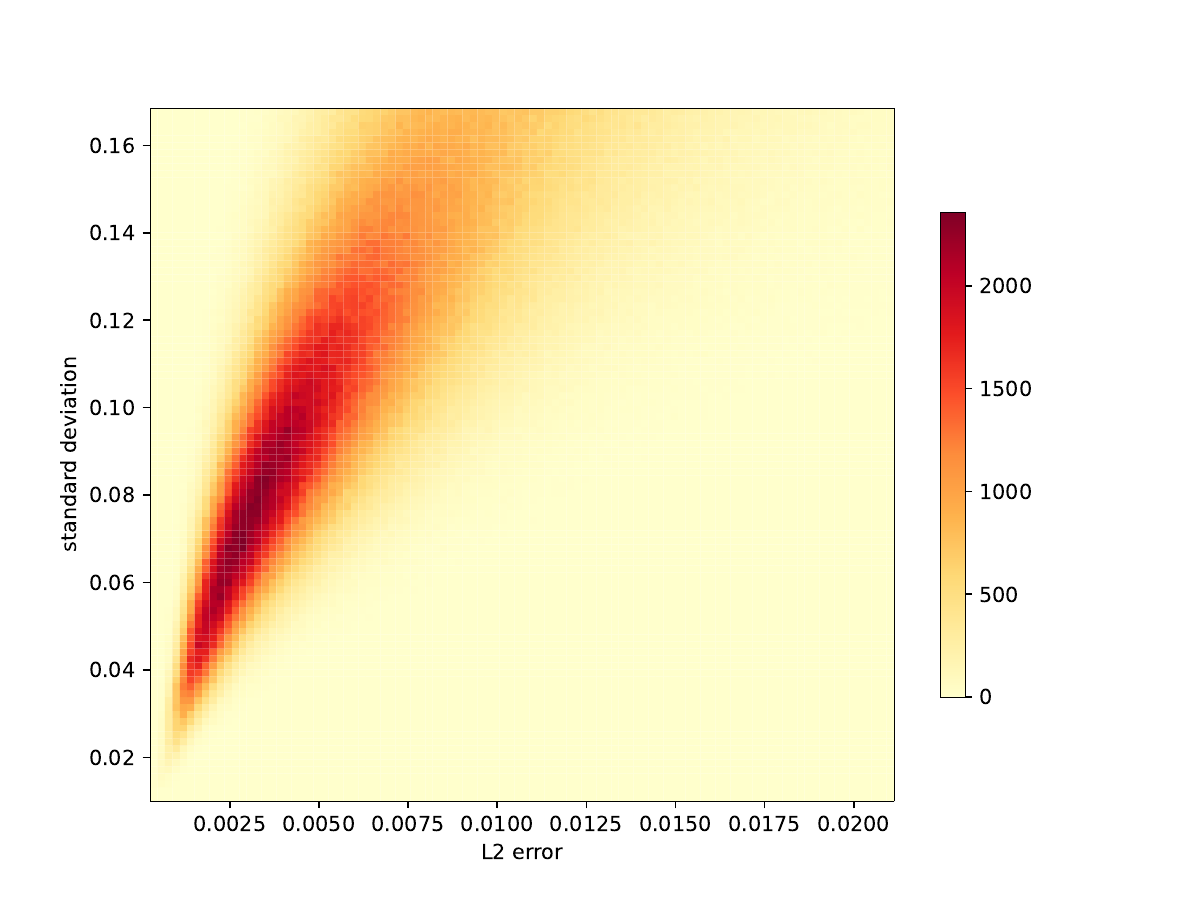}
  \caption{2D histograms showing the correlation between error and standard deviation for the AtmoRep ensemble for vorticity, specific humidity and meridional velocity (from left to right).}
  \label{fig:ensemble:err:stddev}
\end{figure}

A challenge with machine learning-based methods is often their unkown reliabilty, for example since many of the guarantees from classical numerical analysis are not available. 
Fig.~\ref{fig:ensemble:err:stddev} above as well as Figs.~2 and~3 in the Extended Data section show that for AtmoRep, the error in a prediction is strongly correlated to the standard deviation of the ensemble. 
In a practical application, the error is not available while the standard deviation is provided with any prediction.
We hence believe that the standard deviation could be used as a proxy for the error to provide, e.g., information on the reliability of AtmoRep predictions.

\paragraph{Missing field reconstruction}

For text and images, generative AI models can be used for conditional state generation, e.g. to obtain an image from a label or a descriptive text only~\cite{Ramesh2021-Dall-e,OpenAI2023}.
In the context of AtmoRep $p_{\theta}(y \vert x ; \alpha)$, analogous capabilities would allow to reconstruct a statistically consistent atmospheric state from just the external conditions $\alpha$, e.g. by specifying a date and a location.
In fact, AtmoRep's ability to reconstruct states with up to $90\%$ of the information being masked is already close to such a state reconstruction.
A special case of relevance would also be to reconstruct one physical field form other ones that are provided.  
Such a field reconstruction would be of great utility for model compression that is a major challenge for state-of-the-art weather and climate models.

%
%

\paragraph{Extensions of the AtmoRep model}

An important and exciting direction for future work is a further extension of the AtmoRep model itself.
Through the training on the ERA5 reanalysis, AtmoRep inherits the biases of the dataset, e.g.~\cite{Hersbach2020,foli_evaluation_2022,hoffmann_assessment_2022,choudhury_evaluation_2023}.
As we demonstrated with the bias correction in Main, observational data can be used to reduce these. 
Pursuing this at scale and with a wide range of data streams from the observational record could lead to a substantially improved model $p_{\theta}( y \vert x , \alpha)$.
In turn, this would likely also lead to substantial benefits in applications that build on AtmoRep.
A principal challenge is thereby how to train a model like AtmoRep with multiple potentially incoherent and often local data sources to obtain an effective improvement in the the learned representations.
The heterogeneity of the data used for training very large text and image models suggests, however, that this can be achieved.
A long term objective would be to continuously update AtmoRep, or a similar model, as new observations become available. 
Such an ingestion of the very large amounts of data (on the order of terabytes) that today's observational network produces would provide a means to process the data into a coherent and informationally rich form that is readily amenable for applications.

The current AtmoRep model work on the $721 \times 1440$ equi-angular grid that is the default for the ERA5 reanalysis. 
Due to the strong distortions at the pole, this is sub-optimal. 
Furthermore, having the ability to process data from different (grid) representations and at different resolutions would  provide a significant benefit for AtmoRep and its applicability.
This is challenging in particular at very high-resolutions, e.g. with km-scale data, due to the quadratic scaling in the dense attention that is currently used with AtmoRep.
Likely, a multi-resolution approach is required for a model that can flexible handle data up to this resolution.
Working with different grids is, in our opinion, significantly easier and can be achieved with representation-specific embedding networks and a suitable training protocol.
This would likely also be required for the training on observations where a remapping to a standard grid can already introduce substantial artifacts.

A question that is important to all of the above is the size of the AtmoRep model and how it has to scale with the volume of the training data set. 
A first important aspect in this context would be the required model size to absorb all the information in ERA5 into AtmoRep and how this depends on the accuracy required in the output.


\section{Data}

In this section we provide more detailed information on the most important dataset used in our work.

\subsection{Data normalization}

In our work, we used both global and local data normalization, in each case with the aim to map the data for a physical field to a standardized form with zero mean and unit variance.
For global normalization, the mean and variance over the entire per-field data set per month was computed, irrespective of the spatial location, and the data was normalized using these values. 
This removed, e.g., seasonal variations from the data. 
For the local normalization, mean and variance were computed independently per grid point in the data but irrespective of the time.
Global normalization worked well for vorticity and divergence but when used for example with temperature one obtained predictions with biases that did not improve with further training. 
The use of the local normalization alleviated the problem substantially. 
We believe that the different behavior results from the different frequency characteristics of the physical fields, see Fig.~\ref{fig:data:spectra_levels}, but more work is required to obtain a fuller understanding of the observations.

\subsection{ERA5}

ERA5 is the fifth generation global atmospheric reanalysis produced by the European Centre for Medium Range Weather Forecasting (ECMWF)~\cite{Hersbach2020}. It provides the most consistent and coherent global estimate for the state of the atmosphere, land surface, and ocean waves that is currently available. 
ERA5 is based on ECMWF's operational Integrated Forecast System (IFS) model version CY41R2. This was the operational weather model of ECMWF in 2016.
Observations are assimilated with a hybrid incremental 4D-var system into the atmospheric state that is represented on a reduced Gaussian grid with an approximate grid spacing of $31\,$km on 137 model levels for the high resolution realisation (HRES). The Copernicus Climate Data Store service provides the ERA5 reanalysis data on a regular (lat,lon)-grid with $0.25^\circ$ resolution. 
In addition to the analysis product, short-range forecasts with a lead time of 18 hours are initialized twice a day at 06 and 18 UTC. These short-range forecasts provide estimates of surface fluxes, accumulated precipitation, and other subgrid-scale variables.
ERA5 data ranges from the year 1940 to almost real time.
For AtmoRep, only data after 1979 was used because these are better constrained from the inclusion of satellite observations.

ERA5 has a comprehensive documentation~\cite{Hersbach2020} and has been extensively evaluated in the scientific literature, for example~\cite{foli_evaluation_2022,hoffmann_assessment_2022,choudhury_evaluation_2023,gomis-cebolla_evaluation_2023,hou_evaluation_2023,jiao_evaluation_2021,jourdier_evaluation_2020,lan_evaluation_2023,tetzner_validation_2019,varga_evaluation_2023}. While mean statistics of several atmospheric variables are generally captured well, some biases have also been noted. For example, \cite{choudhury_evaluation_2023} found an underestimation of daily maximum temperatures and an overestimation of daily minimum temperatures over Australia as well as an underestimation of the decadal warning trend. With respect to precipitation, ERA5 has a tendency to overestimate low and medium rainfall, while it underestimates heavy precipitation~\cite{gomis-cebolla_evaluation_2023}.

As described in~\cite{Hersbach2020}, the quality of the ERA5's data assimilation varies over time because of the changing global observations that are available. In combination with certain model errors (ERA5 uses an IFS model version from 2016), biases may vary over time. It is noteworthy for assessing the quality of AtmoRep and other deep learning models that have been trained on ERA5 data that improvements in model physics, resolution and in the data assimilation scheme of ECMWF result in better quality atmospheric state analyses in recent IFS analyses and forecasts compared to ERA5 results, i.e. after 2019~\cite{ben-bouallegue_rise_2023}.

\subsection{COSMO REA6}

The COSMO REA6 constitutes a reanalysis dataset of the atmospheric state for the European CORDEX-domain~\citep{Bollmeyer2015}. 
The smaller spatial domain allows for much higher spatial resolution of the reanalysis product. The COSMO REA6 grid spacing $\Delta x_{CREA6} \simeq 6\,$km is about five times finer than the ERA5-grid ($\Delta x_{ERA5} \simeq 31\,$km ). The regional NWP model COSMO v4.25 \citep{Baldauf2011}, operationally deployed from September to December 2012 at the German Weather Service DWD, was used to generate the dataset. The initial and boundary conditions were provided by the ERA-Interim reanalysis, the predecessor of ERA5. The COSMO REA6 reanalysis data is available between 1995 and August 2019. 
A continuous nudging scheme (as opposed to the incremental 4D-Var in ERA5) is used for data assimilation which relaxes the model's prognostic variables towards observations over a predefined time window. 
Due to the rotated pole grid of the COSMO REA6 data, a first-order conservative remapping was applied to align the data with the unrotated ERA5 grid. Specifically, we used the Climate Data Operators (CDO) software \citep{Schulzweida2020} to reproject the COSMO data from the rotated pole grid with $\Delta x_{CREA6}^{rot} = 0.055^{\circ}$ onto an unrotated (lat,lon)-grid with $\Delta x_{CREA6} = 0.0625^\circ$. While the horizontal wind vector components  $(u,v)$ are directly processed, the 2m temperature $T_{2m}$ is expressed in terms of the dry static energy $s = c_{pd}T + g(z_{sfc} + 2m)$ to account for the change in the surface topography $z_{sfc}$ during remapping. 

ERA5 and COSMO REA6 have been compared in a few studies. For example, \cite{urraca_evaluation_2018} investigated irradiance estimates from both products and found an overestimation in ERA5 and an underestimation of COSMO REA6. Added value of the COSMO REA6 reanalysis product have been diagnosed mainly over mountainous regions such as the Alps due to the increased spatial resolution, see~\cite{Kaspar2020} for a comprehensive evaluation. Notable examples in this context constitute the 2m temperature \citep{Scherrer2020} and the near-surface wind \citep{jourdier_evaluation_2020, Kaspar2020} confirming the suitability of COSMO REA6 as a target dataset for statistical downscaling. 

\subsection{RADKLIM}

The RADKLIM dataset provides precipitation observations derived from the German radar network operated since 2001 by the German Weather Service DWD. The reflectivity data measured at 17 ground-based C-Band radar stations of the network is converted to precipitation rates based on the well-known Z-R relation \citep{Marshall1947}. A sophisticated calibration procedure based on rain gauge observations is deployed to remove systematic observations errors, cf.~\cite{Wagner2018}. Initially, the RADar OnLine AdjustmeNt (RADOLAN) procedure~\citep{Bartels2004} applies several corrections by eliminating clutter pixels due to backscattered signal of non-meteorological targets (e.g. insects, buildings or other solid objects), reducing noise with gradient filters and compensating for orographic shading effects. The RADOLAN-procedure is complemented by further processing techniques to improve correction of clutter artifacts and to account for signal reduction with increasing distance and altitude of the radar beam \citep{winterrath_overview_2019}. Evaluations of RADKLIM have concluded that it provides a reliable dataset for gridded precipitation \citep{Kreklow2020}.
In this study, we make use of the so-called YW-product which provides precipitation rates at a temporal resolution of 5 minutes. The precipitation rates are then accumulated to a 3h-precipitation product. 
Similar to the COSMO REA6-data, first-order conservative remapping with CDO was performed to map the RADKLIM data from a polar-stereographic grid with $\Delta x_{RDK}^{polar} = 1\,$km onto the regular, spherical $0.25^\circ$ grid of the ERA5-data.


\newpage
\bibliography{bibliography}


\end{document}